\documentclass[onecolumn]{aa} 

\usepackage{geometry}
\usepackage{pdflscape}
\usepackage{graphicx}
\usepackage[fleqn]{amsmath}
\usepackage{txfonts}
\newcommand{\tco}{$^{13}$CO}
\newcommand{\co}{$^{12}$CO}
\newcommand{\ceo}{C$^{18}$O}
\newcommand{\tcoone}{$^{13}$CO $J$=1$-$0}
\newcommand{\coone}{$^{12}$CO $J$=1$-$0}

\newcommand{\cotwo}{$^{12}$CO $J$=2$-$1}
\newcommand{\tcotwo}{$^{13}$CO $J$=2$-$1}
\newcommand{\ceotwo}{C$^{18}$O $J$=2$-$1}

\newcommand{\cothree}{$^{12}$CO $J$=3$-$2}

\newcommand{\hco}{HCO$^+$}
\newcommand{\hcoone}{HCO$^+$~$J$~=~1$-$0}

\newcommand{\hcothree}{HCO$^+$~$J$~=~3$-$2}
\newcommand{\hcofour}{HCO$^+$~$J$~=~4$-$3}

\newcommand{\hcn}{HCN}
\newcommand{\hcnone}{HCN~$J$~=~1$-$0}

\newcommand{\cstwo}{CS~$J$~=~2$-$1}

\newcommand{\csfour}{CS~$J$~=~4$-$3}
\newcommand{\csfive}{CS~$J$~=~5$-$4}

\newcommand{\cth}{C$_{\rm{2}}$H}
\newcommand{\cthone}{C$_{\rm{2}}$H~$N$~=~1$-$0}

\newcommand{\siotwo}{SiO~$J$~=~2$-$1}

\newcommand{\chcnsix}{CH$_{\rm{3}}$CN(6-5)}
\newcommand{\chcnten}{CH$_{\rm{3}}$CN(10-9)}

\newcommand{\hncofive}{HNCO~$J_{k,k'}$~=~5$_{0,4}$~-~4$_{0,4}$}

\newcommand{\tc}{\element[][13]{C}}
\newcommand{\car}{\element[][12]{C}}
\newcommand{\eo}{\element[][18]{O}}

\newcommand{\lfir}{$L_{\rm{FIR}}$}
\newcommand{\lir}{$L_{\rm{IR}}$}

\newcommand{\lsol}{$L_{\sun}$}
\newcommand{\msol}{$M_{\sun}$}
\newcommand{\mdyn}{$M_{\rm{dyn}}$}

\newcommand{\mmol}{$M_{\rm{H_{2}}}$}
\newcommand{\kms}{km s$^{-1}$}
\newcommand{\tkin}{$T_{\rm{kin}}$} 
\newcommand{\nhtwo}{$n_{\rm{H_{2}}}$}
\newcommand{\alphacou}{$M_{\odot}$ (K km s$^{-1}$ pc$^{2}$)$^{-1}$}
\newcommand{\alphaco}{$\alpha_{\rm{CO}}$}

\newcommand{\nco}{$N_{\rm{^{12}CO}}$}
\newcommand{\ff}{$\Phi_{\rm{A}}$}
\newcommand{\nx}{$N_{\rm{X}}$}
\newcommand{\xco}{[$^{12}$CO]/[$^{13}$CO]}
\newcommand{\xceo}{[$^{12}$CO]/[C$^{18}$O]}

\newcommand{\arc}{$^{\prime\prime}$}

\begin{document} 

\title{PdBI U/LIRG Survey (PULS):  Dense Molecular Gas in Arp 220 and NGC 6240}
\titlerunning{Dense Molecular Gas in Arp 220 and NGC 6240}
 \author{Kazimierz Sliwa
 \inst{1}
\and
Dennis Downes\inst{2}\fnmsep
          }

   \institute{Max Planck Institute for Astronomy, K\"onigstuhl 17, D-69117 Heidelberg, Germany\\
              \email{sliwa@mpia-hd.mpg.de}
         \and
             Institut de Radio Astronomie Millimetrique, Domaine Universitaire, 38406 St. Martin d'H\`eres, France \\
             \email{downes@iram.fr}
             }

   \date{}

 
  \abstract
   {}
   {We present new IRAM Plateau de Bure Interferometer observations of Arp~220 in HCN, HCO$^{+}$, HN$^{13}$C $J$=1--0, C$_{2}$H $N$=1--0, \siotwo, \hncofive, \chcnsix, CS~$J$=2--1 and 5--4 and \tcoone\ and 2--1 and of NGC~6240 in  HCN, \hcoone\ and \cthone. In addition, we present Atacama Large Millimeter/submillmeter Array science verification observations of Arp 220 in \csfour\ and \chcnten. Various lines are used to analyse the physical conditions of the molecular gas including the \xco\ and \xceo\ abundance ratios. These observations will be made available to the public.}
   {We create brightness temperature line ratio maps to present the different physical conditions across Arp 220 and NGC 6240. In addition, we use the radiative transfer code RADEX and a Monte Carlo Markov Chain likelihood code to model the \co, \tco\ and \ceo\ lines of Arp 220 at $\sim$2\arcsec\ ($\sim$700~pc) scales, where the \co\ and \ceo\ measurements were obtained from literature.}
   {Line ratios of optically thick lines such as \co\ show smoothly varying ratios while the line ratios of optically thin lines such as \tco\ show a east-west gradient across Arp 220. The HCN/HCO$^{+}$ line ratio differs between Arp 220 and NGC 6240, where Arp 220 has line ratios above 2 and NGC 6240 below 1. 
   The radiative transfer analysis solution is consistent with a warm ($\sim$40~K), moderately dense ($\sim$ 10$^{3.4}$~cm$^{-3}$) molecular gas component averaged over the two nuclei. We find \xco\ and \xceo\ abundance ratios of $\sim$ 90 for both. The abundance enhancement of \ceo\ can be explained by stellar nucleosynthesis enrichment of the interstellar medium. }
   {}

   \keywords{Galaxies: interactions - Galaxies: ISM -  Galaxies: starburst - Galaxies: individual: Arp 220, NGC 6240 -  Submillimeter: ISM - ISM: molecules
               }

   \maketitle
%

\section{Introduction}
Ever since the discovery of a class of objects with extremely bright infrared emission called Ultra/Luminous Infrared Galaxies (LIRG: \lfir\ = 10$^{11-12}$\lsol; ULIRG: \lfir\ $>$10$^{12}$\lsol) with the \textit{The Infrared Astronomical Satellite} \citep[$IRAS$; e.g.][]{Houck1984,Houck1985,Soifer1984b,Soifer1984}, these objects have been studied at every wavelength. High-resolution optical imaging show that most of these systems resemble merging/interacting systems \citep[e.g.][]{Murphy1996}. \cite{Larson2016} showed that major mergers play a significant role for all sources with \lir\ $\geq$ 10$^{11.5}$\lsol. In addition, optical wavelengths show that these systems are highly dust obscured. Various star formation tracers have shown extreme star formation rates \citep[e.g.][]{U2012} suggesting that local U/LIRGs are great laboratories to study extreme modes of star formation. Mid-infrared observations are used to reveal the contribution of an active galactic nucleus (AGN), if present, to the far infrared luminosity \citep[e.g.][]{Genzel1998,Veilleux2009}. Molecular gas observations via carbon monoxide (\co) have revealed rich concentrations of fuel for current and future star formation \citep[e.g.][]{Downes1998,Bryant1999,Wilson2008} and also massive, energetic molecular outflows on kiloparsec scales in several sources \citep[e.g.][and references therein]{Cicone2014}. 
 
\citet{Gao2004b} used \hcnone\ observations to show that there exists a tight correlation between infrared luminosity (\lir; proxy for star formation rate) and \hcnone\ (proxy for amount of dense molecular gas). This tight relation was extended to span over 7-8 orders of magnitude of \lir\ by \citet{Wu2005}. These results are interpreted to mean that \hcnone\ is tracing the molecular gas component that is directly related to the star formation.

Single dish observations of the \co\ isotopologue, \tco\ \citep[e.g.][]{Casoli1992,Garay1993,Aalto1991,Aalto1995,Greve2009,Papadopoulos2012a} in U/LIRGs have shown a trend of extremely weak emission relative to \co. The integrated brightness temperature line ratios of \co/\tco\ in U/LIRGs were found to be usually high ($>$20) when compared to normal disk galaxies suggesting different interstellar medium (ISM) environments between the two classes of galaxies. High-resolution observations \citep[e.g][]{Aalto1997,Casoli1999,Sliwa2012,Sliwa2013,Sliwa2014} show a similar trend with values evening exceeding 50 for some sources \citep{Sliwa2017c}. Possible explanations for the unusually high ratios include, photodissociation of \tco, excitation and/or optical depth effects, or abundance variations \citep[e.g][]{Casoli1992,Henkel1993,Taniguchi1999}. While photodissociation is likely ruled out as the dominant process due to the strong \ceo\ emission \citep[see][]{Casoli1992}, recent radiative transfer studies are showing that the \xco\footnote{Square brackets denote abundances while no brackets around ratios denote brightness temperature line ratios unless specifically stated} abundance ratio is higher ($\geq$100) than what we perceive as normal \citep{Sliwa2013,Sliwa2014,Sliwa2017a,Papadopoulos2014,Henkel2014,Tunnard2015b}. In this paper, we model the molecular gas of the well studied ULIRG Arp 220 to determine whether it follows the \xco\ trend seen in the literature.

\textit{Arp 220:} \object{Arp 220} (IRAS 15327+2340, UGC 9913, VV 540, IC 1127) is the nearest example of a ULIRG and thus well studied at every wavelength. In this advanced merger, the two nuclei are still distinguishable, separated by 1\arc\ \citep[$\sim$390~pc;][]{Norris1988} and is modelled to finish merging within 6~$\times$~10$^{8}$~years \citep{Konig2012}. With one of the most powerful star-forming environments in the local universe, it is a popular starburst template for dusty, star forming high redshift galaxies. Recently, \citet{Barcos-Munoz2015} observed 33~GHz continuum at very high-resolution ($\sim$30~pc) revealing that synchrotron radiation is dominant at this frequency. Combining the sizes measured from the 33~GHz continuum and infrared observations, \citet{Barcos-Munoz2015} derived very high molecular gas surface densities ($>$2~$\times$~10$^{5}$~\msol~pc$^{-2}$)  and infrared surface luminosities ($>$4~$\times$~10$^{13}$~\lsol~pc$^{-2}$). Although there is no clear evidence of an AGN with the 33~GHz continuum \citep{Barcos-Munoz2015}, \citet{Downes2007} showed that the dust in the western nucleus is hot ($\sim$ 170~K) and the size of the dust source is small that it implies a large surface luminosity that can only be plausible by an AGN. \citet{Wilson2014} used 691~GHz continuum to also show that the western nucleus has a very high luminosity surface density that requires either the presence  of an AGN or a "hot starburst". \citet{Lockhart2016} were able to resolve previously observed H$\alpha$+[NII] emission into a bubble-shaped feature that is aligned with the western nucleus. Either an AGN or extreme star formation within the inner $\sim$100~pc of the nuclei are the likely possibilities for the origin of this bubble. \cite{Zsch2016} confirm an outflow from the western nucleus by comparing their high-resolution Very Large Array (VLA) data of several molecular species to the ALMA data.

Over the last two decades, Arp 220 has been observed in high-resolution CO many times \citep{Scoville1997,Downes1998,Sakamoto1999,Downes2007,Sakamoto2008,Matsushita2009,Martin2011,Konig2012,Rangwala2015,Scoville2017}. The observations reveal a large concentration of molecular gas ($\sim$~5~$\times$~10$^{9}$~\msol; e.g. Downes $\&$ Solomon 1998) within two compact nuclei surrounded by a diffuse kiloparsec-scale disk. Observations of rare CO isotopologues (i.e. \tco\ and \ceo) have been mainly obtained using single dish telescopes resulting in global fluxes \citep[e.g.][]{Greve2009,Papadopoulos2012a}; however, \citet{Matsushita2009} published high-resolution Submillimeter Array (SMA) \tco\ and \ceotwo\ ($\sim$ 3.5\arcsec) observations showing that the two isotopologues have similar intensities where they suggest that either the lines are optically thick or there is an overabundance of \ceo\ compared to \tco.  

Using $Herschel$ Fourier Transform Spectrometer (FTS) spectra, \citet{Rangwala2011} modelled the global \co\ emission from $J$ =1-0 to $J$ = 13-12 with a 2-component molecular gas: cold, moderately dense (\tkin\ = 50~K and \nhtwo\ = 10$^{2.8}$~cm$^{-3}$) and warm, dense (\tkin\ = 1350 K and \nhtwo\ = 10$^{3.2}$ cm$^{-3}$) molecular gas components. The dust continuum from the FTS spectrum was also shown to be consistent with a warm dust ($T_{\rm{dust}}$ = 66~K) with a large optical depth ($\tau_{\rm{dust}}$ $\sim$ 5 at 100~$\mu$m; Rangwala et al. 2011). 

Dense gas tracers such as HCN, HCO$^{+}$, HNC and CS have also been observed in Arp 220 \citep[e.g.][Barcos-Mu\~noz et al. in preparation]{Aalto2009,Aalto2015b,Greve2009,Sakamoto2009,Imanishi2010,Scoville2015,Martin2011,Martin2016,Tunnard2015a}. \cite{Sakamoto2009} reported P-Cygni profiles in the \hcothree\ and $J$~=~3-2 suggesting a $\sim$100~\kms\ outflow originating from the inner regions of the nuclear disks. \cite{Aalto2009} found that HNC $J$~=~3-2 is bright and, perhaps, amplified in the western nucleus and weak in the east suggesting very different physical conditions. \cite{Martin2016} find that HCN and \hcofour\ and $J$~=~3-2 are optically thick and affected by different absorption systems that can hide up to 70$\%$ of the total intrinsic emission from these lines. 

\textit{NGC 6240:} \object{NGC 6240} (IRAS 16504+0228, UGC 10592, VV 617, IC 4625) is unusual among the class of LIRGs in having exceptionally strong emission in the near and mid-IR lines of molecular hydrogen. The consensus of the infrared observers is that the strong H$_{\rm{2}}$ lines are due to shocked gas, not star formation \citep{Rieke1985,Depoy1986,Lester1988,Herbst1990,Elston1990,vanderWerf1993,Mori2014}. Since the cooling time of this shocked gas is short ($\sim$10$^{7}$~yr), we may be seeing NGC 6240 at a privileged moment during the merger process \citep[e.g.][]{Sugai1997}. Recent work of \citet{Meijerink2013} showed that the \co\ spectral line energy distribution (SLED), obtained using the $Herschel$ FTS, is consistent with either an X-ray dominated region (XDR) or with shocked molecular gas; however, the lack of the ionic species OH$^{+}$ and H$_{2}$O$^{+}$, normally found in high abundance in gas clouds near elevated X-ray or cosmic ray fluxes, ruled out the XDR models and once again, shocks are the most likely scenario to explain the observations.

The near and mid-infrared H$_{\rm{2}}$ emission peaks between the two nuclei (e.g., Lester et al. 1988; Herbst et al. 1990), suggesting several possible scenarios. Two of the more recent ideas are those of \citet{Ohyama2003} and \citet{Nakanishi2005}. The preferred model of Ohyama et al. (2003) is that the merger's tidal forces have channeled the molecular gas into the space between the two nuclei, where it is being shocked by a superwind from the southern nucleus, the stronger source in radio continuum, X-rays, and near- and mid-IR. Nakanishi et al. (2005) propose instead that the two nuclei have had a nearly head-on collision, after which the two supermassive black holes, the older bulge stars, and the recent starburst stars all moved onwards to the present sites of the two nuclei. Due to drag and stripping forces, however, the circumnuclear gas was left behind at the original collision site. 
This resembles the well-known Bullet Cluster scenario, except that collision speeds are lower, and NGC 6240's gas densities are 10$^{7}$ times higher than those of the Bullet Cluster, so the gas left behind in the middle is cool molecular gas rather than hot X-ray-emitting plasma.

NGC 6240 has also been observed in high-resolution \co\ many times \citep{Wang1991,Bryant1999,Tacconi1999,Nakanishi2005,Iono2007,Wilson2008,Feruglio2013a,Feruglio2013b}. The CO emission is peaked in between the two nuclei as is observed for the warm H$_{2}$ emission. \cite{Feruglio2013b} detected blueshifted CO emission between -200~\kms\ and -500~\kms\ peaking near the southern AGN position at the same position where an H$_{2}$ outflow was found \citep{Ohyama2000,Ohyama2003}.  A redshifted CO component peaks in between the two nuclei, similar to the CO emission at the systemic velocity, with a large velocity dispersion ($\sim$500~\kms\ at the maximum) suggesting highly turbulent gas \citep{Feruglio2013b}. 

In addition to CO, dense gas tracers have also been observed, however, not to the extent as that for Arp 220. \cite{Nakanishi2005} observed HCN and \hcoone\ at 2-3\arcsec, where both lines peak in between the two nuclei, similar to CO. \cite{Wilson2008} observed part of the \hcofour\ line, where it still peaks in between the two nuclei. \cite{Tunnard2015b} observed the rarer $^{13}$C isotopologues of HCN and \hco\ in NGC 6240 and found line ratios $>$ 30. 
\begin{table*}
\caption{Source Summary}\label{tab:sourcesum}
\centering
\begin{tabular}{lccccccc}
\hline \hline
		&\multicolumn{2}{c}{\underline{(0,0) Position}} 		&& \multicolumn{2}{c}{\underline{Center Velocity}} & & \\
Source	& RA 	& Dec	&	\lfir\ &	cz$_{lsr}$	&  z$_{lsr}$	&D$_{L}$	 & Linear Scale \\
		& (J2000)	& (J2000)	& (\lsol)	&(\kms)		&			& (Mpc)	& (pc arcsec$^{-1}$) \\
\hline
Arp 220		&15 34 57.24	&+23 30 11.2 	 &1.4 $\times$ 10$^{12}$	&5434	&0.018126	&81.3	&	390 \\
NGC 6240	&16 52 58.89	&+02 24 03.7	 &5.4 $\times$ 10$^{11}$	&7340	&0.02448	&108		&	520 \\
\hline
\end{tabular}
\tablefoot{[1] 9-year WMAP parameters \citep{Hinshaw2012}: $H_{o}$ = 69.3, $\Omega_{\rm{matter}}$ = 0.28, $\Omega_{\rm{vacuum}}$ = 0.72.[2] \lfir\ reference: \cite{Sanders2003}. [3] The (0,0) position is the phase center of the observations.}
\end{table*}

In this paper, we present new Plateau de Bure Interferometer (PdBI) observations of \hcn\ and \hcoone\ and \cthone\ for both Arp 220 and NGC 6240 (Table \ref{tab:sourcesum}). In addition, we present new PdBI observations of \tcoone\ and $J$ = 2-1, \cstwo\ and $J$ = 5-4, \hncofive, \chcnsix, SiO $J$ = 2-1, and HN$^{13}$C $J$ = 1-0 and Atacama Large Millimeter/submillimeter Array (ALMA) Science Verification (SV) observations of \csfour\ and \chcnten\ for Arp 220. These observations will be made available to the public. The paper is broken down into the following sections: In Section 2, we describe the observations and reduction. In Section 3, we present integrated brightness temperature line ratio maps to show the varying conditions of the ISM across these sources. In Section 4, we present a radiative transfer analysis of \co, \tco\ and \ceo\ for Arp 220 at $\sim$700~pc scales. In Section 5, we discuss the molecules detected in both sources, the results of the radiative transfer analysis of Arp 220, the \xco\  and \xceo\ abundance ratios found in Arp 220 and a comparison of the HCN/HCO$^{+}$ line ratios. In Section 6, we summarize the major results. We end the paper with an appendix describing the release of the data online and the continuum from the observations.

\section{Observations} 
\subsection{PdBI}
For the line-ratio investigations in this paper, we used
\coone\ and $J$~=~2--1
PdBI data sets on Arp~220 and NGC~6240, that were  
partly data from earlier publications (for Arp 220: Downes \& Solomon 1998;
Downes \& Eckart 2007; K\"onig et al.\ 2012; and for NGC~6240: 
Taconni et al.\ 1999). We also use the SMA observations of \cothree\ in Arp~220 published in \cite{Sakamoto2008} and the PdBI observations of \coone\ in NGC~6240 published in \cite{Feruglio2013a}.

Other, previously unpublished, data sets are, for Arp~220, 
\tcoone\ at 3\,mm and \tcotwo\ at 1.4\,mm observed simultaneously,  
and also \cstwo\ at 3\,mm and \csfive\ at 1\,mm observed simultaneously.
Additional data sets included
\hcn, HN$^{13}$C, \hcoone, \cthone,
and SiO $J$~=~2--1 (v=0) all observed simultaneously.

For most of these observations, the six 15\,m antennas
were arranged with spacings from 24\,m to 400\,m.  The longer baselines,
observed in winter, had phase errors $\leq 40^\circ$ at 1.4\,mm, and
$\leq 15^\circ$ at 3\,mm.  Short spacings ($\leq 80$\,m), observed in
summer at 3\,mm, had r.m.s.\ phase errors $\leq 20^\circ$.

The SIS receiver noise plus spillover
and sky noise gave typical equivalent system temperatures outside the
atmosphere of 150\,K at 3\,mm (86\,GHz) in the lower sideband, and 250
to 400\,K at 1.3\,mm in upper and lower sidebands separated by 3\,GHz,
with the upper band typically at 215 to 225\,GHz.  The spectral
correlators covered 1700\,\kms \ at 3\,mm and 800\,\kms \ at 1.4\,mm,
with instrumental resolutions of 8 and 4\,\kms , respectively.

These raw data were then smoothed to channels of 10, 20, and 40\,\kms .  
The primary amplitude calibrators were 3C273 (a variable source, but
typically 18\,Jy at 3\,mm and 13\,Jy at 1.4\,mm, at the 
time of the observations), and MWC349
(a mostly non-varying source, 
with 1.0 and 1.7\,Jy at 3 and 1.4\,mm during the years of
the observations). The uncertainties
in the flux scales are typically $\pm 5$\% at 3\,mm and $\pm 10$\% at 1.4\,mm.

In some of the observing epochs, 
the observing program monitored phases every 20\,min, at 3 and 1.4\,mm
simultaneously, with the same phase calibrators used in earlier
observations of these sources (see Table~1 of Downes \& Solomon 1998).  Prior to 2004, the
data processing program used the 1.4\,mm total power to correct
amplitudes and phases at 3 and 1.4\,mm for short-term changes in
atmospheric water vapor.  After 2004, this was done with 
water-vapour monitoring receivers at 22 GHz on each antenna.
A post-observation calibration program took the 3\,mm curve of phase
versus time, scaled it to 1.4\,mm, then subtracted it from the observed
1.4\,mm calibrator phases, and then fit the phase difference between
the two receivers.  All visibilities are weighted by the integration
time and the inverse square of system temperature.  Most maps were
made with this ``natural weighting'' of the $uv$ data.

Some sources were observed with improved IRAM receivers in February
2008, in a $2\times 1$\,GHz spectroscopic mode that simultaneously covered the lines of HCN, \hcoone\ and \cthone, 
in the newer PdBI extended configuration with antenna
spacings up to 760\,m.  For these later observations, typical receiver 
temperatures were 50\,K at 3\,mm.

All data reductions were done with the MAPPING program in the standard
IRAM GILDAS\footnote{http://www.iram.fr/IRAMFR/GILDAS} package.

The datacubes were converted to FITS files and imported into \verb=CASA= \citep{McMullin2007} v4.7.1 for data analysis. We created integrated intensity maps using a 1$\sigma$ cutoff for weak lines such as, for example, HN$^{13}$C and a 5$\sigma$ cutoff for strong emission lines such as, for example, \co, HCN and \hco. These integrated intensity maps are presented in Figures \ref{fig:arp220maps} and \ref{fig:ngc6240maps}. Table \ref{tab:observations} presents the various molecular lines detected and their observational properties. We also present spectra in Figures \ref{fig:arp220spec} and \ref{fig:n6240spec}. 

\subsection{ALMA}
\begin{figure*}[!htbp] 
\centering
$\begin{array}{c@{\hspace{0.1in}}c@{\hspace{0.1in}}c}
\includegraphics[scale=0.2]{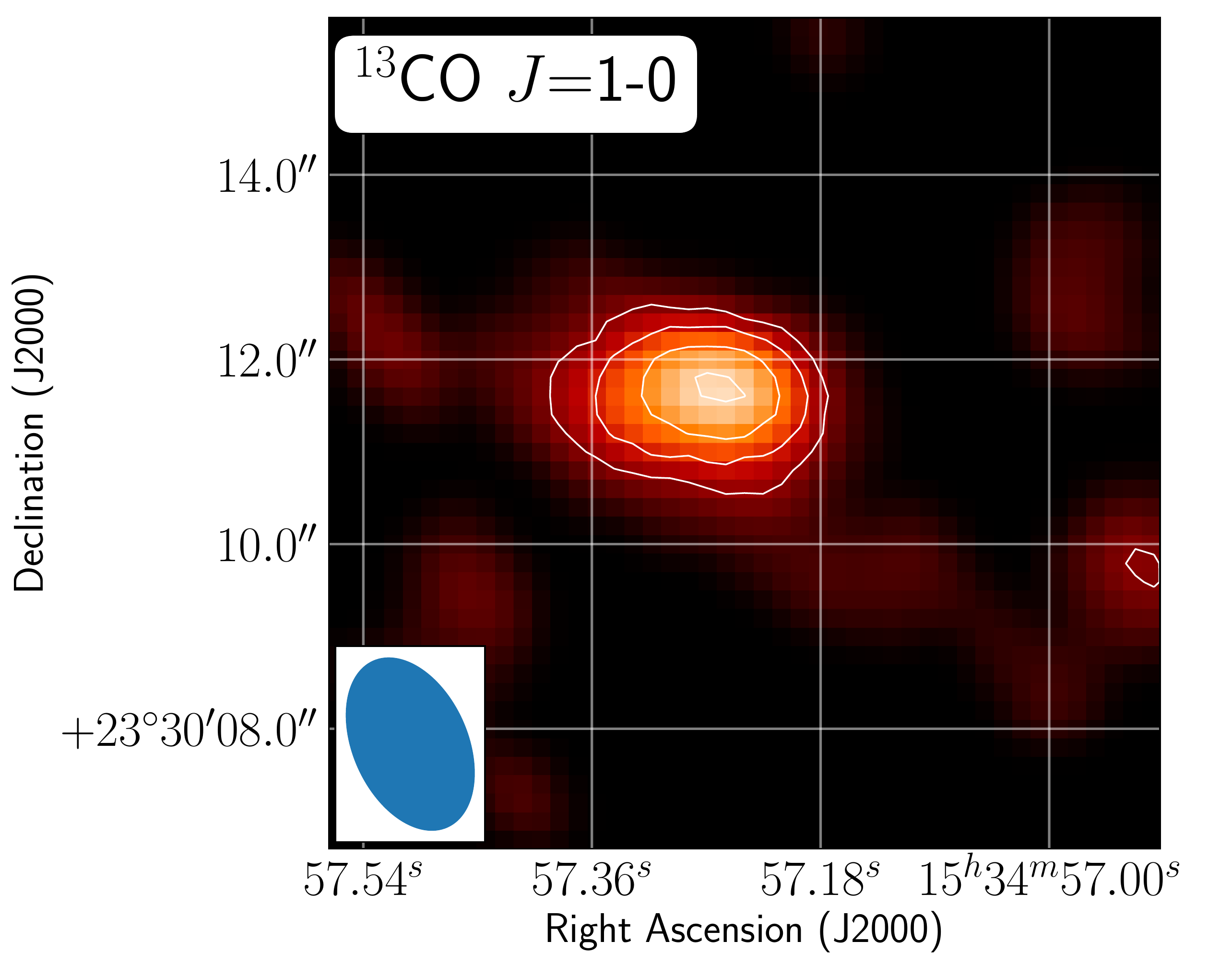}  &\includegraphics[scale=0.2]{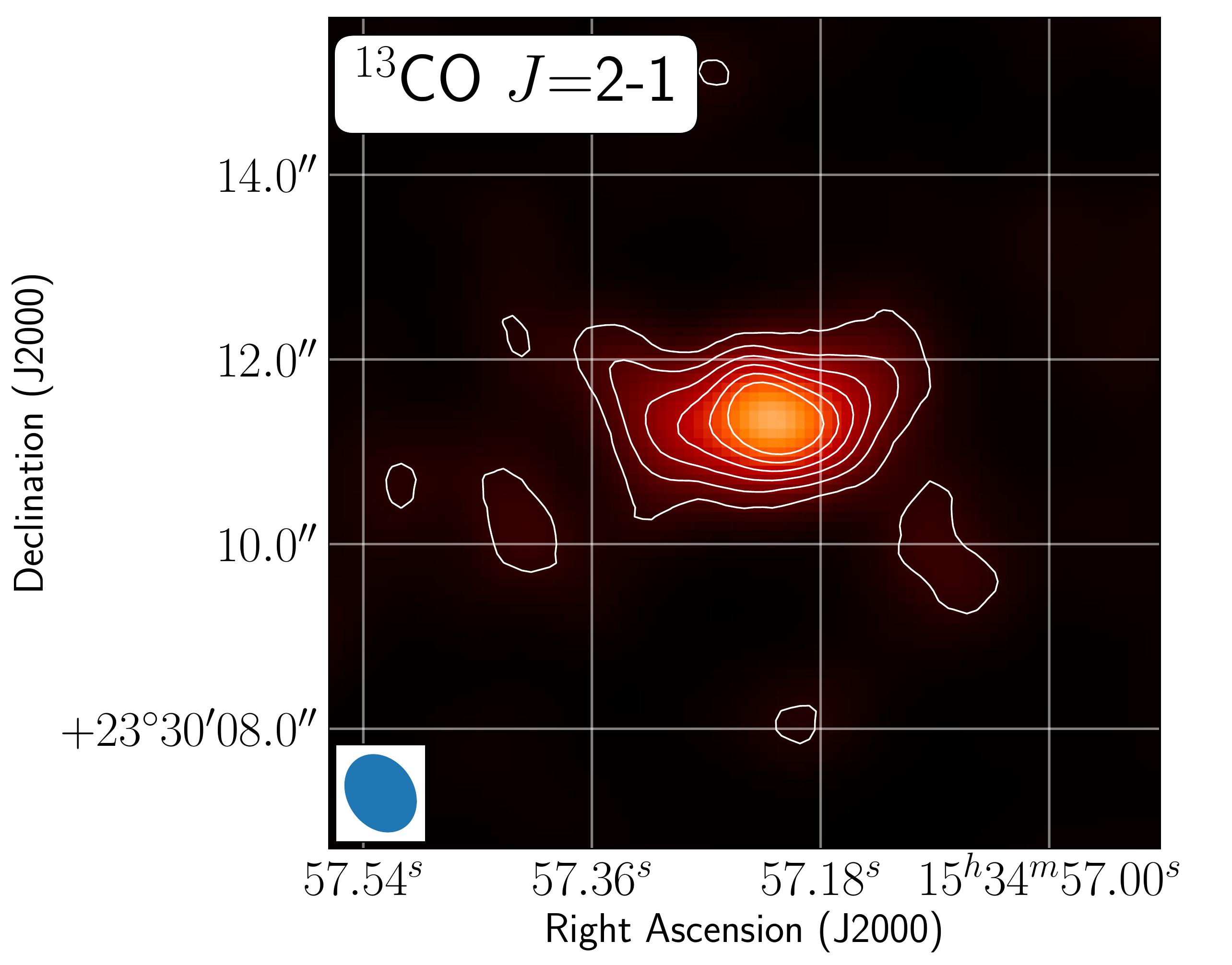}  & \includegraphics[scale=0.2]{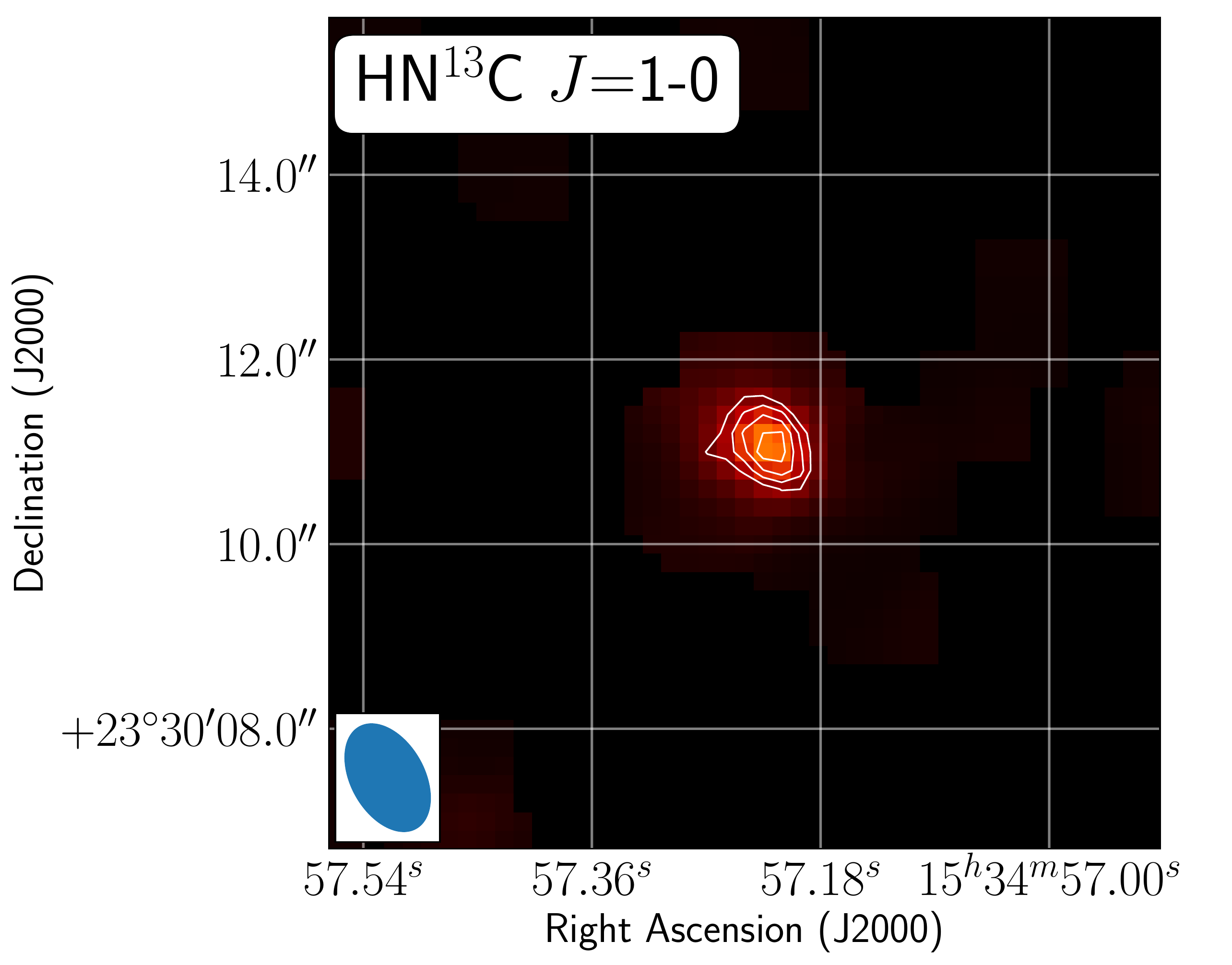}  \\
\includegraphics[scale=0.2]{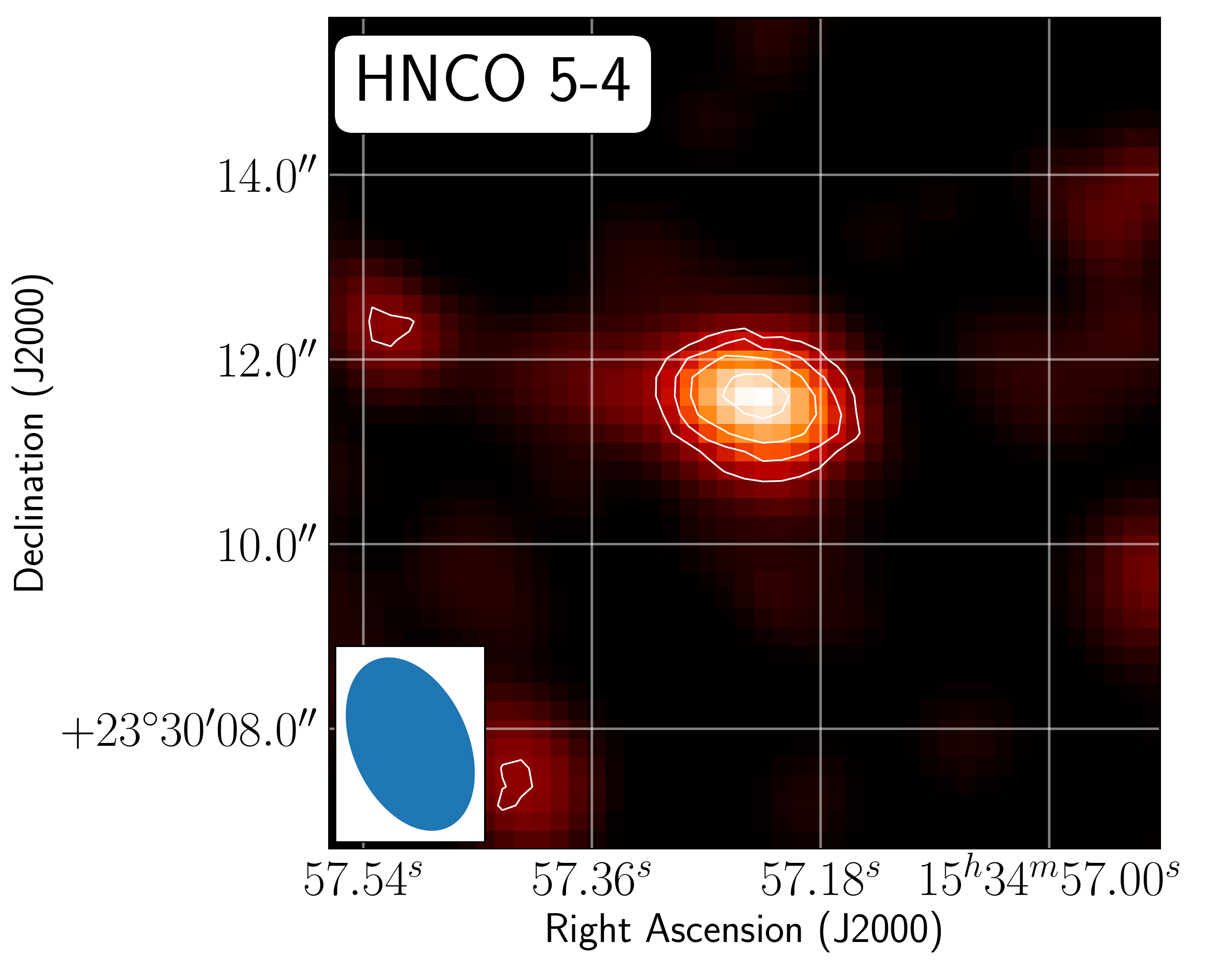} &\includegraphics[scale=0.2]{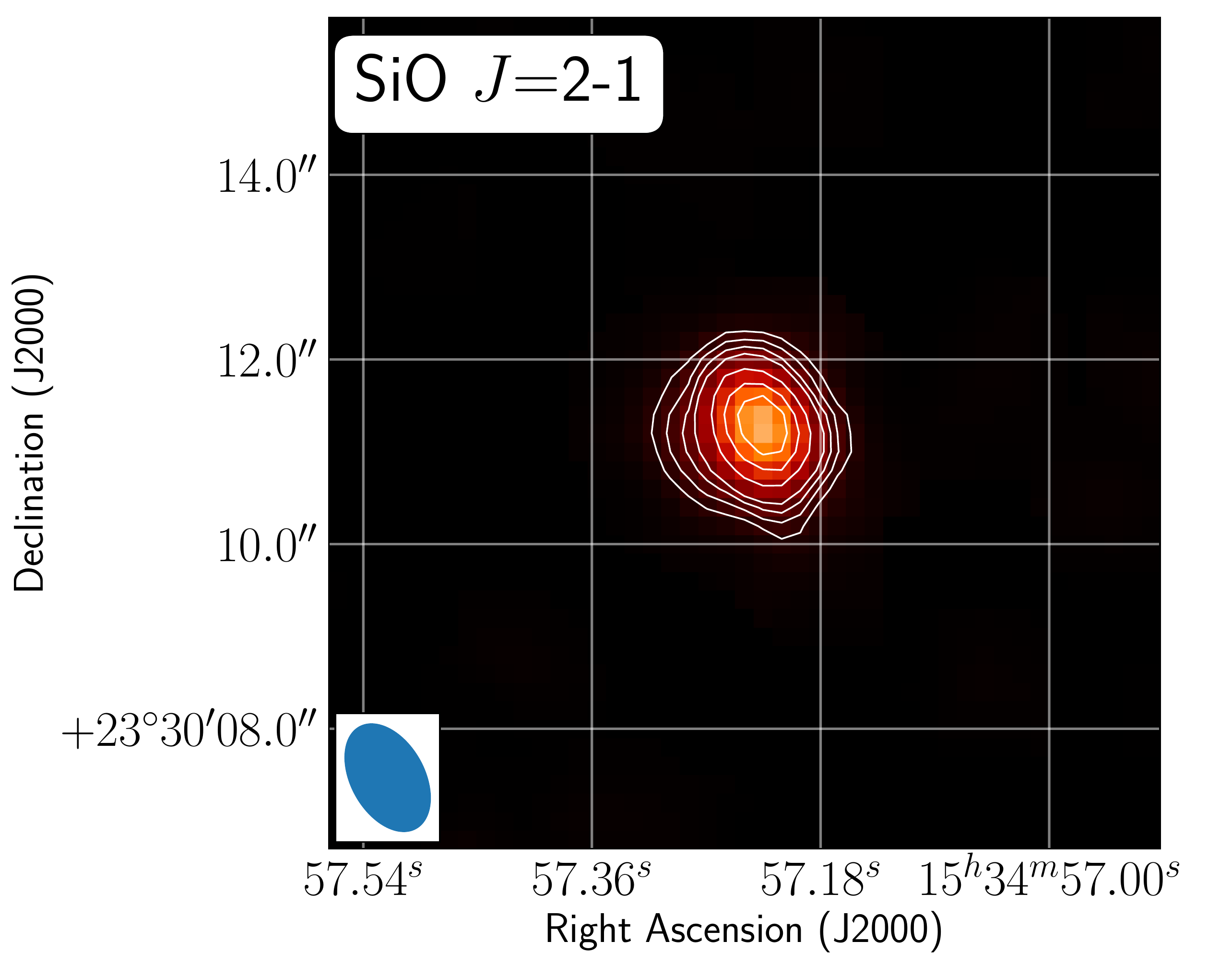} &\includegraphics[scale=0.2]{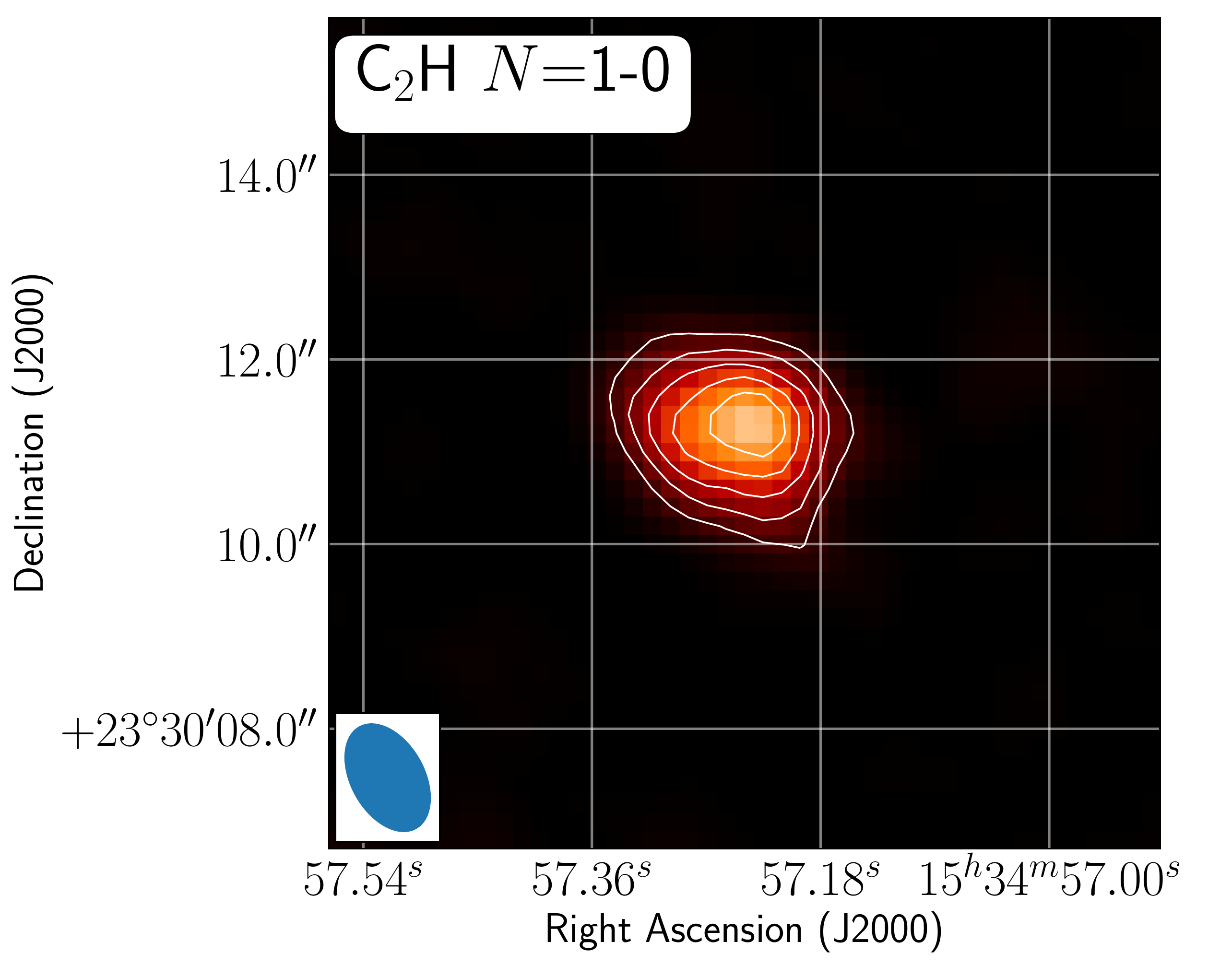}  \\
 \includegraphics[scale=0.2]{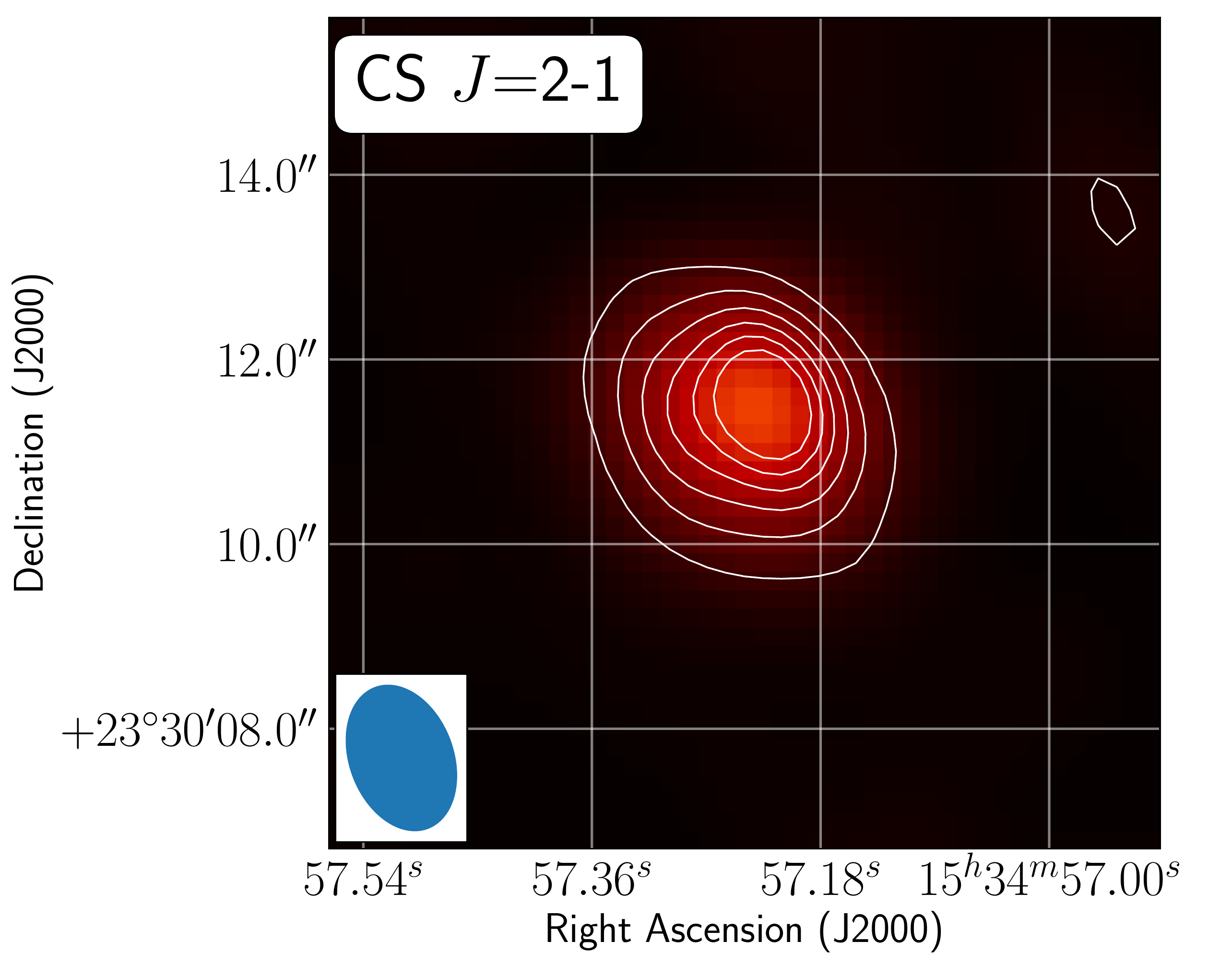} & \includegraphics[scale=0.2]{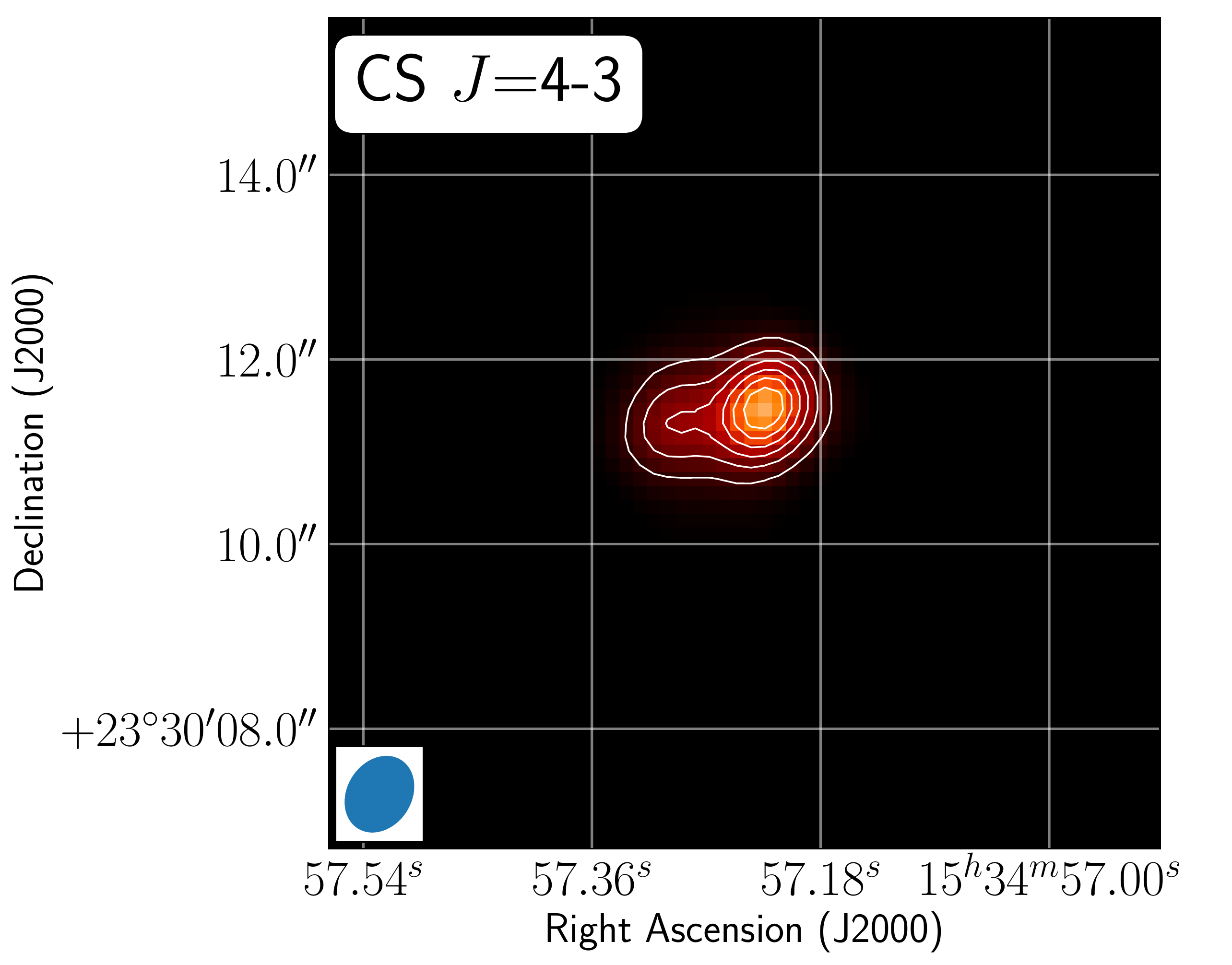} & \includegraphics[scale=0.2]{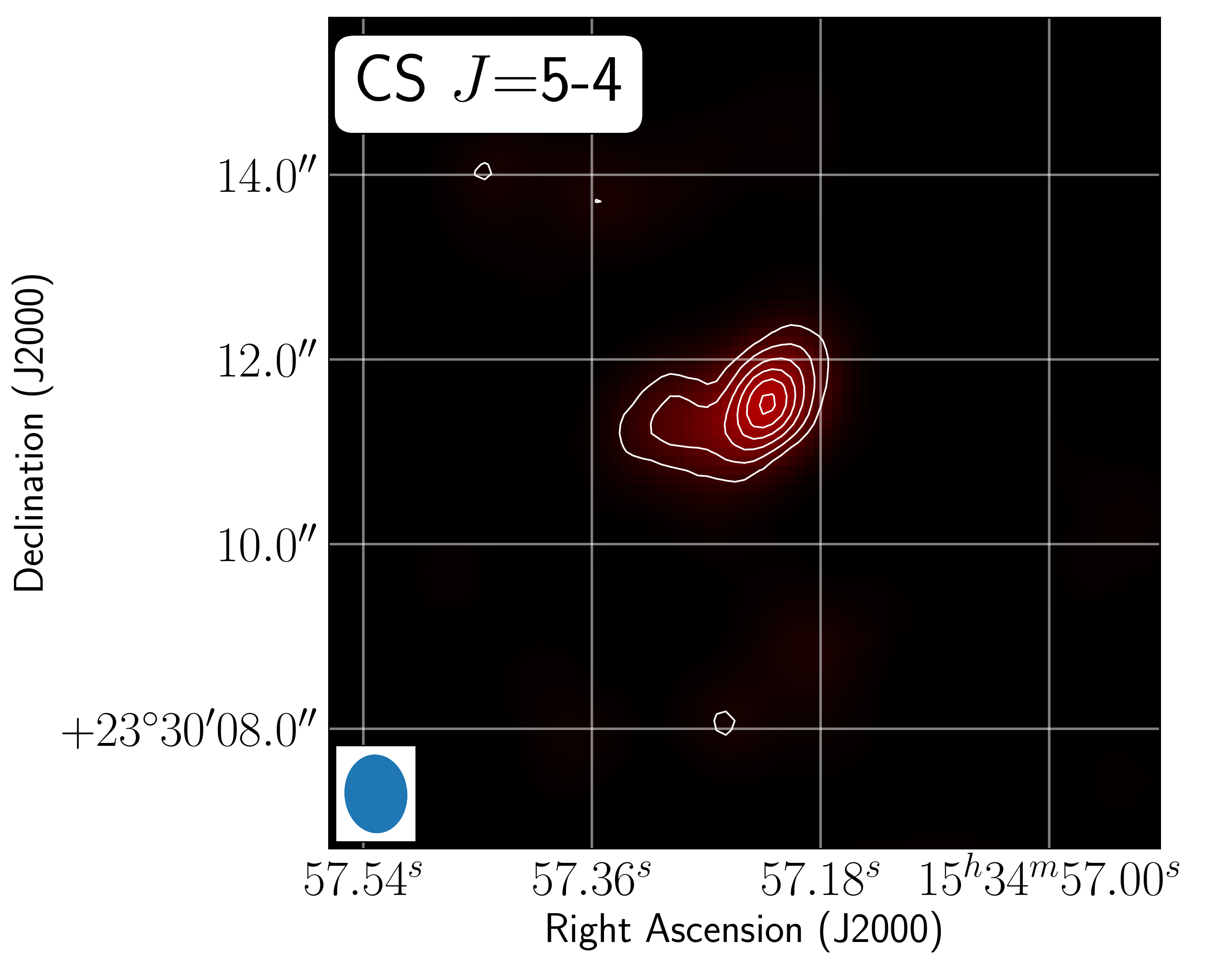}   \\
 \end{array}$
 $\begin{array}{c@{\hspace{0.1in}}c}
  \includegraphics[ scale=0.2]{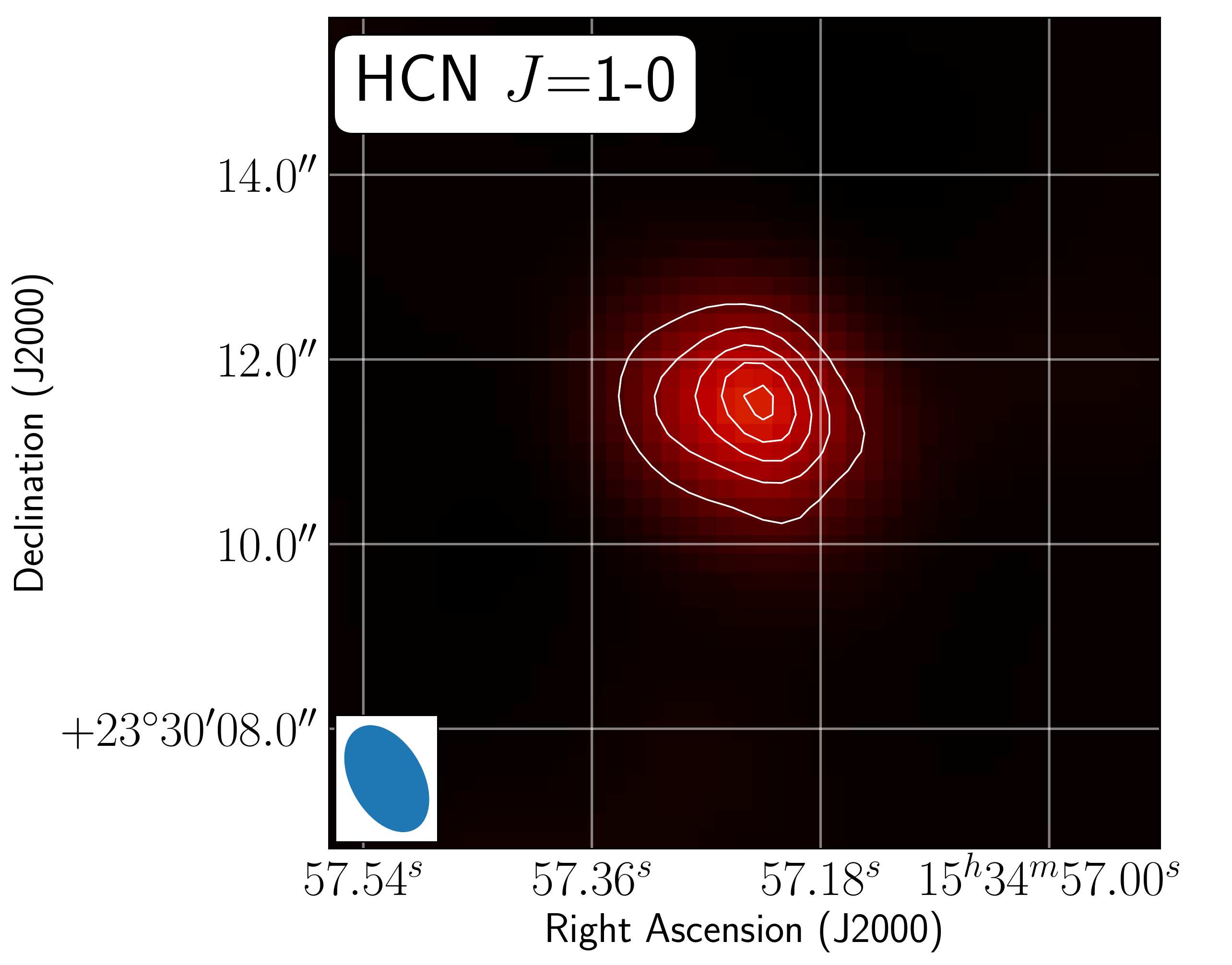}  & \includegraphics[scale=0.2]{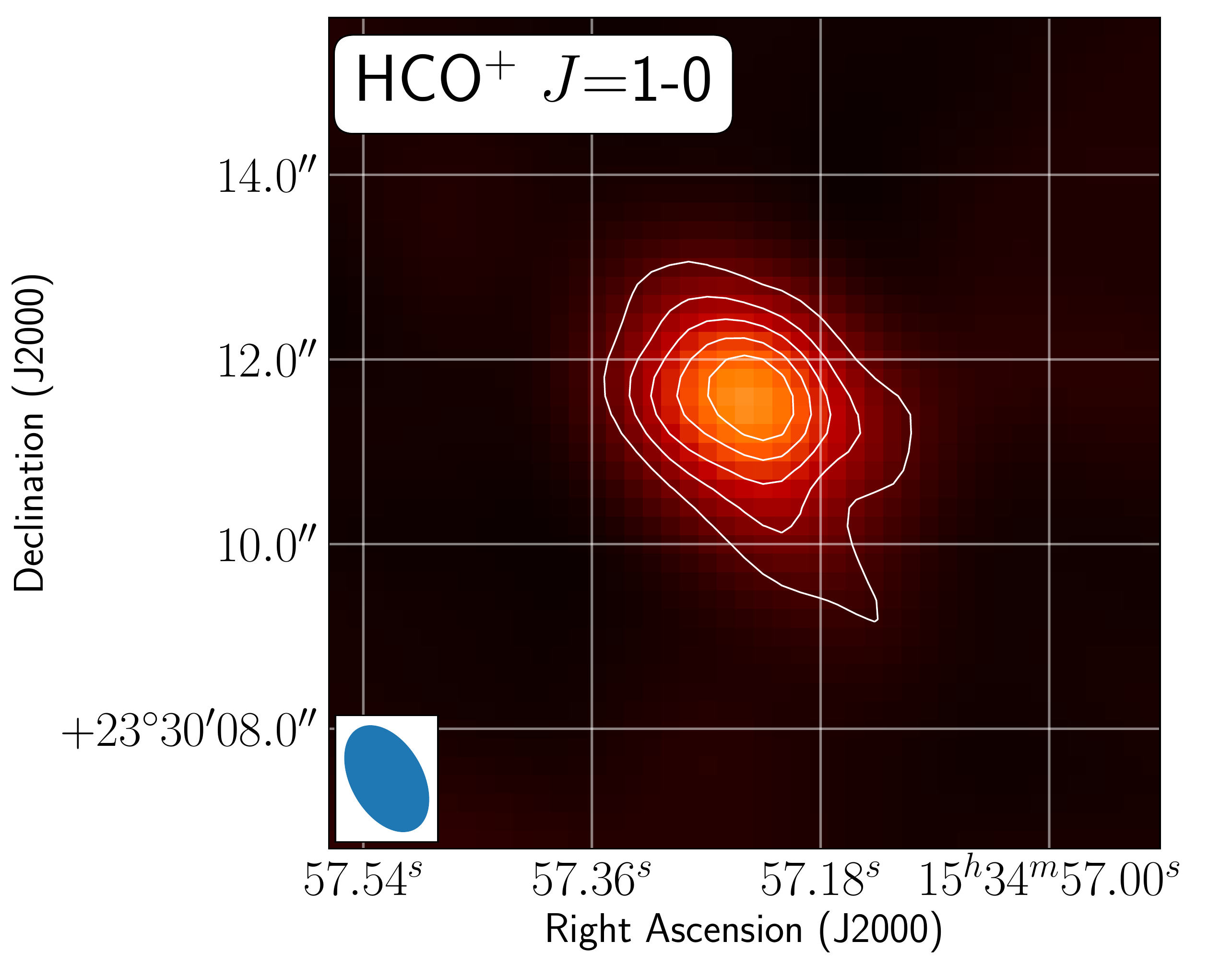}  \\
\includegraphics[scale=0.2]{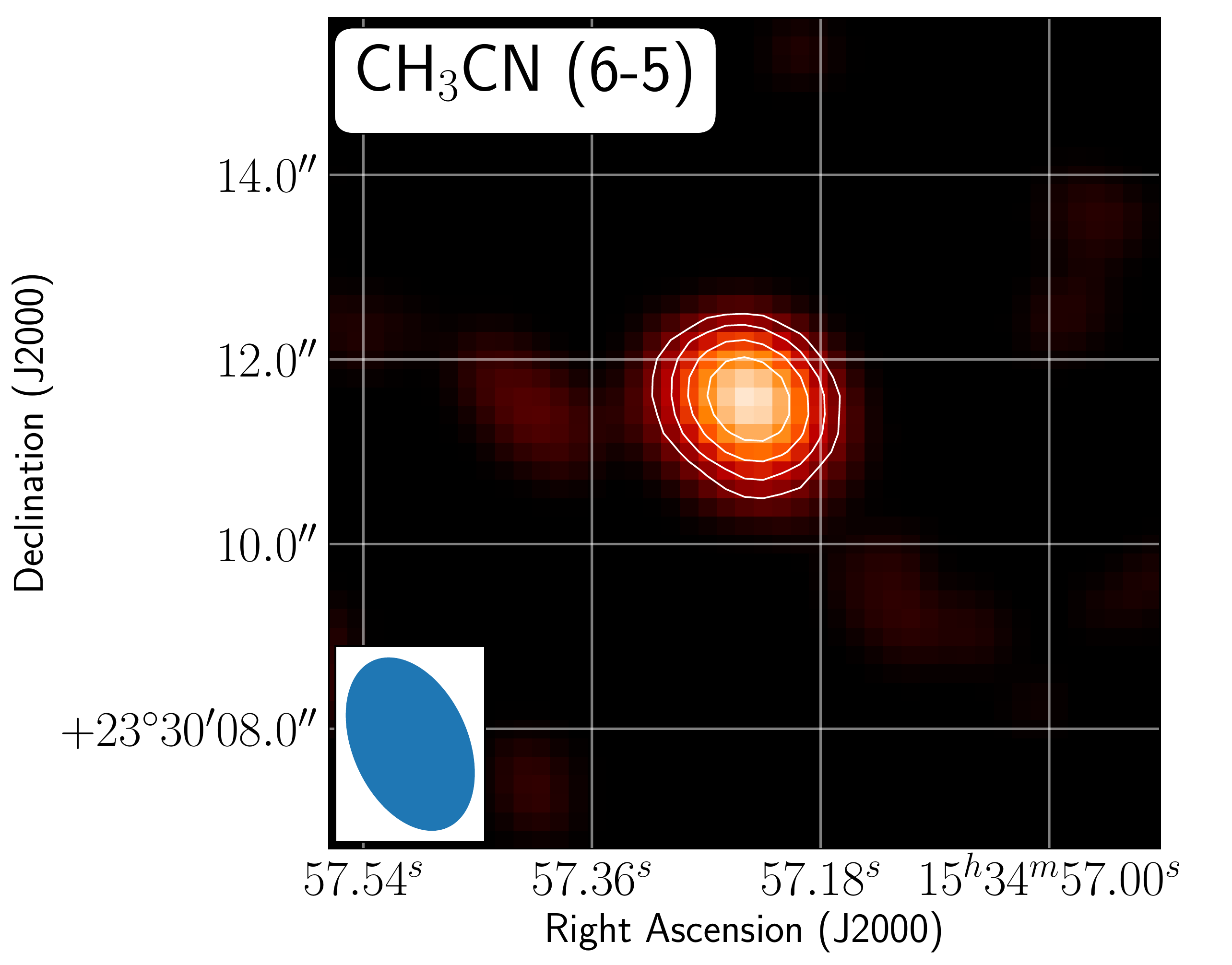} &\includegraphics[scale=0.2]{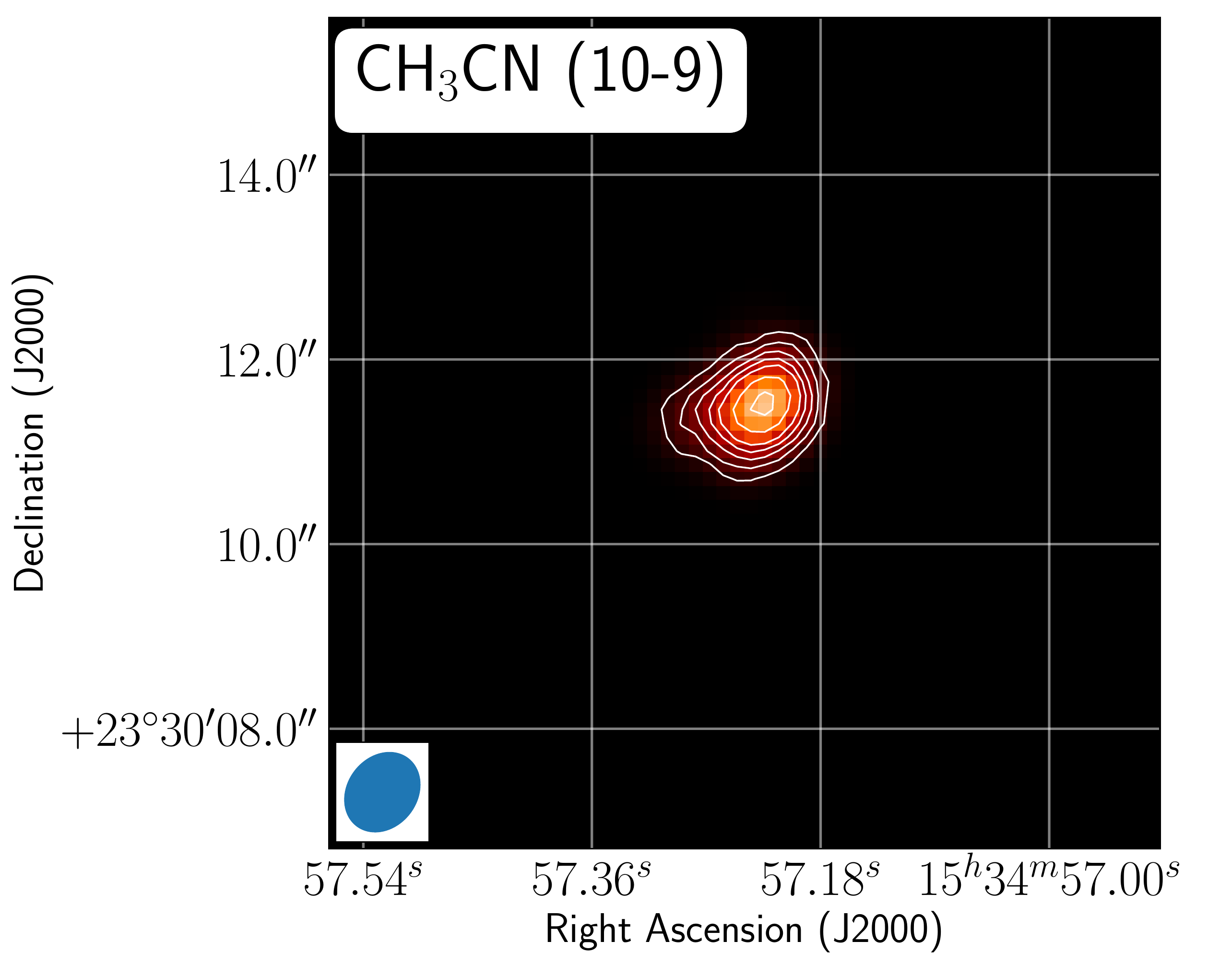} 
\end{array}$
\caption[]{Arp 220: The ellipse in the bottom left corner of each map represents the synthesized beam size. (TOP ROW) \coone, $J$=2-1 and HN$^{13}$C $J$=1-0 with contours corresponding to [4, 6, 8, 10] $\times$ 0.45 Jy beam$^{-1}$ \kms, [4, 8, 12, 16, 20, 24] $\times$ 0.5 Jy beam$^{-1}$ \kms\ and [3, 4, 5, 6] $\times$ 0.135  Jy beam$^{-1}$ \kms, respectively. (2$^{nd}$ ROW) \hncofive, SiO $J$=2-1 and \cthone\ with contours corresponding to [4, 6, 8,10] $\times$ 0.36Jy beam$^{-1}$ \kms,  [4, 6, 8, 10, 15, 20, 25, 30] $\times$ 0.135 Jy beam$^{-1}$ \kms\ and [5, 10, 15, 20, 25] $\times$ 0.135 Jy beam$^{-1}$ \kms, respectively. (3$^{rd}$ ROW) \cstwo, $J$=4-3 (ALMA) and  $J$=5-4 with contours corresponding to [4, 8, 12, 16, 20, 24] $\times$ 0.5, 1.2 and 0.29 Jy beam$^{-1}$ \kms, respectively. (4$^{th}$ ROW) \hcnone\ and \hcoone\ with contours corresponding to [5, 10, 15, 20, 25] $\times$ 0.52 and 0.15 Jy beam$^{-1}$ \kms, respectively. (BOTTOM ROW) \chcnsix\ and (10-9) with contours corresponding to [4, 6, 8, 10]  and [6, 12, 18, 24, 30, 40, 50] $\times$  0.33 Jy beam$^{-1}$ \kms, respectively.
}
\label{fig:arp220maps}
\end{figure*}

Arp~220 was a target for Band 5 SV observations on 16 July 2016. The four spectral windows, of 1.875GHz bandwidth each, were placed on H$_{2}$O (3$_{13}$ - 2$_{20}$), HNC $J$=2-1, CS $J$=4-3 and CH$_{3}$OH (4$_{31}$ - 3$_{30}$). After checking for any obvious calibration flaws, we use the once phase-only self-calibrated delivered visibility dataset. We image the \csfour\ and what we have identified as \chcnten\ lines. The H$_{2}$O line has been presented in \cite{Koenig2016b}. Using \texttt{CASA} v4.7.1, we create datacubes of 20 \kms\ channel widths with a natural weighting for maximum sensitivity. We created integrated intensity maps using a 2$\sigma$ cutoff (Figure \ref{fig:arp220maps}.)

\subsection{Short-Spacings Flux for \co}
\cite{Greve2009} present single dish fluxes for both Arp~220 and NGC~6240. Comparing the fluxes measured for Arp~220 in \citet[][;\coone\ and $J$=2-1]{Downes1998} to that of \cite{Greve2009} shows that the PdBI maps have recovered nearly all flux (\coone: 410 Jy \kms\ vs 420 Jy \kms; \cotwo: 1100 Jy \kms\ vs 1130 Jy \kms). \cite{Sakamoto2008} made comparisons to single dish observations for \cothree\ and concluded that $\sim$10\% of the total flux is missing. Since 10\% of the total flux is likely spread over the entire source, each point in Arp~220 will have insignificant missing \cothree\ flux when compared to the calibration uncertainty ($\pm$ 15\%). 

For NGC~6240, \cite{Feruglio2013a} made a comparison with single dish observations for \coone\ and found an agreement in fluxes. The \cotwo\ observations of \cite{Tacconi1999} are missing $\sim$30\% of the total flux when compared to the flux measured in \cite[][1220 Jy \kms\ vs 1740 Jy \kms]{Greve2009}.

\begin{table*}
\caption{Line Observations Summary}\label{tab:observations}
\centering
\begin{tabular}{llccccc}
\hline \hline
Source 	& Line 	& $\nu_{\rm{rest}}$	&Resolution 	& rms 			&Channel Width		& Flux \\
		&		& (GHz)			&(arcsec)		&(mJy beam$^{-1}$)	&(\kms) 	& (Jy km s$^{-1}$) \\
\hline 
Arp 220	&\tcoone\	 &110.201	&2.0 x 1.3	 & 1.5	&40	&8.0  \\
		&\tcotwo\	 &220.399	&0.9 x 0.7	 & 1.9	&20	&47.0 \\
		&\hcoone\	 &89.189	&1.3 x 0.8	 &0.7	 	&20	& 17.8 \\
		&\hcnone\	 &88.631	&1.3 x 0.8	 &0.7	 	&20	& 45.0  \\
		&\cstwo\	 &97.981	&1.6 x 1.1	 &1.3	 	&20	&17.1  \\
		&\csfour\	&195.954	&0.9 x 0.7 & 1.4	&20	&61.0 \\
		&\csfive\	 &244.936	&0.9 x 0.7	 &2.5	 	&40	&40.5  \\
		&\cthone\	 &87.3/4	&1.3 x 0.8	 &0.8	 	&20  &8.8 \\
		&\chcnsix\ & 110.328 - 110.385	&2.0 x 1.3	& 1.5		&40	& $>$3.9 \\
		&\chcnten\ & 183.674 - 183.964	&0.9 x 0.8	& 2.4		&20	& 19.7 \\		
		&\hncofive\ & 109.906	&2.0 x 1.3		&1.5	&40	& $>$2.0\\
		&HN$^{13}$C $J$ = 1-0 &	87.091	&1.3 x 0.8& 0.8	&40	& 0.7 \\
		&SiO $J$=2-1 & 86.847	&1.3 x 0.8& 0.8	&40	& 5.5 \\
		&	 	 &			 &	 	&	&  \\ 
NGC 6240&\hcoone\	 &89.189	&1.1 x 1.1	 &1.0	 	&20& 	9.9  	\\	
		&\hcnone\	 &88.63	&1.1 x 1.1	 &1.0	 	&20& 	6.3	\\	
		&\cthone\	 &87.3/4	&1.1 x 1.1	 &0.3	 	&40& 	1.3	\\		
\hline
\end{tabular} \\
\textbf{NOTE}: [1] Splatalogue (http://www.splatalogue.net/) was used to obtain frequencies. [2] CH$_{3}$CN is composed of several lines that span the stated frequency range. [3] Calibration uncertainties are $\sim$5$\%$ and $\sim$10$\%$ for 3mm and 1mm observations, respectively.
\end{table*}
\begin{figure*}[htbp] 
\centering
$\begin{array}{c@{\hspace{0.1in}}c@{\hspace{0.1in}}c}
\includegraphics[ scale=0.2]{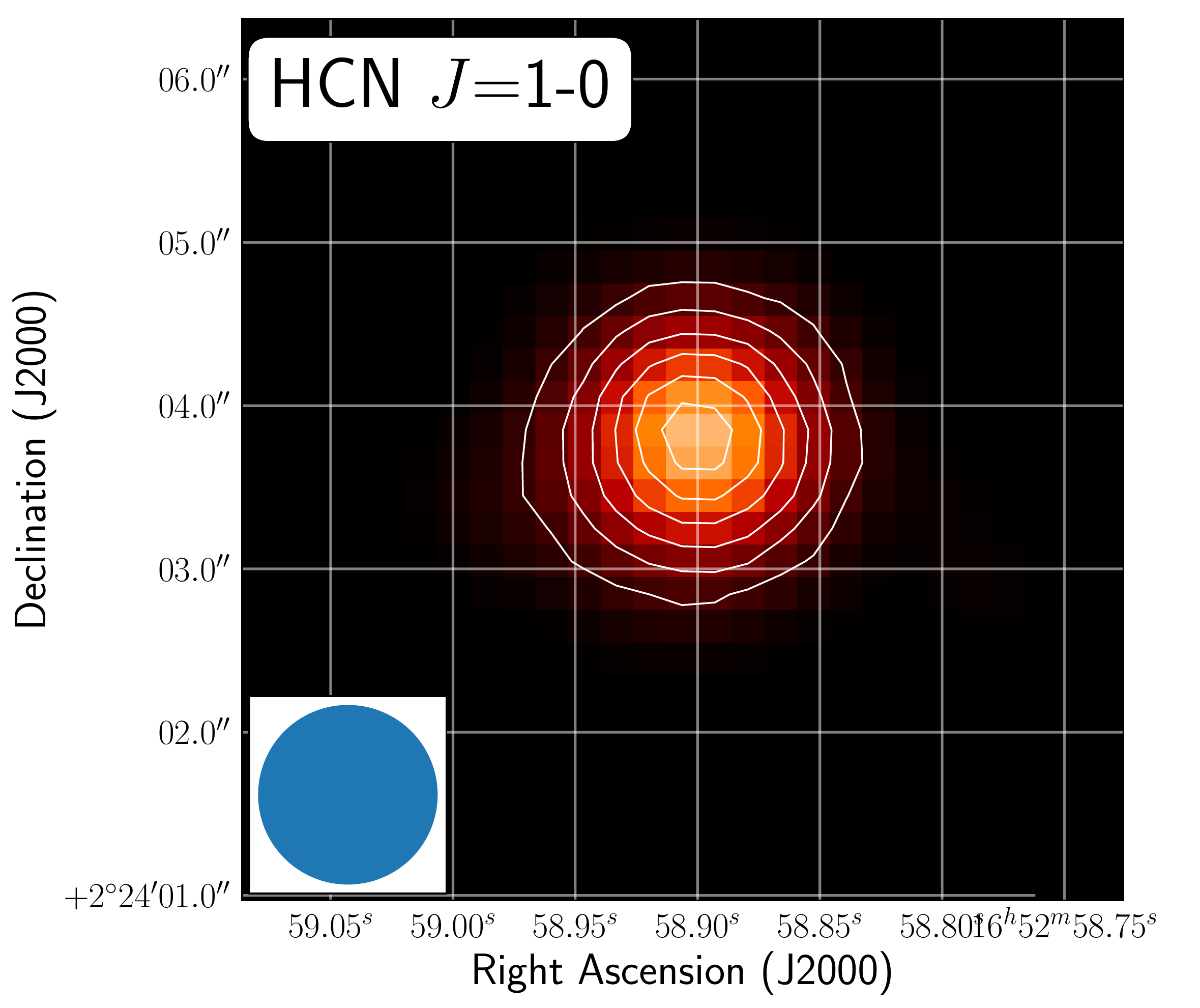}  & \includegraphics[scale=0.2]{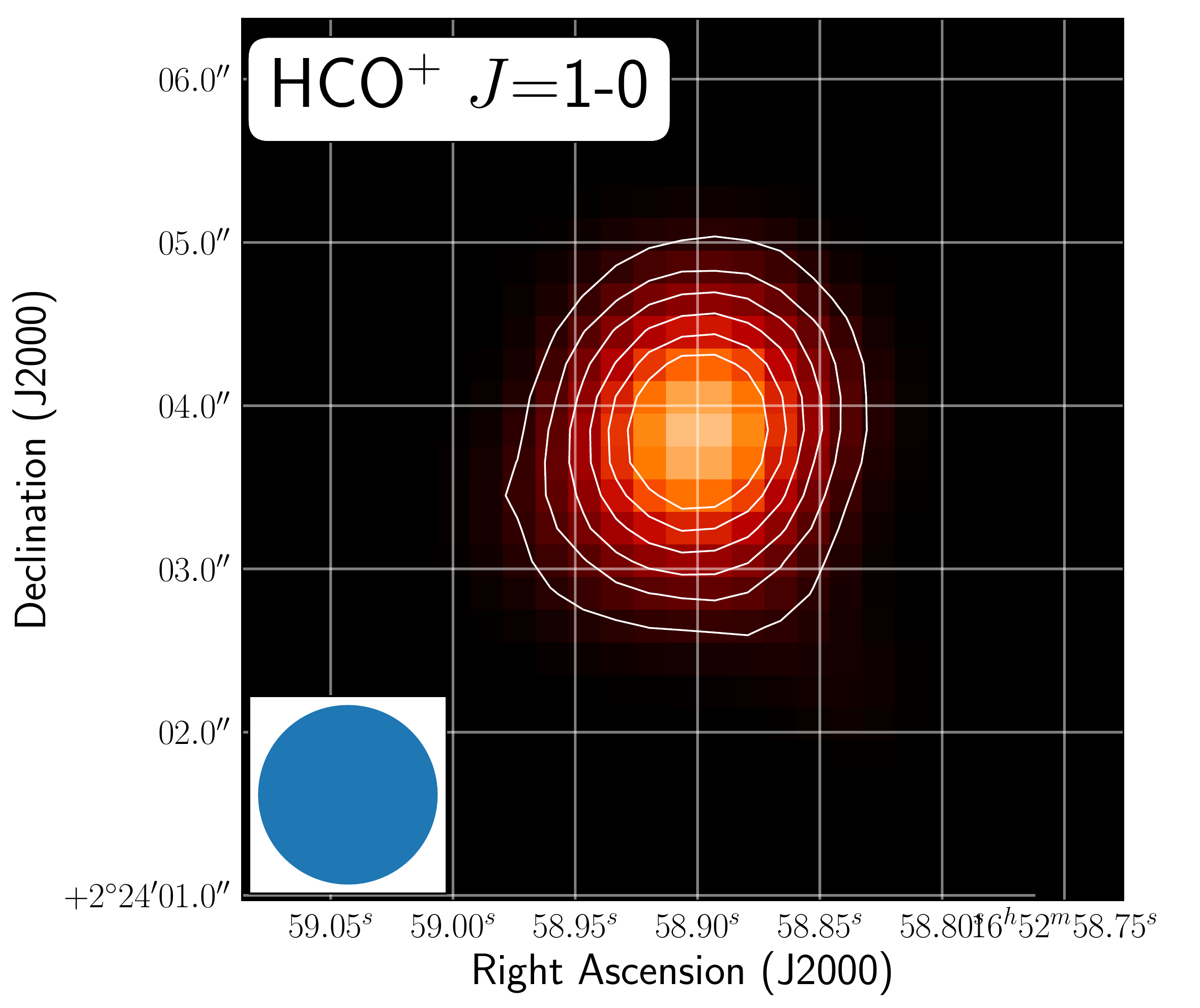}  &  \includegraphics[scale=0.2]{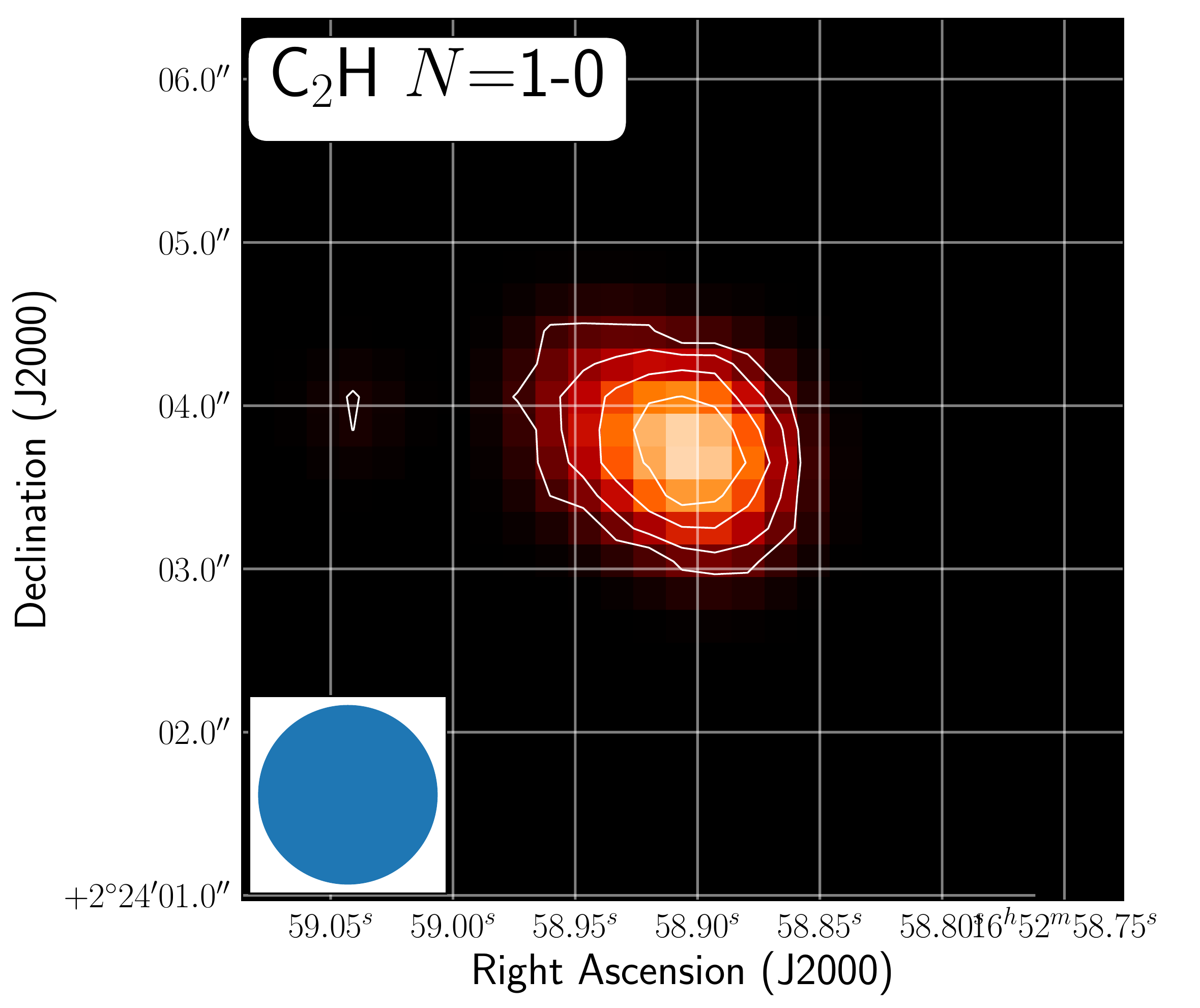}\\ 
\end{array}$
\caption[]{NGC 6240: The ellipse in the bottom left corner of each map represents the synthesized beam size. HCN $J$=1-0 and \hcoone\ with contours corresponding to [3, 6, 9, 12, 15, 18] $\times$ 0.184 Jy beam$^{-1}$ \kms\ and \cthone\ with contours corresponding to [1, 2, 3, 4, 5] $\times$ 0.2 Jy beam$^{-1}$ \kms.}
\label{fig:ngc6240maps}
\end{figure*}
\begin{figure*}[!htbp] 
\centering
$\begin{array}{c@{\hspace{0.1in}}c}
\includegraphics[ scale=0.55]{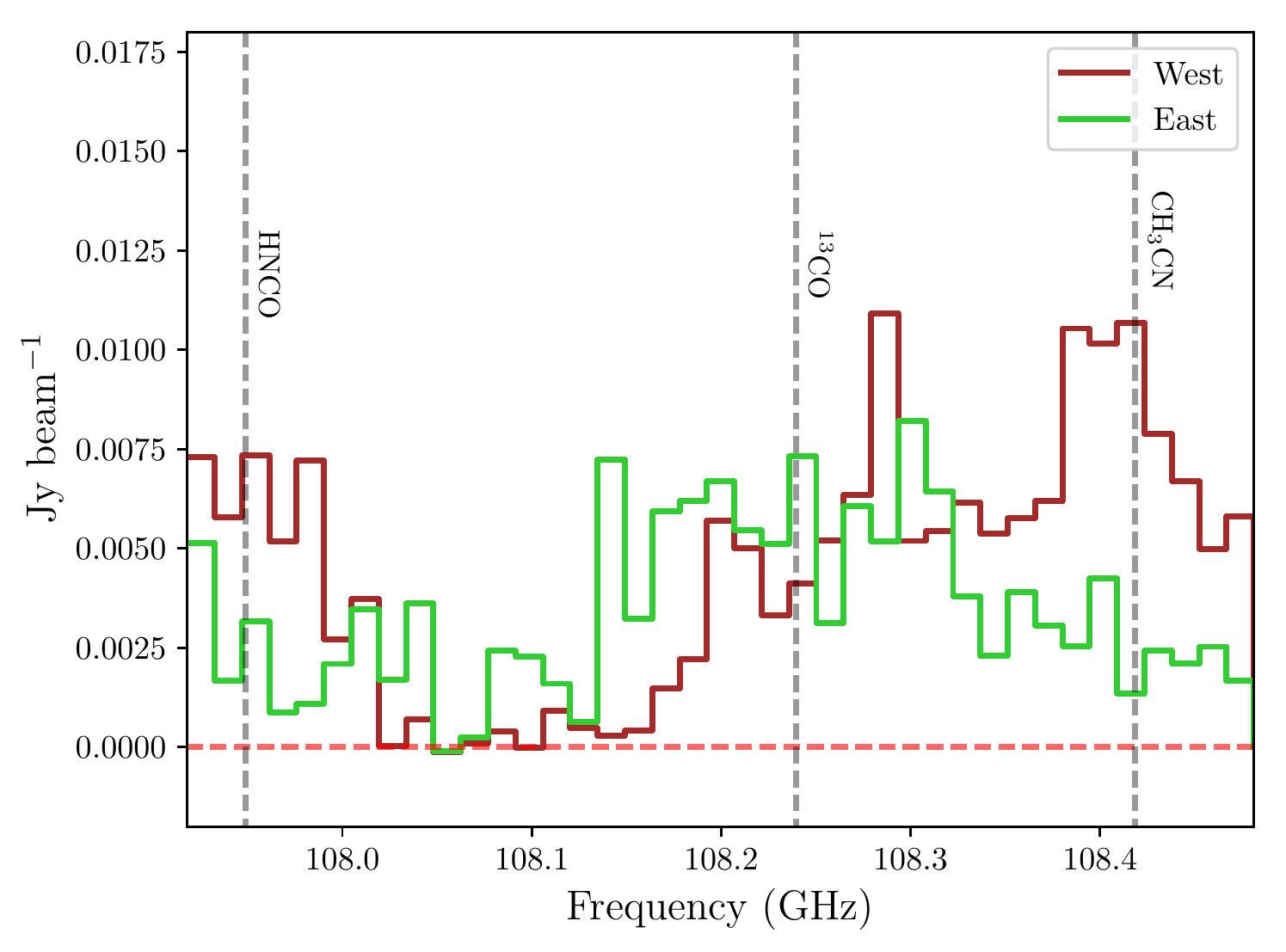}  & \includegraphics[scale=0.55]{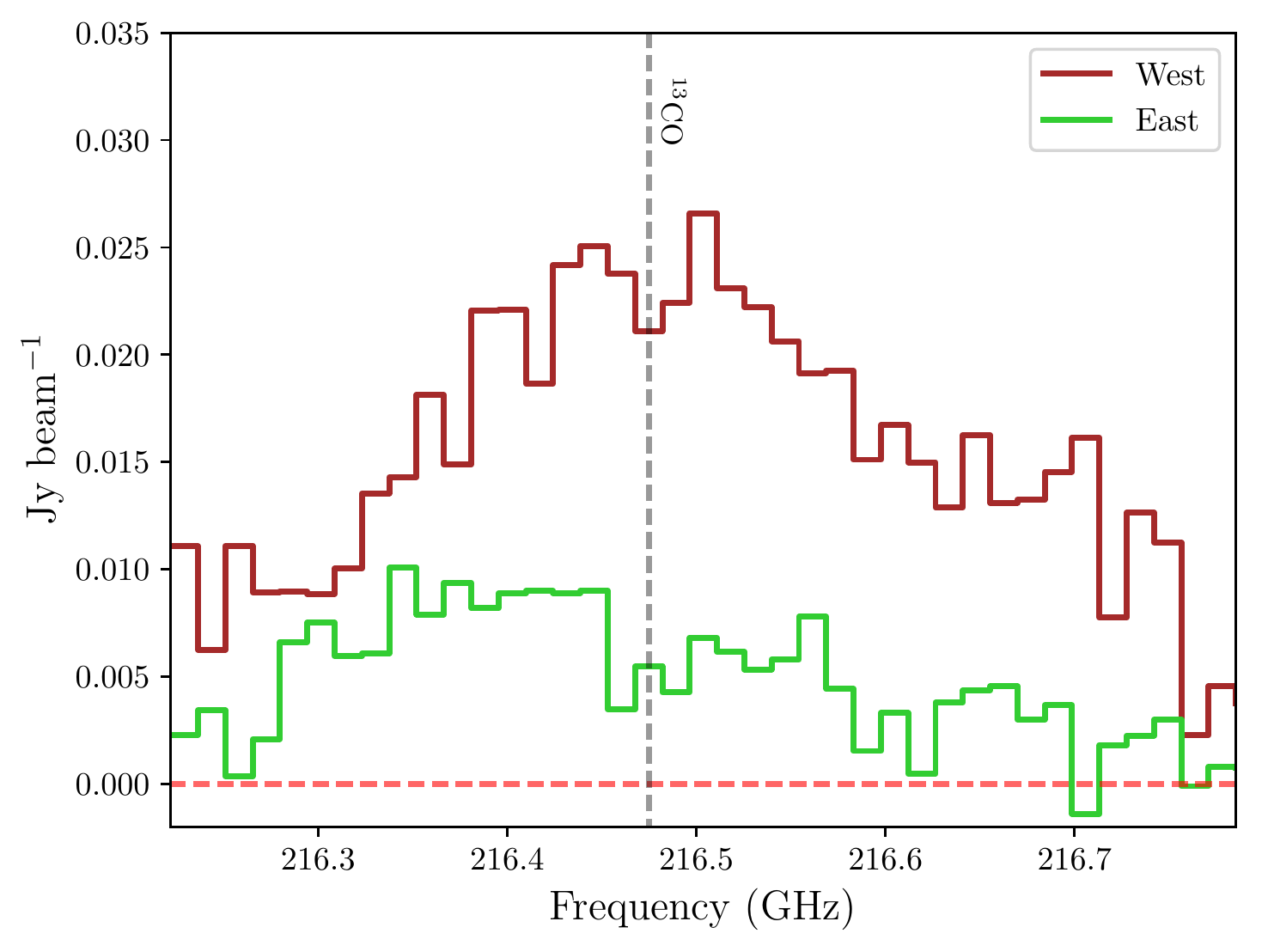} \\
  \includegraphics[scale=0.55]{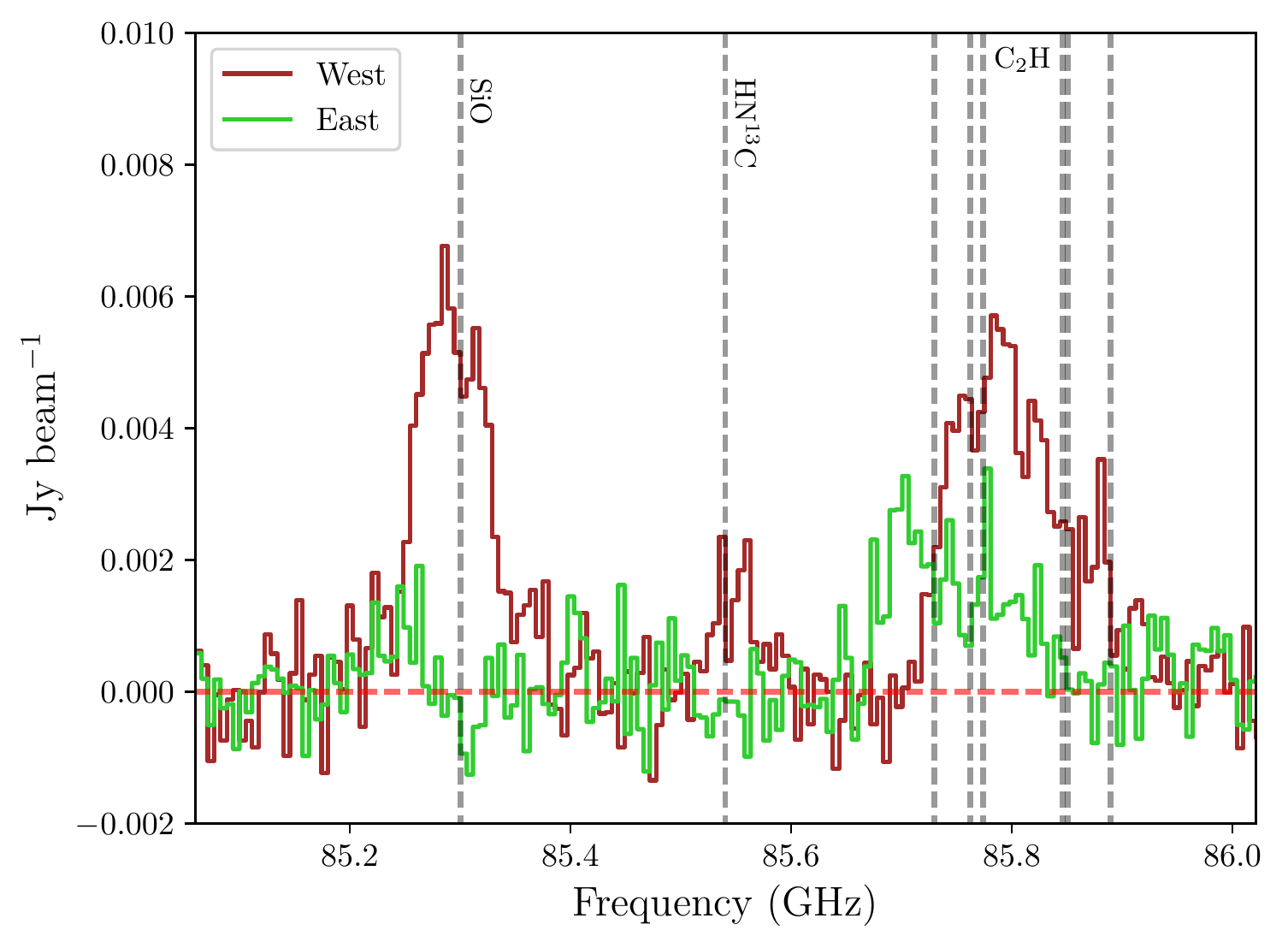} & \includegraphics[scale=0.55]{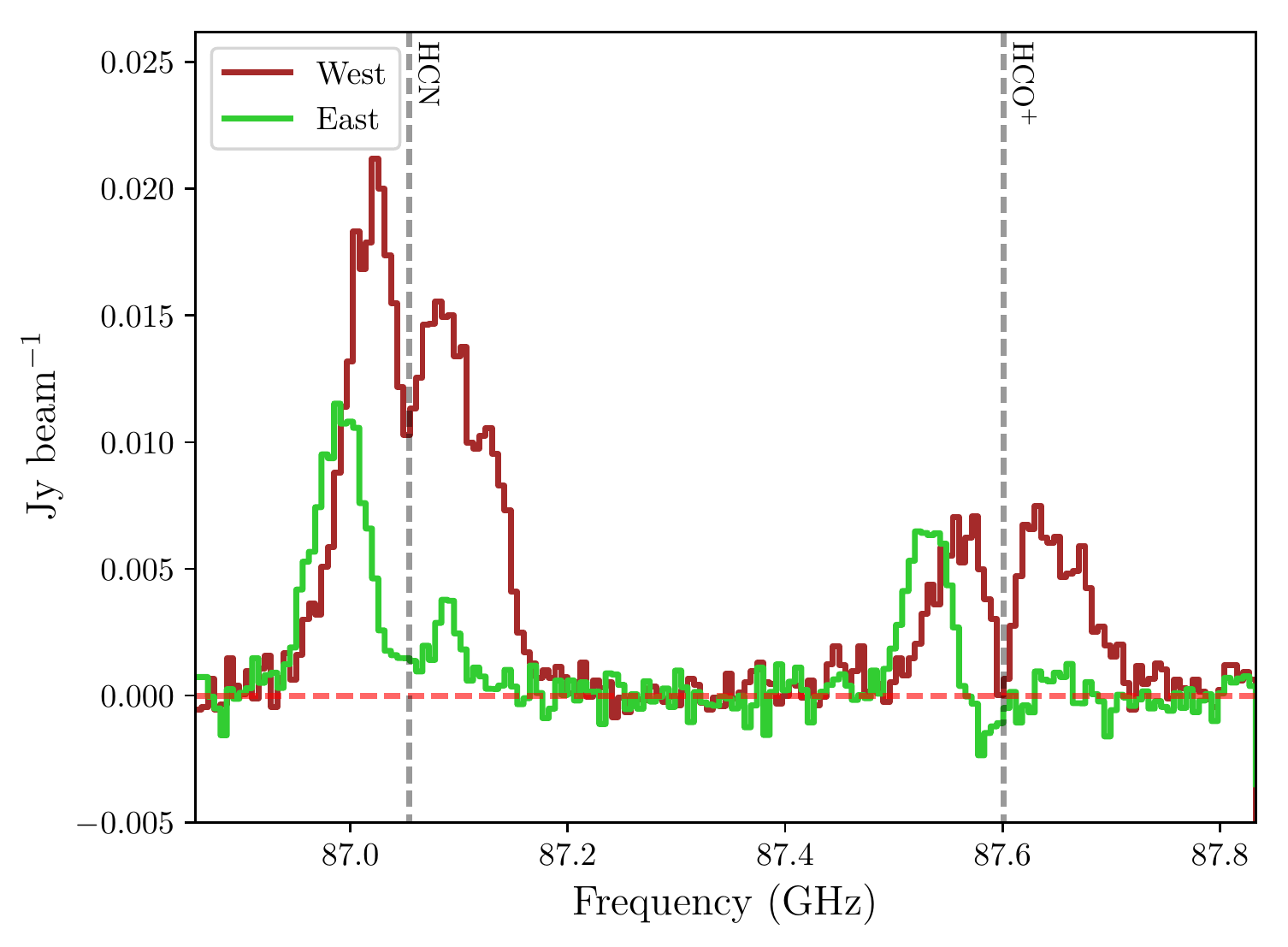} \\

  \includegraphics[scale=0.55]{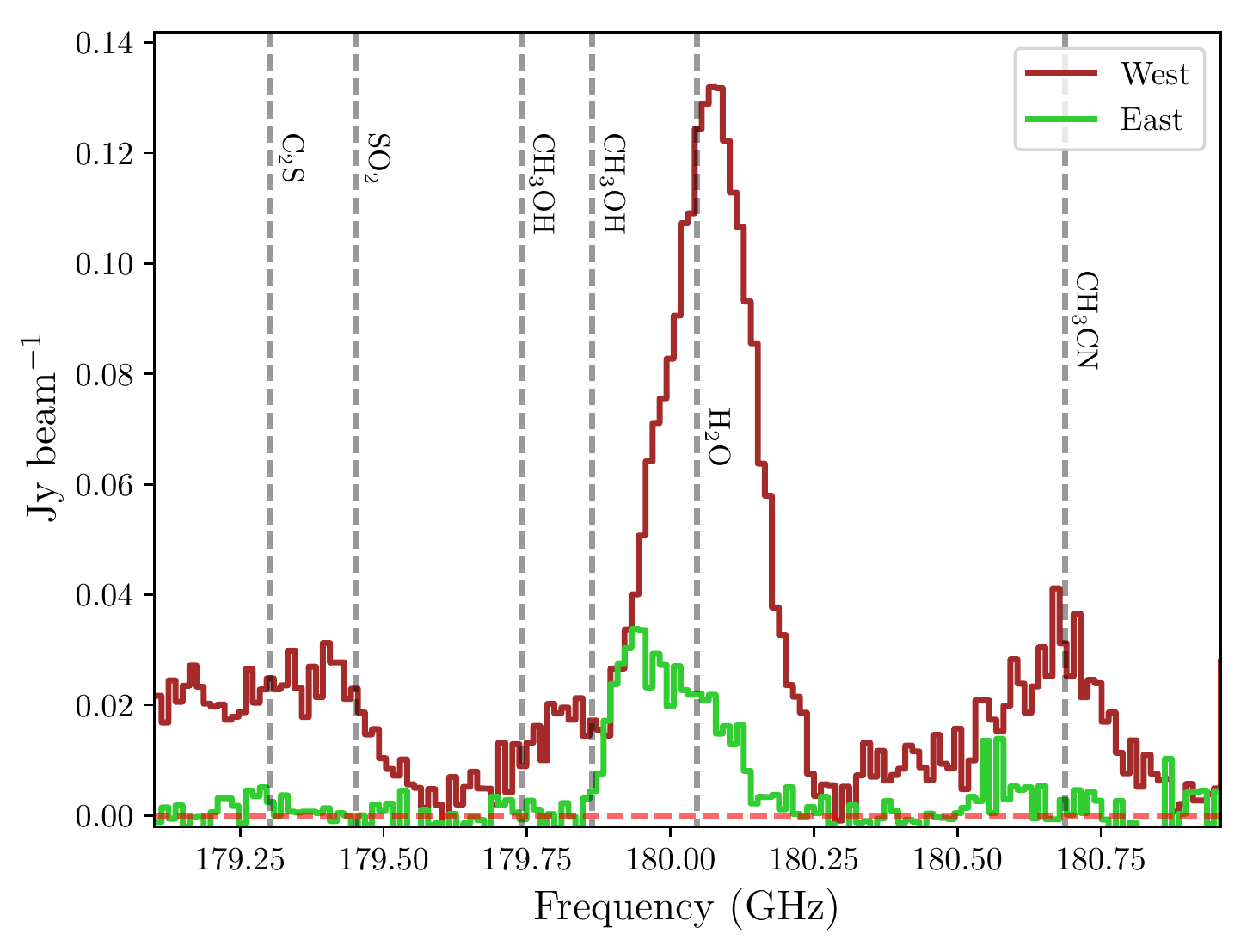} &\includegraphics[scale=0.55]{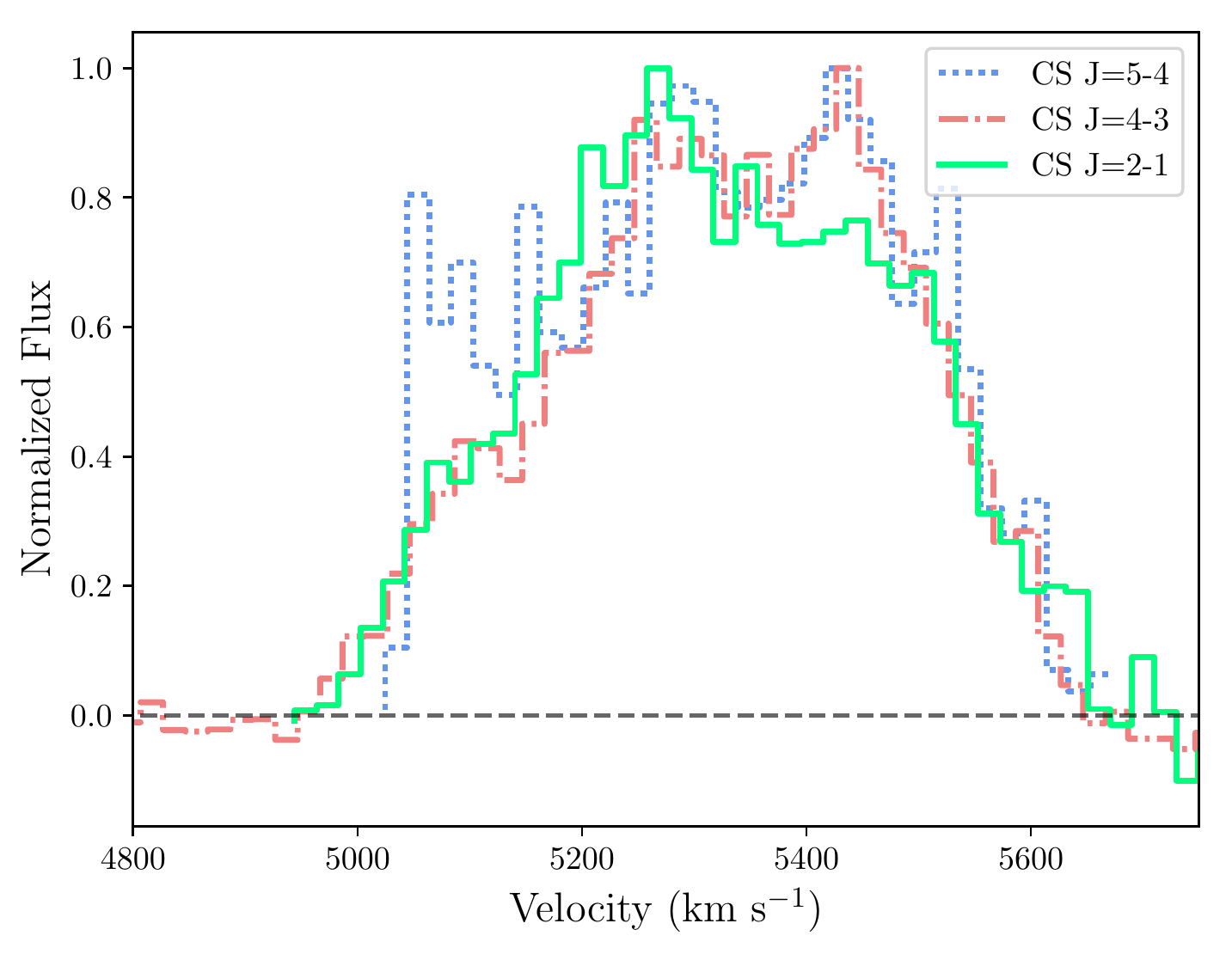}
    \end{array}$
\caption[]{\textbf{($Top$ and $Middle$) Spectra averaged over a 1\arcsec\ diameter aperture centered at Arp 220W and Arp 220E (see Table \ref{tab:lineratios} and Section \ref{sec:rad}) and ($Bottom$) Normalized spectra of each of the CS lines averaged over a 3\arcsec\ diameter aperture. }}
\label{fig:arp220spec}
\end{figure*}
\begin{figure*}[!htbp] 
\centering
$\begin{array}{c@{\hspace{0.1in}}c}
\includegraphics[ scale=0.55]{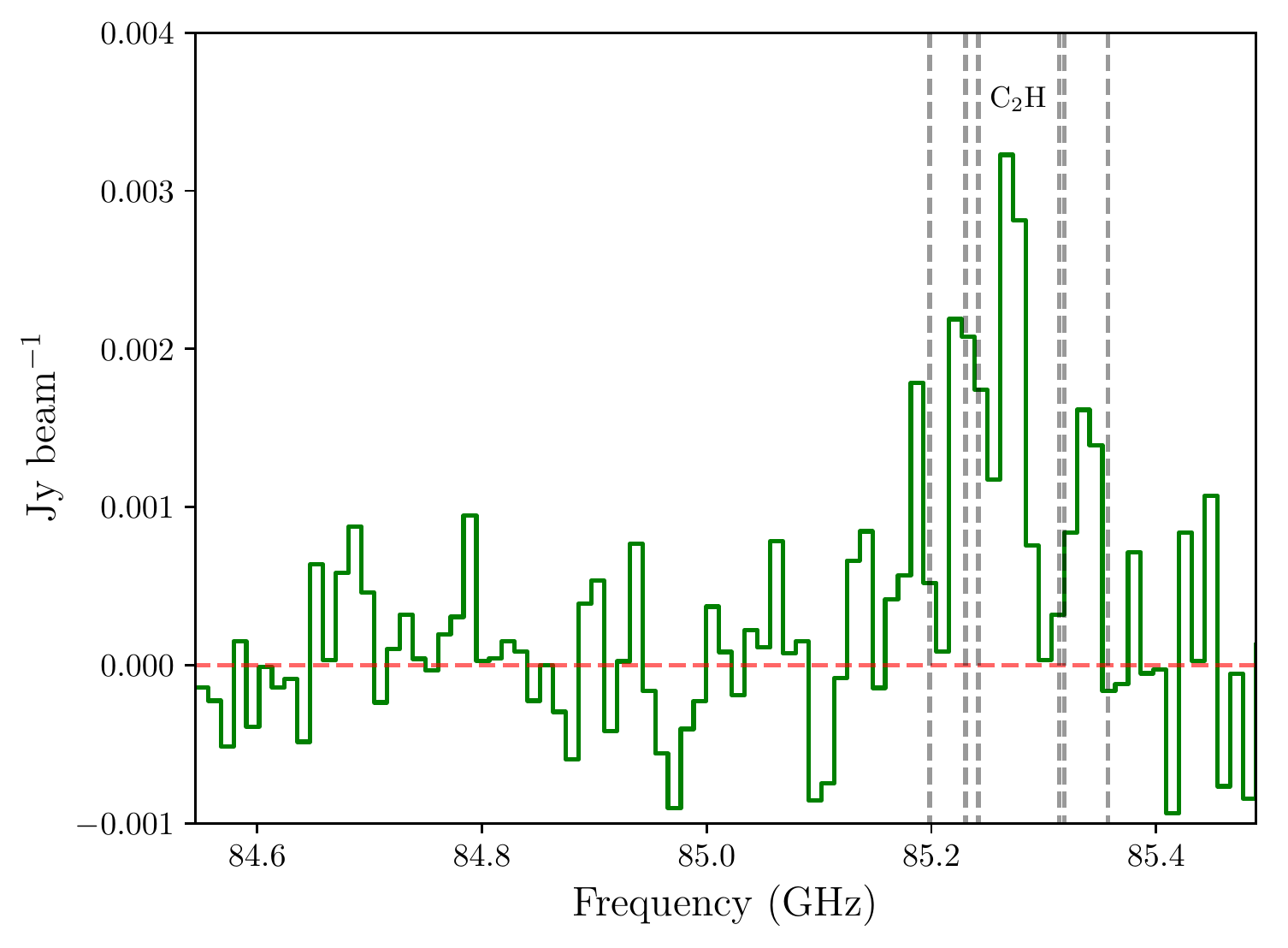}  & \includegraphics[scale=0.55]{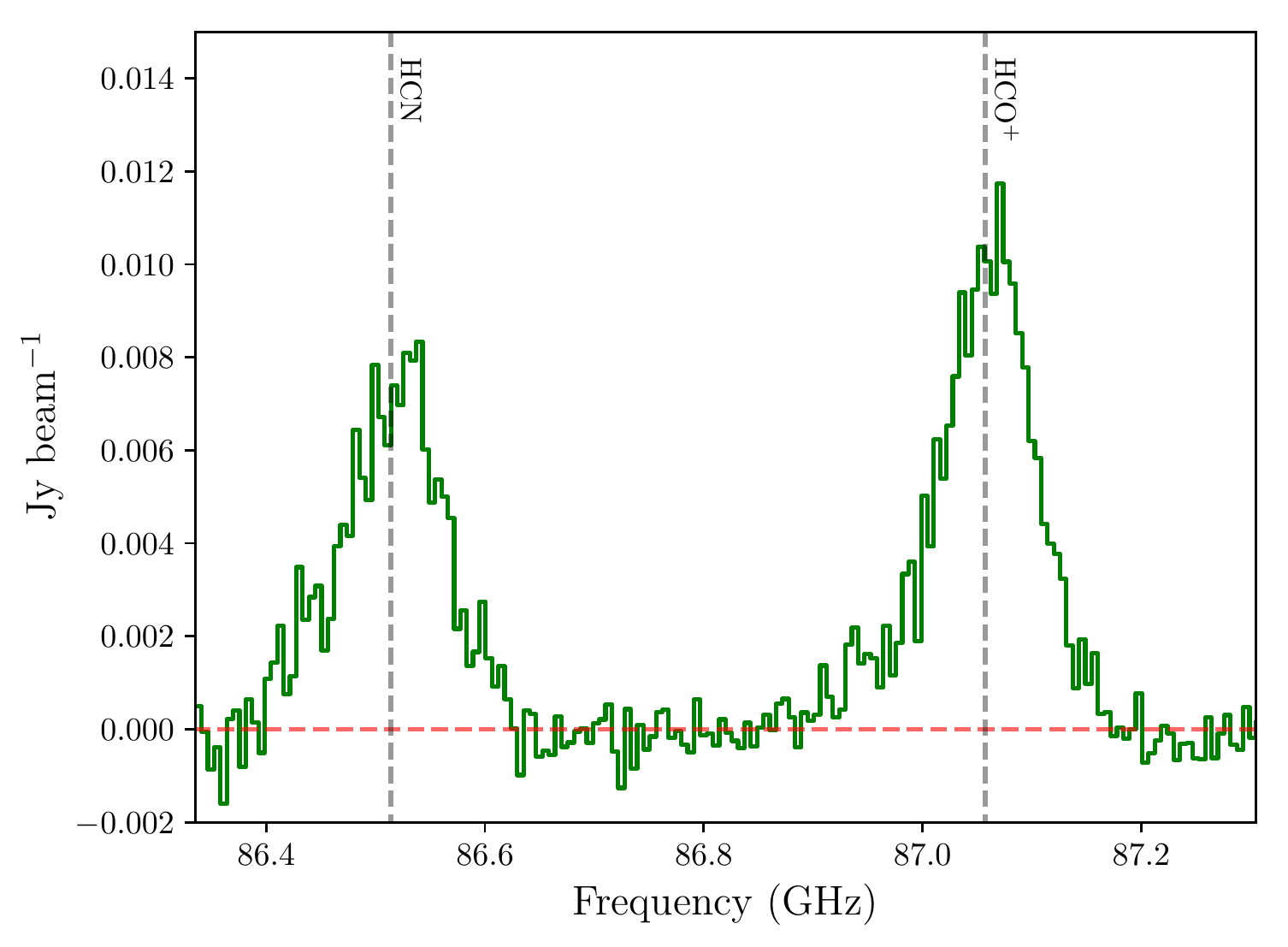} \\
  \end{array}$
\caption[]{ \textbf{Spectra averaged over a 1\arcsec\ diameter aperture for NGC 6240 centered on position (0,0).} }
\label{fig:n6240spec}
\end{figure*}

\section{Line Ratios}
For Arp 220, we create the following integrated line brightness temperature (T$_{B}$) ratio 
maps for  \co\ and \tco:

r$_{21}$ = T$_{\rm{B}} ^{^{12}\rm{CO}(2-1)}$  / T$_{\rm{B}} ^{^{12}\rm{CO}(1-0)}$,  

r$_{32}$ = T$_{\rm{B}} ^{^{12}\rm{CO}(3-2)}$ / T$_{\rm{B}} ^{^{12}\rm{CO}(2-1)}$,  

$^{13}$r$_{21}$ = T$_{\rm{B}} ^{^{13}\rm{CO}(2-1)}$ / T$_{\rm{B}} ^{^{13}\rm{CO}(1-0)}$,

R$_{10}$ = T$_{\rm{B}} ^{^{12}\rm{CO}(1-0)}$ / T$_{\rm{B}} ^{^{13}\rm{CO}(1-0)}$,

R$_{21}$ = T$_{\rm{B}} ^{^{12}\rm{CO}(2-1)}$ / T$_{\rm{B}} ^{^{13}\rm{CO}(2-1)}$.

We also create integrated T$_{B}$ line ratio maps for molecules other than CO as follows:

H$_{10}$  =  T$_{\rm{B}} ^{\rm{HCN}(1-0)}$ / T$_{\rm{B}} ^{\rm{HCO^{+}}(1-0)}$ and

\begin{figure*}[!htbp] 
\centering
$\begin{array}{c@{\hspace{0.5in}}c}
\includegraphics[ scale=0.3]{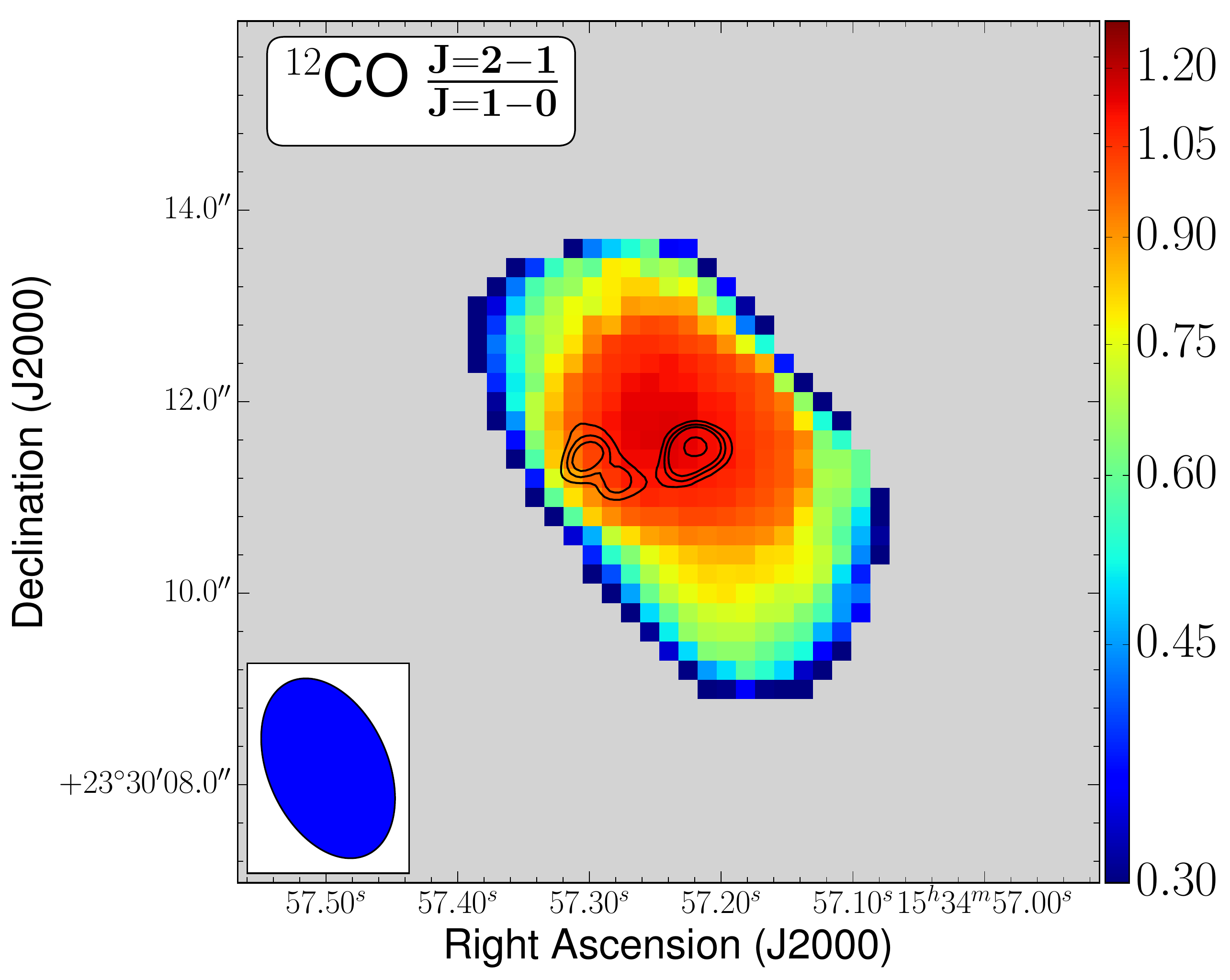}  & \includegraphics[scale=0.3]{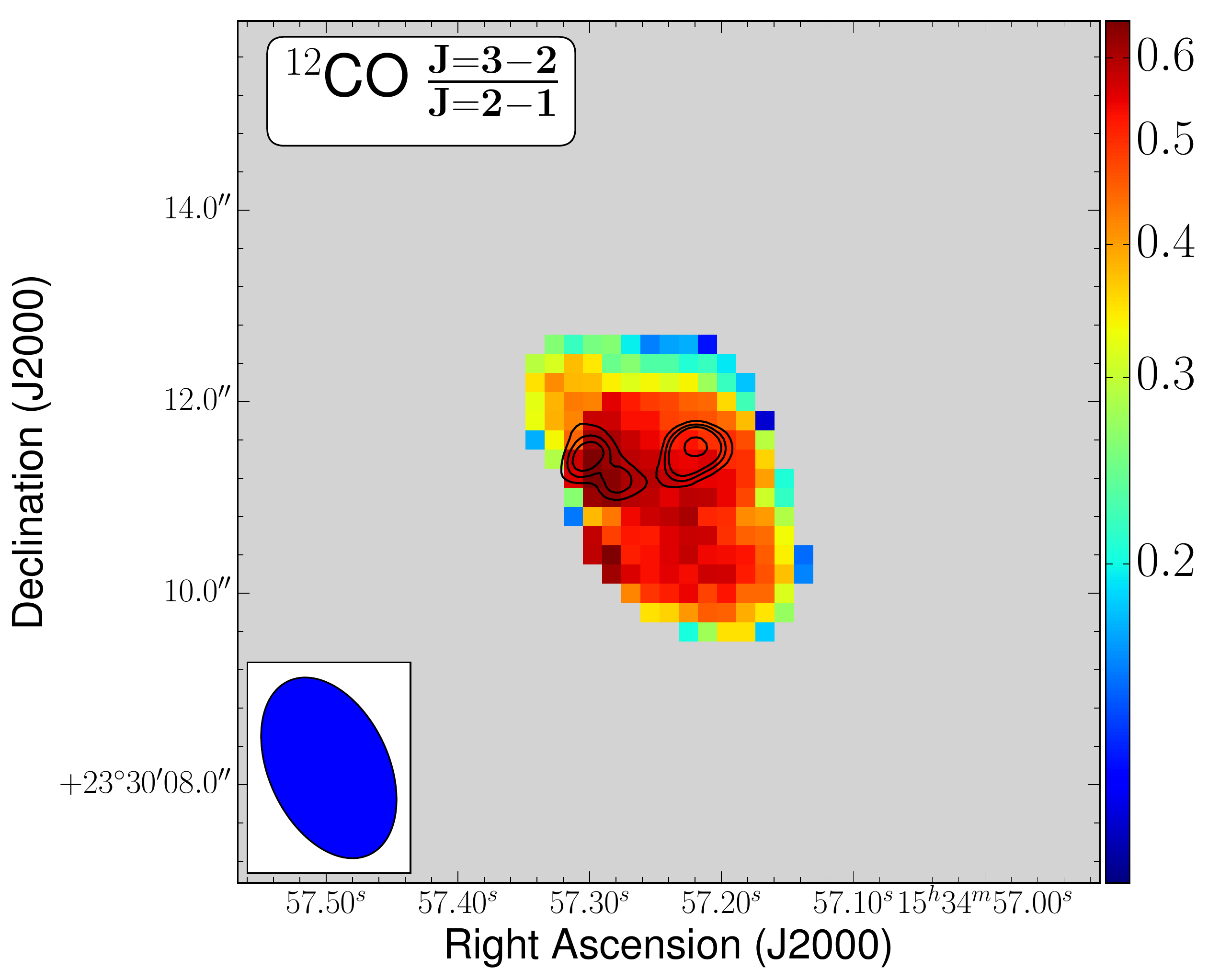} \\
\includegraphics[ scale=0.3]{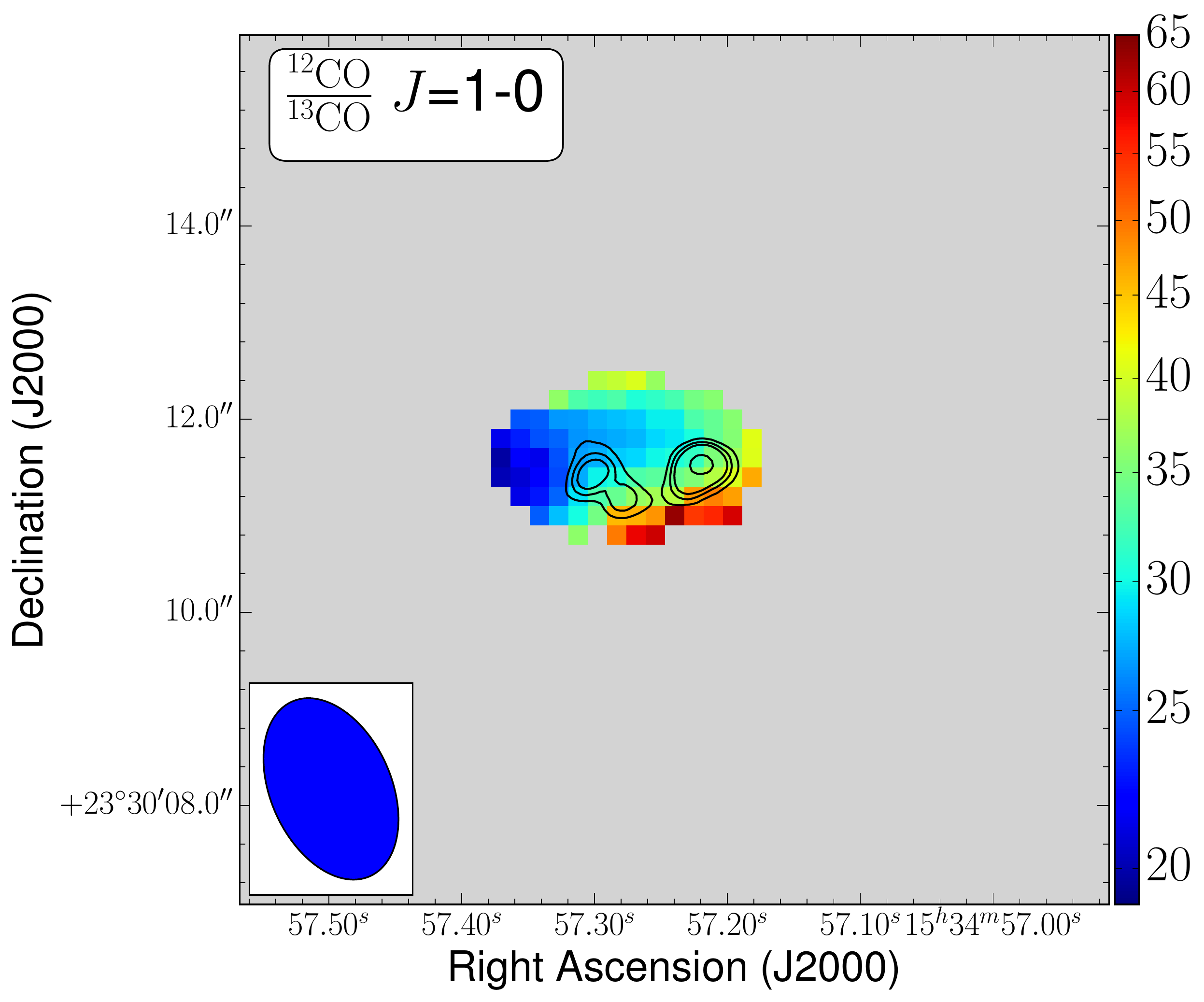}  & \includegraphics[ scale=0.3]{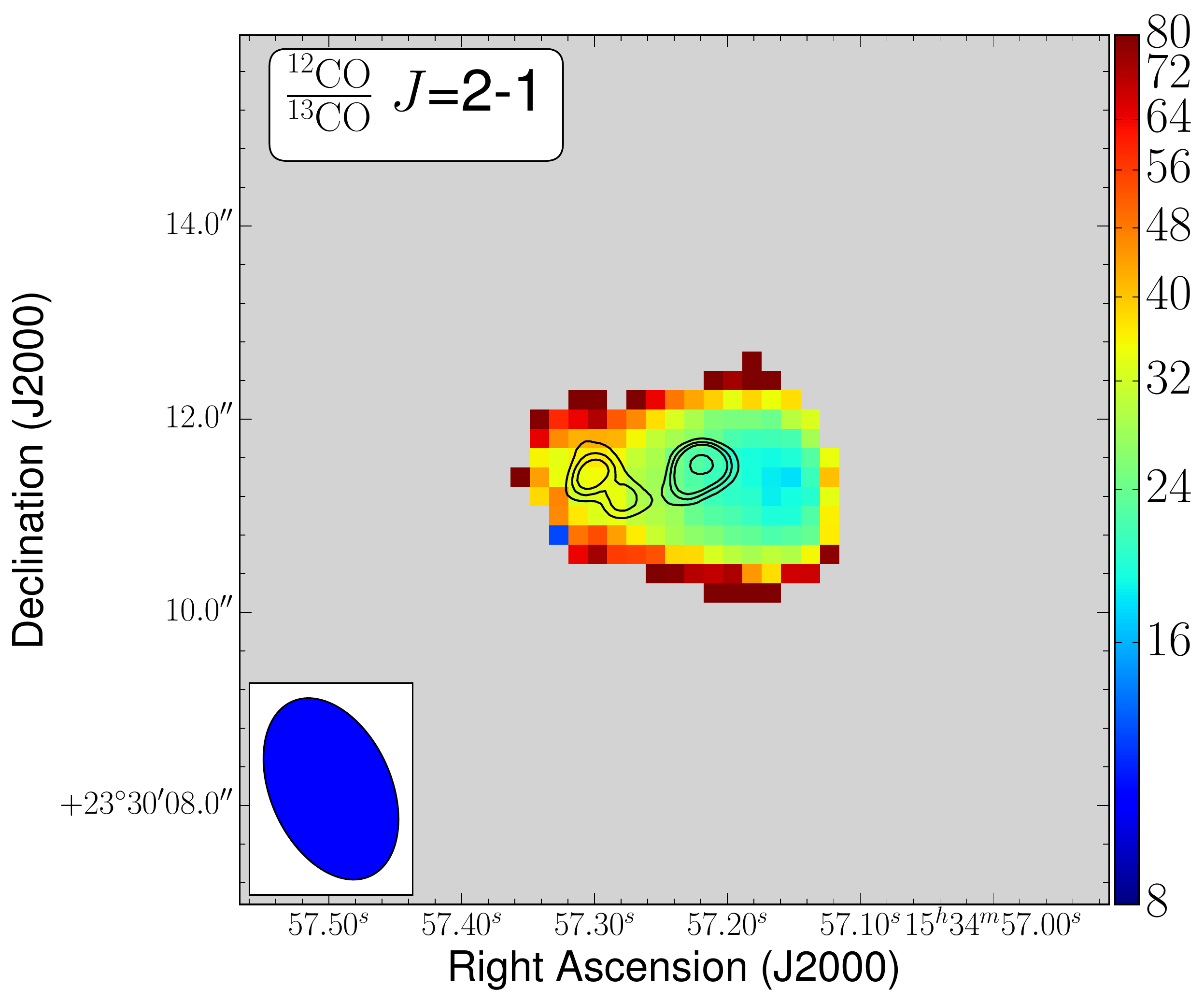}\\
\includegraphics[ scale=0.3]{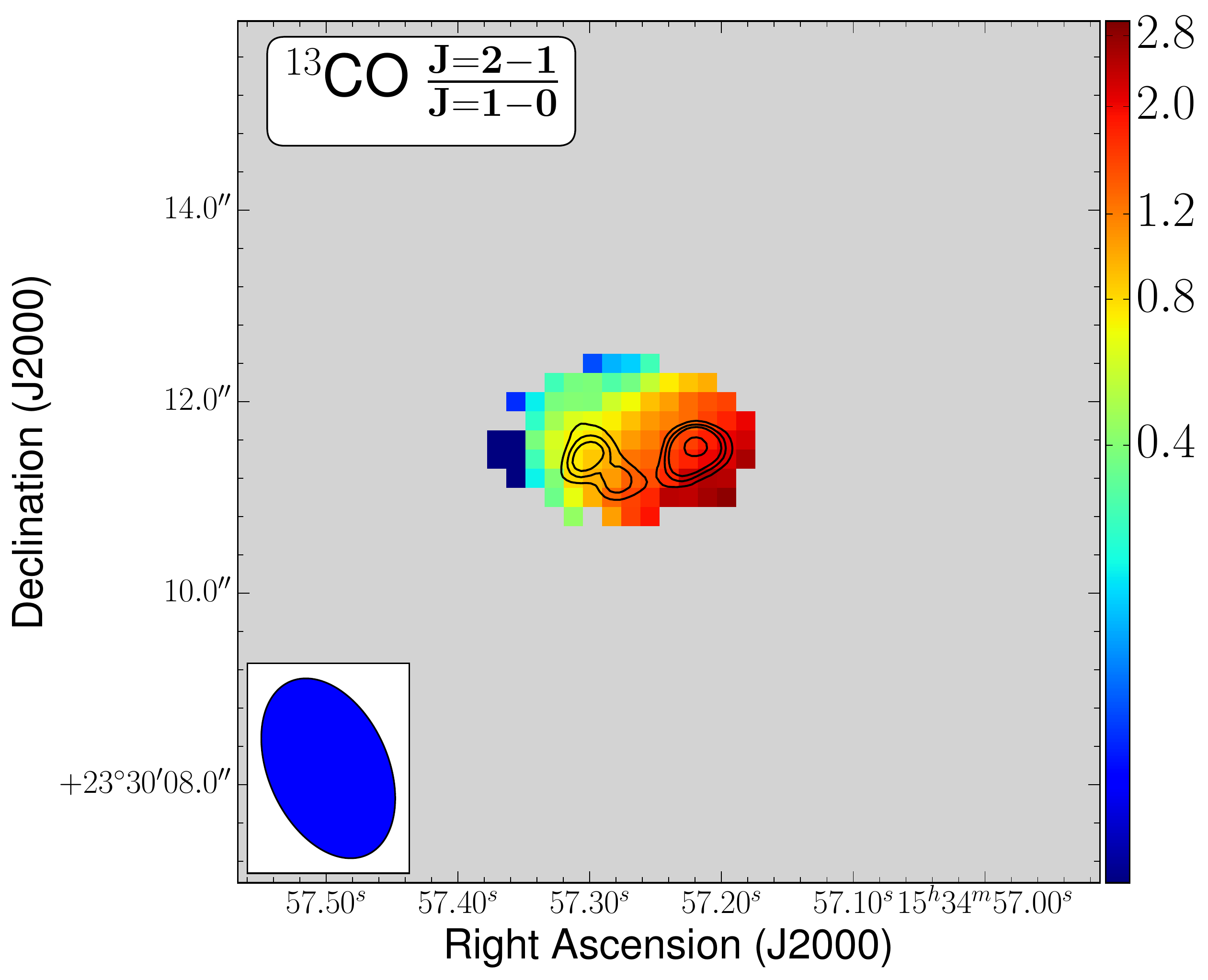}  & \includegraphics[scale=0.3]{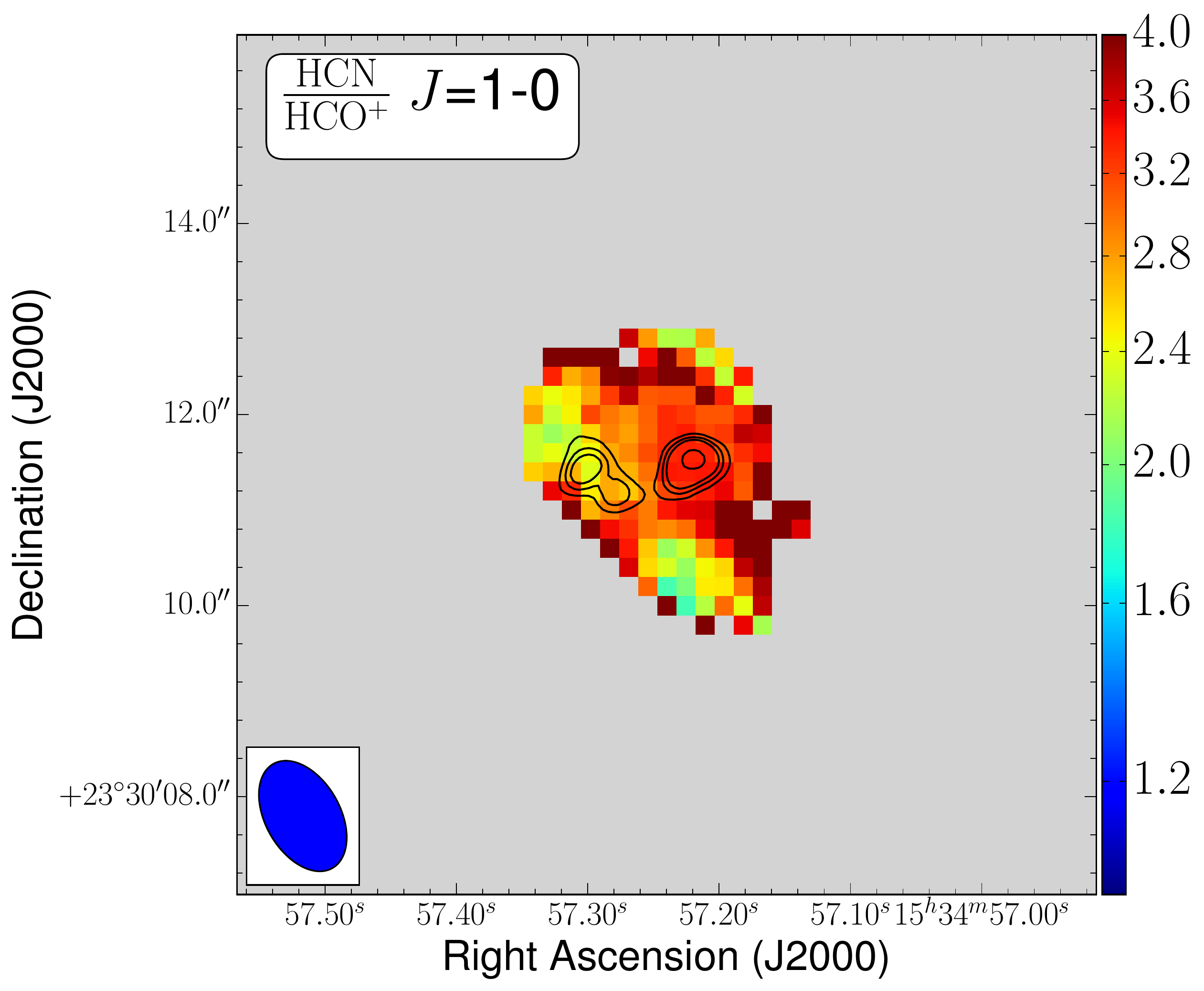} \\
\includegraphics[ scale=0.3]{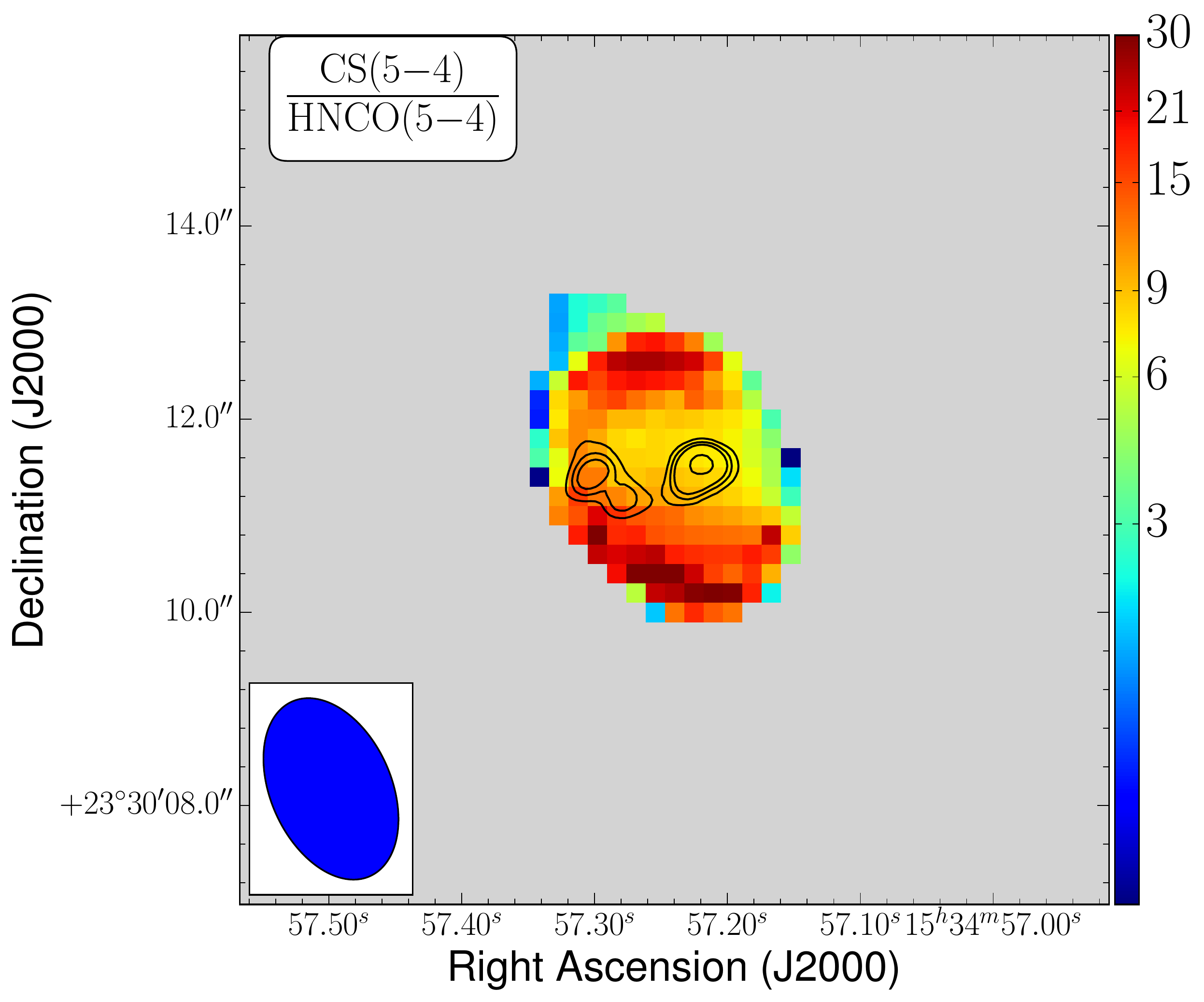} \\
\end{array}$
\caption{Arp~220 line ratios: (TOP) r$_{21}$ and r$_{32}$, (2nd ROW) R$_{10}$ and R$_{21}$, (3rd ROW) $^{13}$r$_{21}$ and H$_{10}$, (BOTTOM) CS / HNCO. The ellipse in the bottom left corner of each map represents the synthesized beam size. The black contours represent the high-resolution \cotwo\ emission published in \cite{Downes2007} to guide the eye to the positions of the two nuclei of  Arp~220. Note that the resolution of our observations do not spatially resolve the two nuclei. }
\label{fig:arp220lineratios}
\end{figure*}

CS / HNCO  = T$_{\rm{B}}^{\rm{CS(5-4)}}$  / T$_{\rm{B}}^{\rm{HNCO(5-4)}}$     (Figure \ref{fig:arp220lineratios}).

\begin{figure*}[!htbp] 
\centering
$\begin{array}{c@{\hspace{0.5in}}c}
\includegraphics[ scale=0.3]{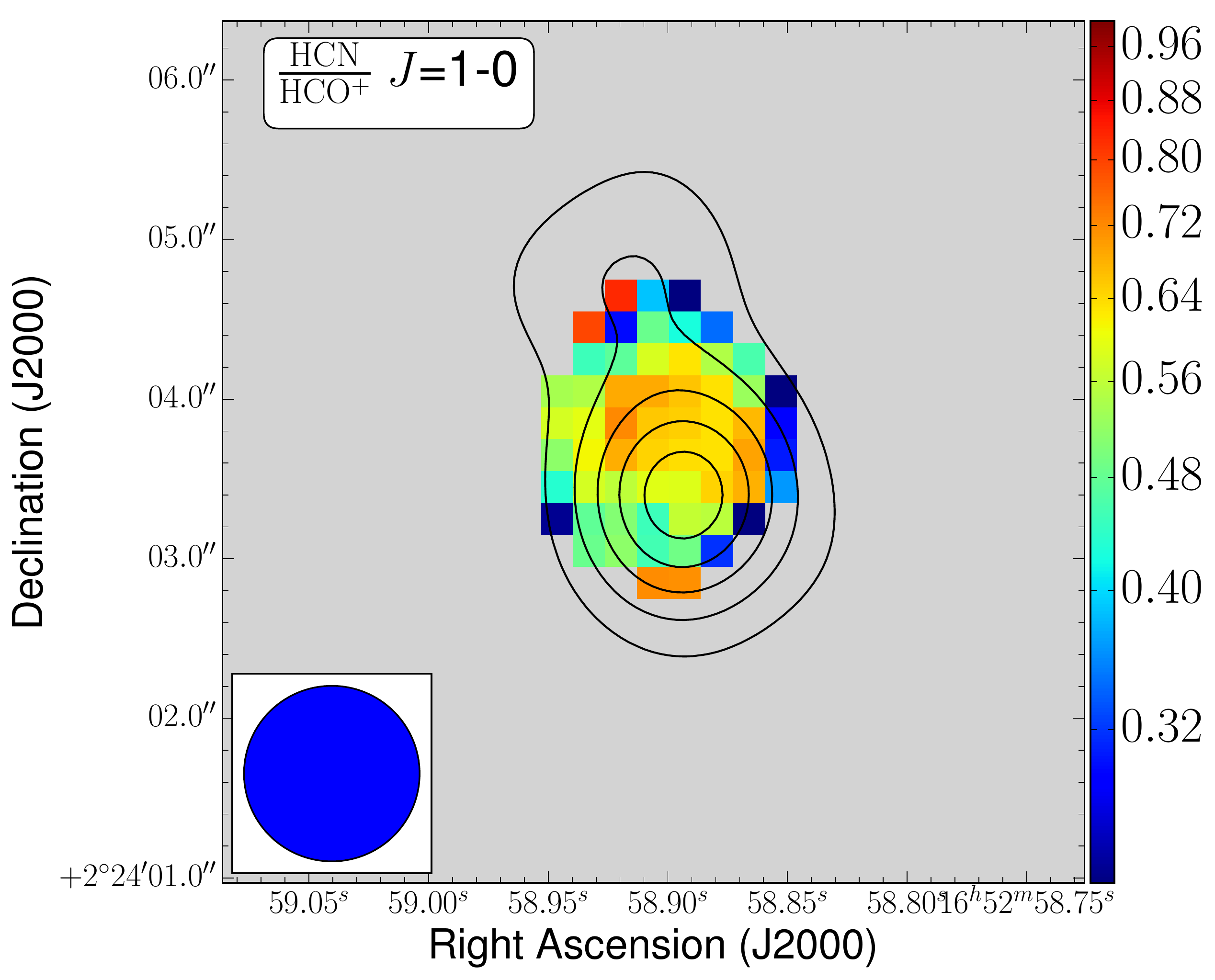}  &  \includegraphics[ scale=0.3]{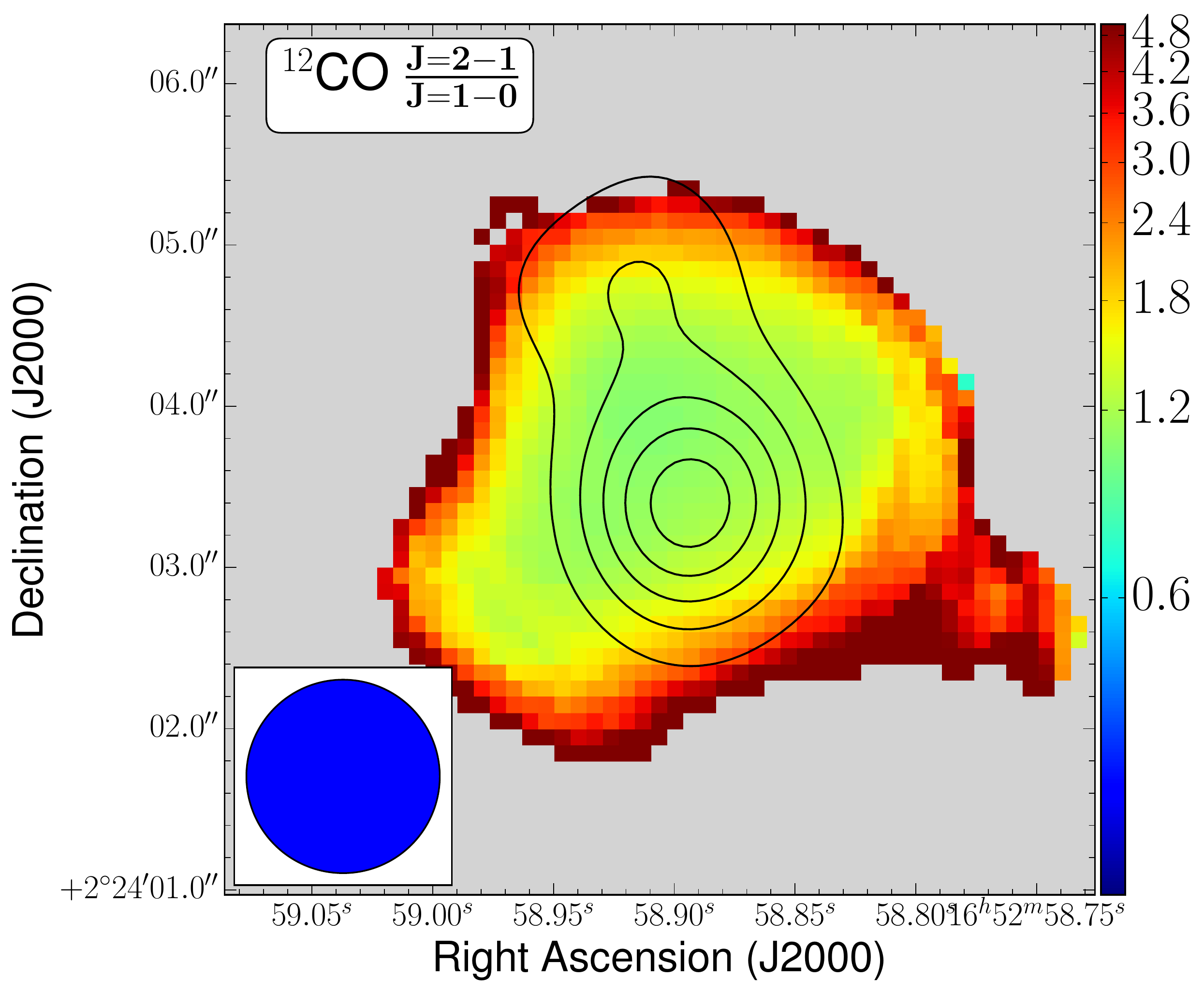} \\
\end{array}$
\caption[]{NGC 6240 Line ratios: (LEFT) H$_{10}$ and (RIGHT) r$_{21}$. The ellipse in the bottom left corner of each map represents the synthesized beam size. The black contours represent the 86~GHz continuum emission to guide the eye to the positions of the two nuclei of NGC~6240.}
\label{fig:ngc6240lineratios}
\end{figure*}

For NGC 6240, we show maps of the T$_{B}$ line ratios of 
r$_{21}$ and H$_{10}$ (Figure \ref{fig:ngc6240lineratios}; with both quantities defined as for Arp 220). Table \ref{tab:lineratios} gives a summary of the observed line ratios. 

To match the physical scales that we analyze, we smooth the data cubes using a Gaussian kernel to match angular resolution. For Arp 220, we smooth the data to the resolution of \tcoone\ (Table \ref{tab:observations}) and for NGC 6240, we use a compromised resolution of 1.2\arcsec, limited by the resolution of the \coone\ observations of \citet{Feruglio2013a}. We applied a 5$\sigma$ cut to each map used to produce the the ratio maps. 

Since \co\ is believed to be optically thick, the R$_{10}$ and R$_{21}$ line ratios give a lower limit to the true \xco\ abundance ratio. To illustrate this point, we start with the most general equation for the R line ratios,
\begin{eqnarray}\label{eqn:ratio}
\begin{aligned}
\rm{R} &= \frac{T_{\rm{B}}^{^{12}\rm{CO}}}{T_{\rm{B}}^{^{13}\rm{CO}}}  \\
\rm{R} &= \frac{T_{\rm{EX}}^{^{12}\rm{CO}}}{T_{\rm{EX}}^{^{13}\rm{CO}}} \frac{(1-\rm{e}^{-\tau_{^{12}CO}})}{(1-\rm{e}^{-\tau_{^{13}CO}})}
\end{aligned}
\end{eqnarray}
where T$_{\rm{EX}}$ is the excitation temperature, $\tau_{\rm{^{12}CO}}$ and $\tau_{\rm{^{13}CO}}$ are the optical depths of \co\ and \tco, respectively. If both \co\ and \tco\ were to be in local thermal equilibrium (LTE) so that their excitation temperatures were equal, then Equation \ref{eqn:ratio} simplifies to 
\begin{equation}\label{eqn:ratio2}
\rm{R} = \frac{(1-\rm{e}^{-\tau_{^{12}CO}})}{(1-\rm{e}^{-\tau_{^{13}CO}})} 
\end{equation}.
In addition, if the \co\ emission is sufficiently optically thick, so that $(1-\rm{e}^{-\tau_{^{12}CO}})$ $\rightarrow$ 1 and if \tco\ is sufficiently optically thin, so that $(1-\rm{e}^{-\tau_{^{13}CO}})$  $\rightarrow$  $\tau_{^{13}CO}$, then Equation \ref{eqn:ratio2} can be further simplified to 
\begin{eqnarray}\label{eqn:ratio}
\begin{aligned}
\rm{R} &= \frac{1}{\tau_{^{13}CO}}\\
\rm{R} &= \frac{\tau_{^{12}CO}}{\tau_{^{13}CO}} \frac{1}{\tau_{^{12}CO}}\\
\rm{R} &= \frac{[^{12}CO]}{[^{13}CO]}\frac{1}{\tau_{^{12}CO}}
\end{aligned}
\end{eqnarray}
where the ratio of the optical depths of the isotopologues are equivalent to the abundance ratio ($\tau$ $\propto$ column density). In this case, the observed ratio R of line brightness temperatures, will always be less than the true abundance, \xco, because of the attenuating factor of the \co\ optical depth. 

Only if \co\ and \tco\ are both optically thin, will the observed ratio of line brightness temperatures directly trace the abundance ratio, providing that the \co\ and \tco\ excitation temperatures are equal.

As is well-known, however, because of resonant trapping in the CO lines, a more realistic approach is to use one of the standard non-LTE approximations, which allows for different excitation temperatures in the different CO lines, and this is what we actually do (Section 4).

\begin{table*}
\caption{Line Ratios}\label{tab:lineratios}
\centering
\begin{tabular}{lcccccccc}
\hline\hline
			& \multicolumn{2}{c}{\underline{Position}} & & & & &  \\
			&RA (J2000)&Dec (J2000)&r$_{\rm{21}}$	& r$_{\rm{32}}$	 & R$_{\rm{10}}$	&R$_{\rm{21}}$	&$^{13}$r$_{\rm{21}}$ &  H$_{10}$		 \\
\hline
Arp 220E		&15 34 57.342 	&+23 30 11.610	 &0.8 $\pm$ 0.1	&0.8 $\pm$ 0.1		&23 $\pm$ 3	&19 $\pm$ 2	&0.9 $\pm$ 0.1 &2.4 $\pm$ 0.2	 \\
Arp 220W		&15 34 57.197	&+23 30 11.595	 &1.1 $\pm$ 0.1	&0.60 $\pm$ 0.08	&38 $\pm$ 4	&19 $\pm$ 2	&2.3 $\pm$ 0.3&	2.7 $\pm$ 0.2\\
Arp 220C		&15 34 57.258	&+23 30 11.592	 &1.2 $\pm$ 0.1  &0.60 $\pm$ 0.08	&31$\pm$ 3	&26 $\pm$ 3	&1.3 $\pm$ 0.1&2.5 $\pm$ 0.2	\\
NGC 6240	&16 52 58.900 & +02 24 03.810	 &1.0 $\pm$ 0.1  &...	&...	&...	&...	&0.80 $\pm$ 0.09	 \\
\hline 
\end{tabular}
\newline
\textbf{NOTE}: The positions for Arp 220E/W do not correspond to the positions of the two nuclei. Our resolution is too poor to distinguish the nuclei spatially, therefore, the Arp 220E/W positions are $\sim$ 2\arcsec\ apart (greater than the synthesized beam major axis) with Arp 220C as the central position. Position error = $\pm$0.1\arcsec
\end{table*}

\section{Radiative Transfer Analysis}\label{sec:rad}
To model the \co\ emission (and that of the rarer CO isotopologues), we use the escape-probability radiative transfer program RADEX \citep{vanderTak2007}.  To find the most likely RADEX solution, given the observed line strengths, and other constraints, we use the Monte Carlo Markov Chain code\footnote{https://github.com/jrka/pyradexnest} of \citet{Kamenetzky2014}.  This code implements the nested sampling algorithm MultiNest \citep{Feroz2009} using its Python wrapper PyMultiNest \citep{Buchner2014} to constrain parameters. As stated in \citet{Kamenetzky2014}, ``The algorithm `nests inward' to subsequently smaller regions of high-likelihood parameter space. Unlike calculating the likelihood using a grid method, the algorithm can focus on high likelihood regions and thus estimate parameter constraints more efficiently." Model points are generated using  RADEX with the following input parameters: kinetic temperature (\tkin), column density of molecular species X per unit line width (\nx/$\Delta$V) and volume density of the collision partner, molecular hydrogen (\nhtwo). In addition, the resulting flux is allowed to scale uniformly down by an area filling factor, \ff\ $\leq$1. We also implement three priors: 
\begin{enumerate}
\item A column length to constrain the diameter of the molecular emission region. This prior assists in constraining the column and volume densities. We estimate the upper limit to the column length to be the diameter of the synthesized beam ($\sim$~700~pc). 
\item A dynamical mass (\mdyn) as an upper limit to the total mass that can be contained within the column density. We use the equation from \cite{Wilson1990} assuming a diameter of $\sim$700~pc and velocity FWHMs from literature presented below.
\item An optical depth range of 0 to 100. An optical depth below 0 implies maser emission and the upper limit of 100 is recommended by the RADEX documentation.
\end{enumerate}
 We refer the reader to Kamenetzky et al. (2014) for more details. 

\begin{figure}[!htbp] 
\centering
\includegraphics[ scale=0.6]{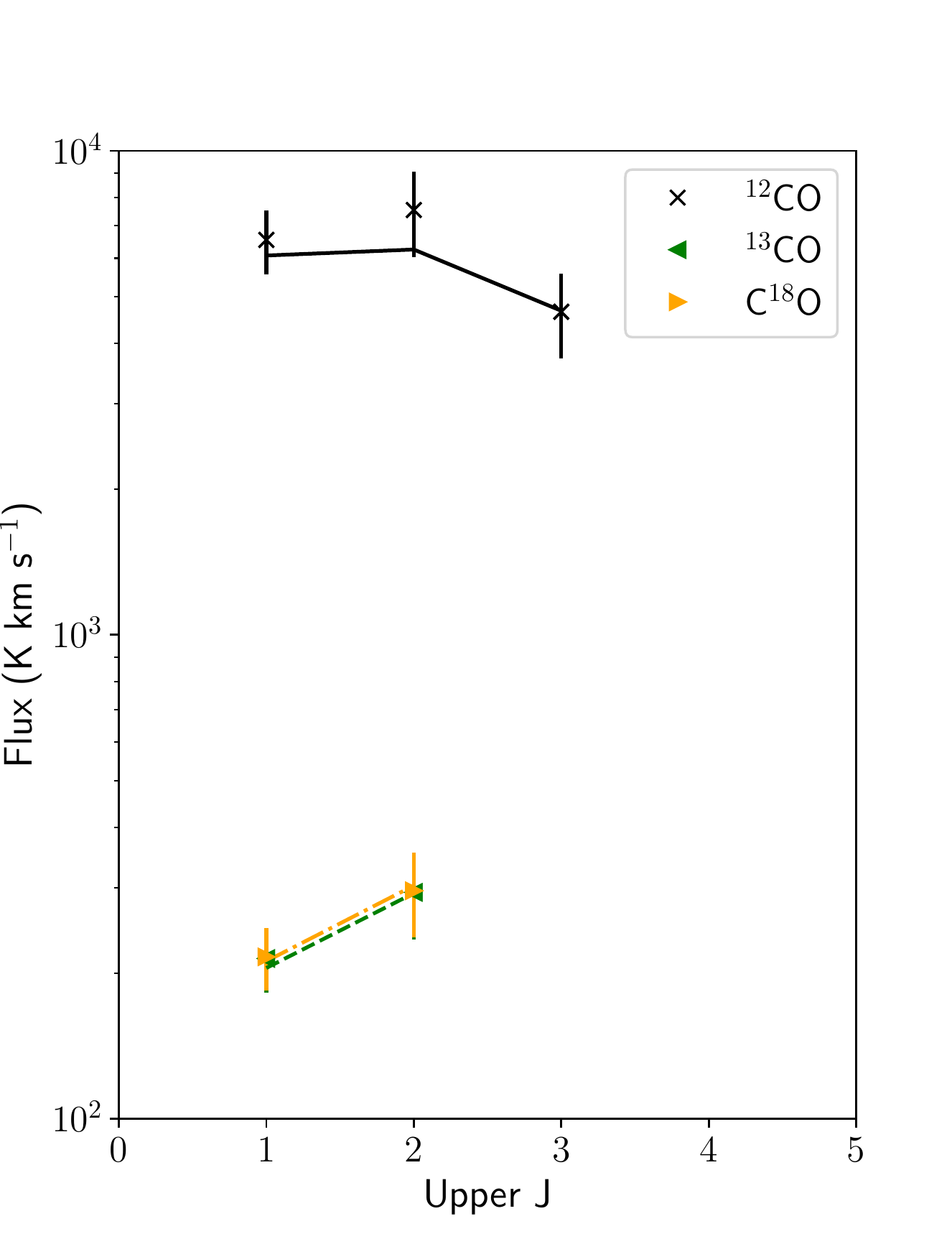}  \\
\caption[]{Arp 220C SLED for \co\ (black x), \tco\ (green left triangle) and \ceo\ (orange right triangle). Solid and dashed lines represent the most probable SLED solution for each molecule. }
\label{fig:sleds}
\end{figure}

\begin{figure*}[!htbp] 
\centering
$\begin{array}{c@{\hspace{0.5in}}c}
\includegraphics[ scale=0.4]{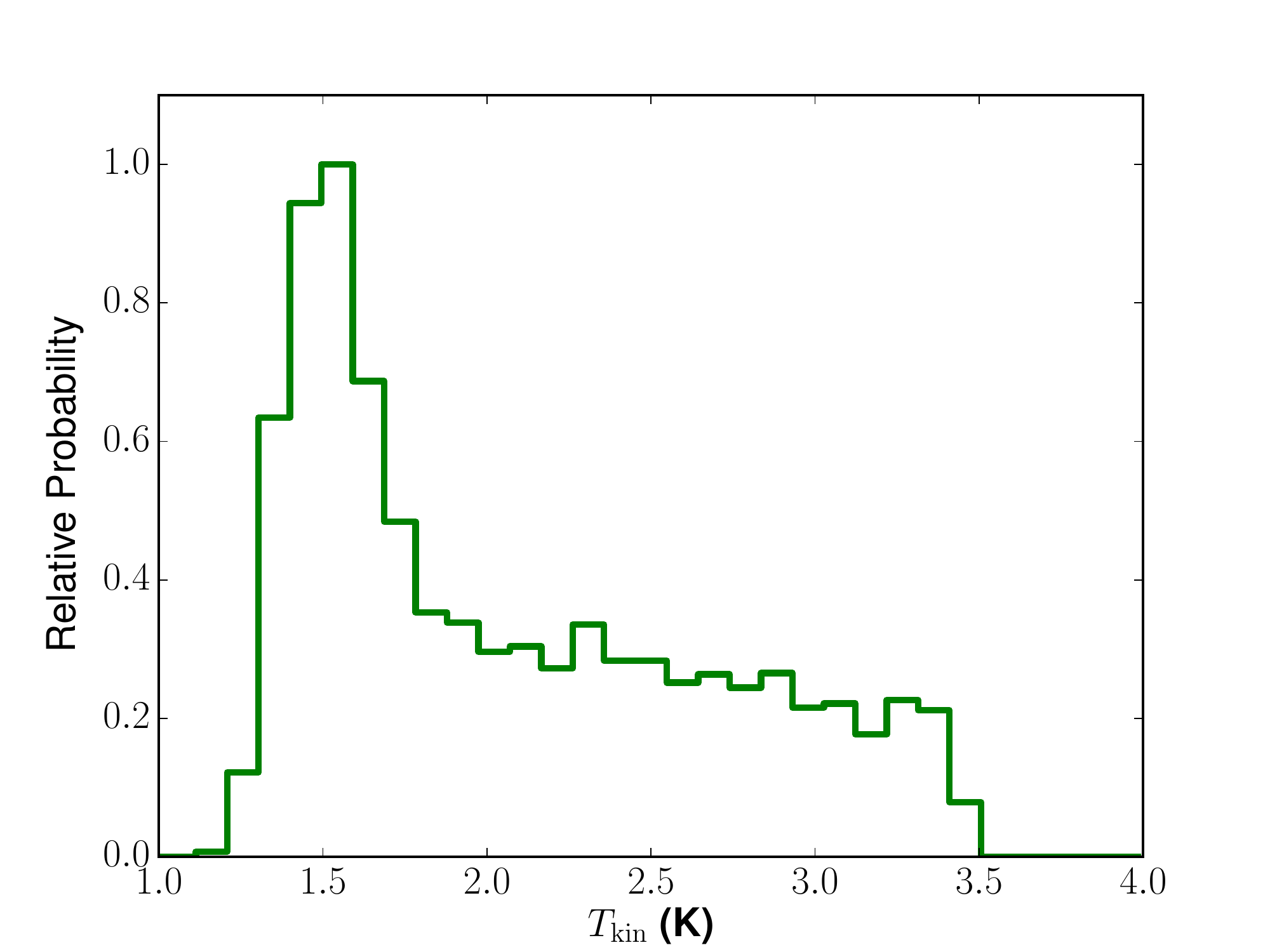}  & \includegraphics[scale=0.4]{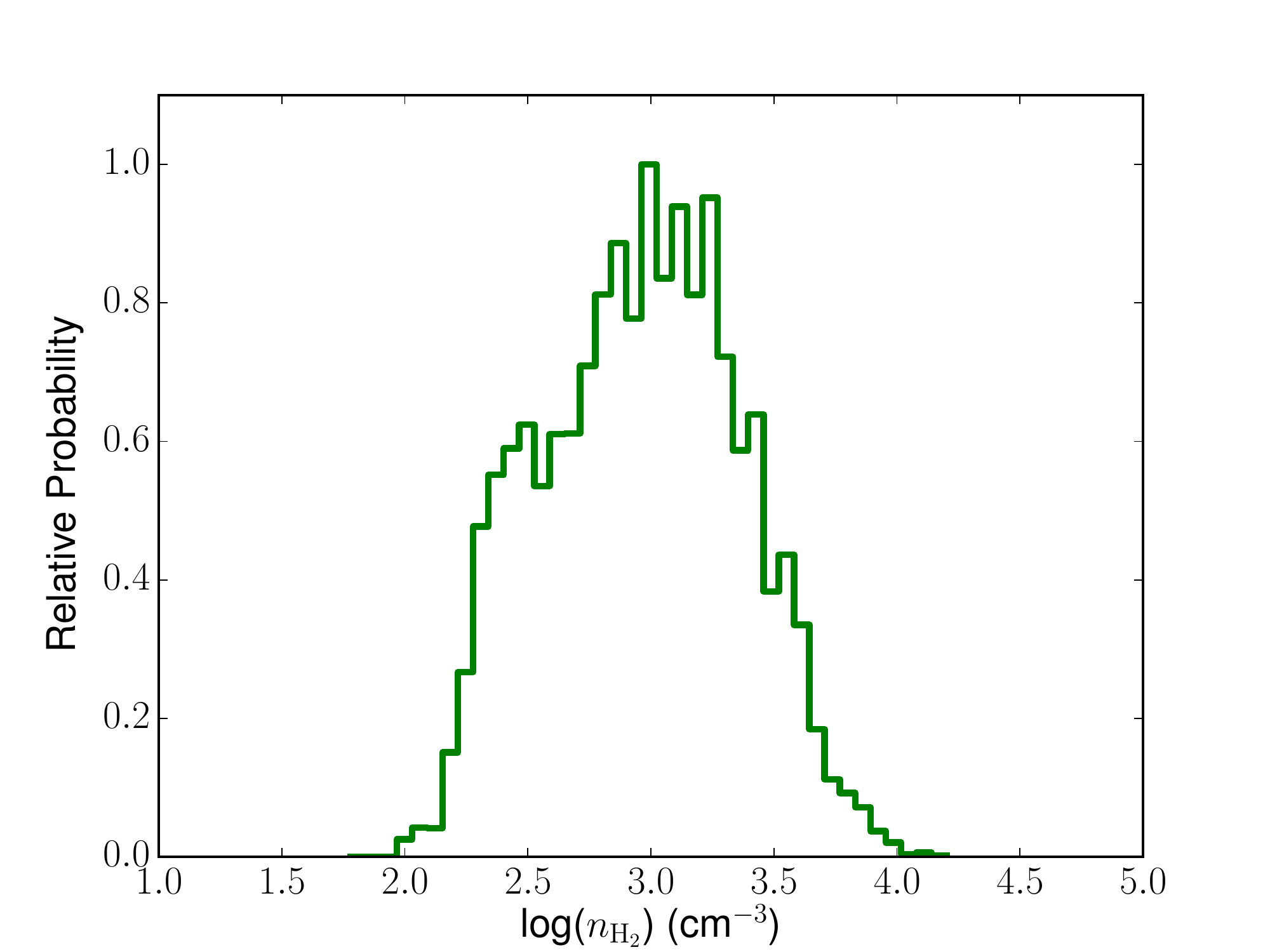} \\
\includegraphics[ scale=0.4]{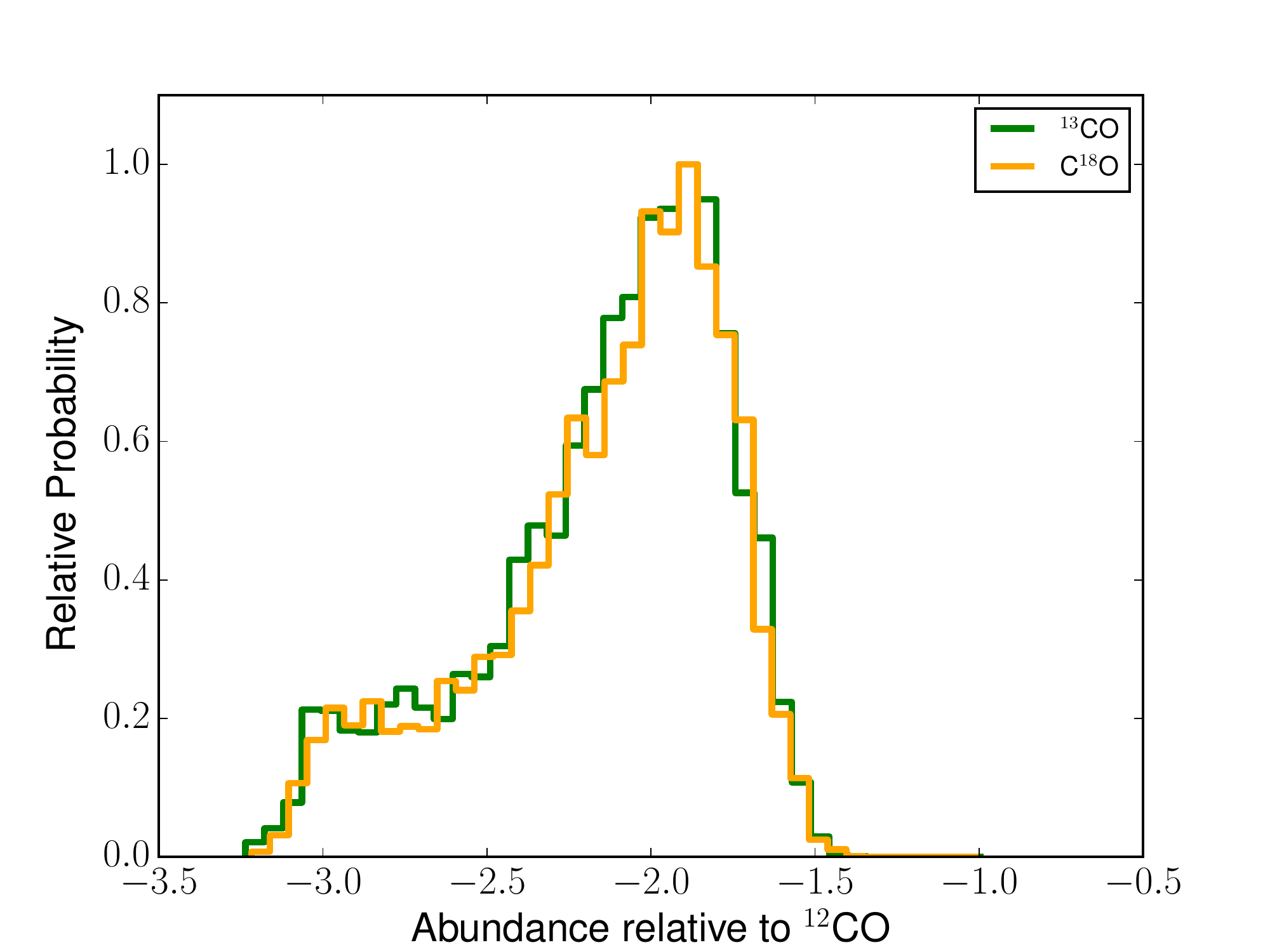}  & \includegraphics[ scale=0.4]{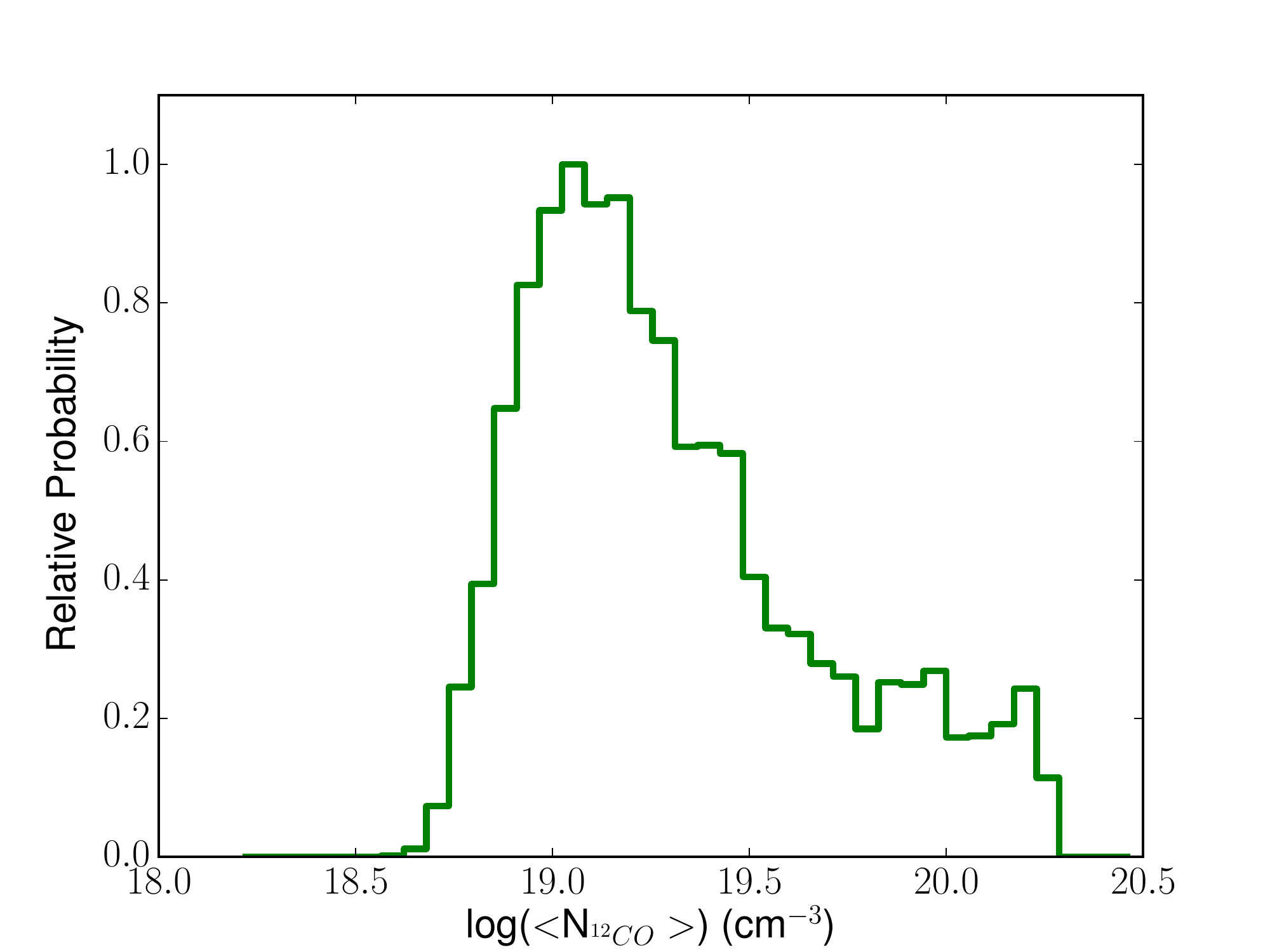} \\
\end{array}$
\includegraphics[scale=0.4]{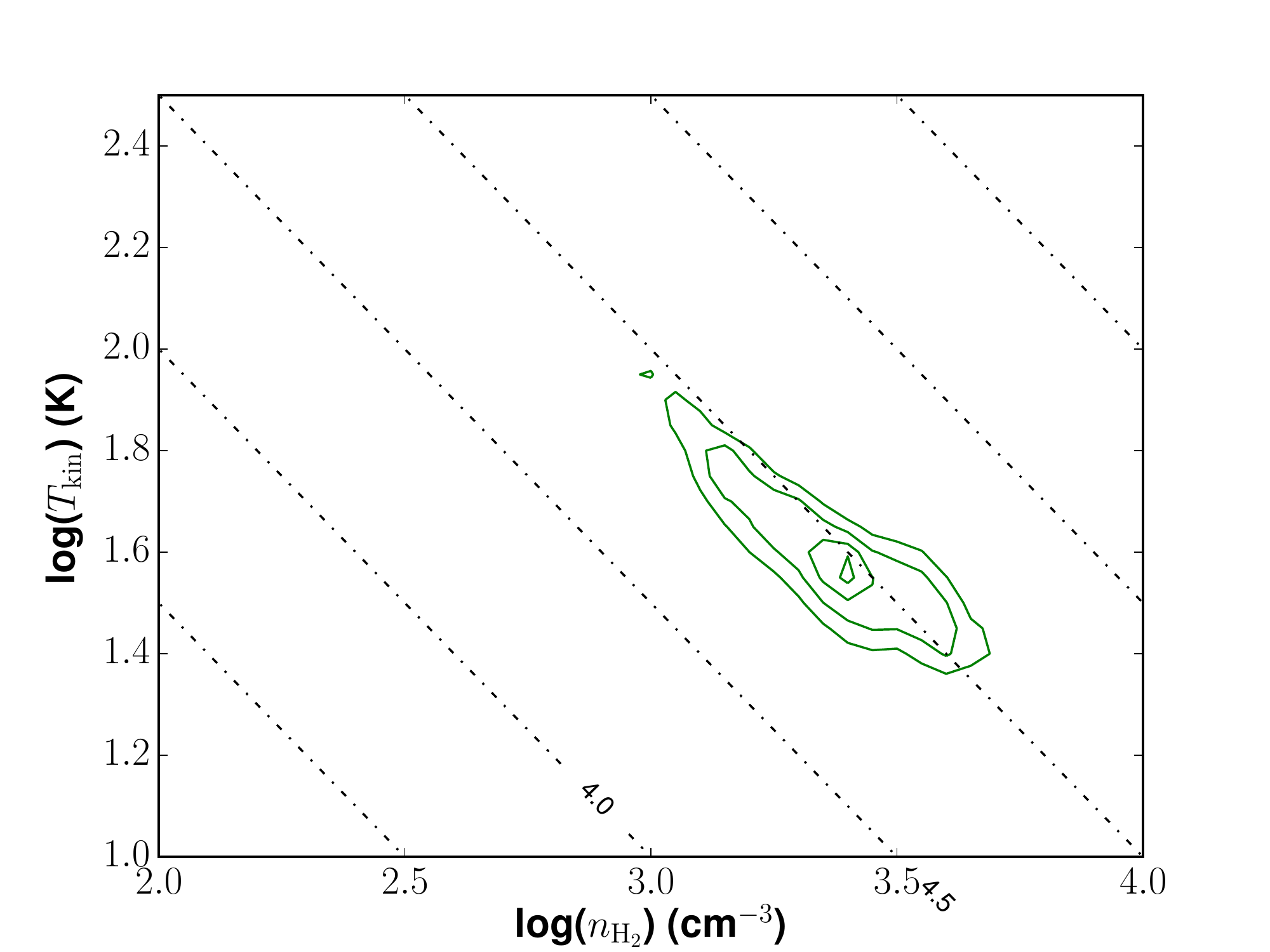} \\
\caption[]{Probability distributions for Arp 220C. (\textit{Top Left}) Temperature,  (\textit{Top Right}) Volume density,  (\textit{Middle Left}) Abundance of \tco\ and \ceo\ relative to \co.  (\textit{Middle Right}) Average column density of \co. (\textit{Bottom}) Temperature versus volume density. Contours are 40, 60, 80 and 95$\%$ of the most probable solution. Dashed lines represent log(pressure).}
\label{fig:arp220prob}
\end{figure*}

We model three molecular species simultaneously: \co, \tco\ and \ceo. The \co\ lines modelled are the $J$=1-0, 2-1 \citep{Downes1998} and 3-2 \citep{Sakamoto2008}. We use a line ratio of \tco/\ceo\ = 1 \citep{Matsushita2009,Greve2009} to estimate the emission of \ceo\ at our angular resolution ($\sim$2\arcsec).  Due to the lack of resolution elements across the \tco\ emission of Arp 220, we model the molecular gas at the central peak position of \tcoone\ (see Table \ref{tab:observations}; Arp 220C) since the peak emission falls in between the two nuclei  and 1\arcsec\ to each side of Arp 220C, which will place Arp 220E/W more than one resolution beam apart. We stress that Arp 220E/W are not the positions of the nuclei and are only modelled for completeness. For Arp 220W, we used only the $J$=1-0 and 2-1 lines because the addition of the $J$=3-2 line resulted in poor SLED fits, likely indicating the presence of a second molecular gas component that cannot be fit with our data. We adopt the following linewidths: 120 \kms\ and 250 \kms\ for Arp~220W and Arp~220E measured using ALMA \coone\ observations at 0.09\arcsec\ (37~pc) \citep{Scoville2017}, respectively, and 320 \kms\ for Arp 220C \citep{Downes1998} measured from PdBI \coone\ observations. The mean, best fit and $\pm$1$\sigma$ results of \tkin, \nhtwo, \nco, \ff, thermal pressure, molecular gas mass, CO-to-H$_{2}$ conversion factor (\alphaco) and the relative abundances of \tco\ and \ceo\ to \co\ per model point are presented in Table \ref{tab:results}. The best fit SLED for Arp 220C is presented in Figure \ref{fig:sleds}. The (marginal) probability of a single parameter is computed by integrating over all other parameters. We present the marginal probability distributions of \tkin, \nhtwo, \nco\ and relative abundances of \tco\ and \ceo\ as well as the 2D probability distribution of log$_{10}$(\tkin) versus log$_{10}$(\nhtwo)  in Figure \ref{fig:arp220prob}. 

We do not model NGC 6240 due to the lack of available high-resolution observations of \tco\ and the significant missing \co\ flux in the existing high-resolution maps. We refer the reader to \citet{Tunnard2015b} for an extensive radiative transfer modelling of the molecular gas in NGC 6240. 

\begin{table*}
\scriptsize
\caption{Radiative Transfer Results}\label{tab:results}
\centering
\begin{tabular}{clccccccccc}
\hline\hline
& 	&\tkin\	&log$_{10}$(\nhtwo)	&log$_{10}$(P) &log$_{10}$(\ff)	&log$_{10}$($<$\nco\ $>$) &log$_{10}$(\mmol) & \alphaco\ &	\xco\		& \xceo\  \\
&	&(K)	&(cm$^{-3}$)	& (K cm$^{-3}$) &	&(cm$^{-2}$) &(\msol)  &(\alphacou)	&  & \\ \relax
&[1]	&[2]&[3]&[4]&[5]&[6]&[7]&[8]&[9]&[10] \\
\hline
Arp 220C&Mean 	& 130 & 2.99 & 5.2 &-0.30 	& 19.33 	&8.81	&0.4 &125	 & 125		\\
	&Best Fit 		& 38 	& 3.35 &	4.9& -0.036 	&19.20 	&8.69 	&0.3 & 90	 &	91	\\
	&-1$\sigma$ Value	&37 	& 2.55  &	4.9 & -0.53 	&18.97	& 8.46 	&0.2 &58	 &	56	\\
	&+1$\sigma$ Value	&690 & 3.40 &	5.4 & -0.074	&19.74 	& 9.23 	&1.0	&309   & 323\\
		\hline

Arp 220E &Mean 		&240 & 2.54 & 4.9 &-0.22 	&19.04 	&8.53	&0.4 &93	 & 93		\\
		&Best Fit 		&34 	 & 3.10 &	4.6& -0.017 	&19.24	&8.73 	& 0.6 & 159	 &	159	\\
		&-1$\sigma$ Value	&10 	 & 2.8  &	4.3 & -0.17	&19.03	& 8.52 	&0.4 &100	 &	100	\\
		&+1$\sigma$ Value	&110 & 3.4 &	4.9 & 0.0	&19.45 	& 8.94 	&1.0	&255   & 255\\
		\hline

Arp 220W&Mean 	& 105 & 3.78 & 5.8 &-0.32 	& 19.36 	&8.85	&0.5 &151	 & 149		\\
	&Best Fit 		& 300 	& 3.2 &	5.7& -0.7 	&19.34 	&8.83 	&0.5 & 142& 143	\\
	&-1$\sigma$ Value	&90 	& 2.3  &	5.0 & -0.91	&18.96	& 8.45 	&0.2 &62	 &	62	\\
	&+1$\sigma$ Value	&1000 & 4.1 &	6.3 & -0.48	&19.71 	& 9.20	&1.1	&330   & 330\\
		\hline	
		
\end{tabular}
\textbf{NOTES:} Col [1]: Statistic;  Col[2] Kinetic Temperature; Col [3] Volume Density; Col[4]: Thermal Pressure Col[5]: Filling Factor; Col[6]: \co\ Column Density; Col[7]: Molecular Gas (H$_{2}$) mass assuming [\co]/[H$_{2}$] = 3 $\times$ 10$^{-4}$; Col[8]: Conversion Factor \alphaco\ = 1.36\mmol/$L_{\rm{CO}}$); Factor of 1.36 is to account for He; Col[9]: \xco\ abundance ratio; Col[10]: \xceo\ abundance ratio; The $\pm\sigma$ values denote the range of values within 1$\sigma$. 
\end{table*}

\section{Discussion}
\subsection{Molecules Detected}
\textbf{\tco\ (\textit{Second most common isotopologue of carbon monoxide, after $^{12}$CO.} )} -The \tcotwo\ peak falls near the western nucleus of Arp 220 while the \tcoone\ peak falls in between the two nuclei. Both \tcoone\ and $J$ = 2-1 have an east-west morphology instead of a north-east south-west morphology of the \co\ maps which suggests that very little \tco\ emission originates from the diffuse kiloparsec disk surrounding the nuclei. The \tco\ lines (and other optically thin tracers) are important to truly trace the physical conditions of the molecular gas as they can probe deeper into the molecular gas ensemble than the optically thick \co\ (i.e. brick wall effect). This is apparent in the line ratio maps (Figure \ref{fig:arp220lineratios}) where the \co\ line ratios do not vary greatly across the system while the \tco\ line ratios show an east-west gradient. 

\textbf{HCN (\textit{Hydrogen cyanide}) and \hco\ (\textit{Formyl ion})} - \textit{Arp 220:} The \hcn\ and \hcoone\ emission in Arp 220 peaks at the same position near the western nucleus which may suggest more dense gas in the west than found near the eastern nucleus. A 2D Gaussian fit of the emission shows that the size of HCN (1.4\arcsec\ $\times$ 1.0\arcsec; size deconvolved from the beam \textbf{assuming a Gaussian structure}) is more compact than \hco (1.9\arcsec\ $\times$ 1.4\arcsec). The spectra (Figure \ref{fig:arp220spec} show absorption features in both HCN and HCO$^{+}$ similar to that seen in \cite{Martin2016} and \cite{Scoville2017}. The absorption appears to have the same depth in both lines, which seem to be absorbing nearly all of the continuum. The main difference between the two lines is that HCN has more emission. 

\textit{NGC 6240:} The integrated line flux from the \hcn\ and \hcoone\ is roughly half the single-dish flux measured by \citet{Greve2009}. This is evidence of a larger scale HCN and \hcoone\ emission that is filtered out by the interferometer. With our shortest baseline ($\sim$ 80~m), 50\% of the flux is at scales greater than 3\arcsec\ ($\sim$1.5~kpc). Future observations need to include shorter spacings to recover all flux.

As also observed for Arp 220, the HCN emission is more compact than that for \hco\ which is evident in the integrated intensity maps (Figure \ref{fig:ngc6240maps}). A 2D Gaussian fit on the integrated intensity maps shows that HCN and \hcoone\ are unresolved at our angular resolution (1.1\arcsec) with an upper limit to the source size of 0.5\arcsec\ for HCN and 0.9\arcsec\ $\times$ 0.5\arcsec\ for \hco. The upper limits still indicate that \hcoone\ is coming from a more extended component which supports the analysis of \citet{Greve2009} suggesting that HCN traces a denser gas phase than \hco.

\textbf{C$_{2}$H (\textit{Ethynyl})} - The C$_{2}$H $N$ =1-0 transition consists of six hyperfine structure lines ($J$=3/2 -1/2 F=1-1, 2-1, 1-0 and $J$=1/2-1/2 F=1-1, 0-1, 1-0). Due to the low spectral resolution, the six lines are blended in our observations for both sources.
\textit{Arp 220:} The \cth\ emission is peaked near the western nucleus.
\textit{NGC 6240:} The \cth\ emission is peaked in between the two nuclei, similar to the HCN and \hco\ emission.

\textbf{SiO (\textit{Silicon monoxide})} -  The SiO $J$=2-1 emission from Arp 220 has been previously observed by \citet{Tunnard2015a} at a similar spatial resolution. The total flux we measure agrees with \citet{Tunnard2015a}; however, we note that there may be contamination from H$^{13}$CO$^{+}$ $J$=1-0. \citet{Tunnard2015a} explore this possible contamination. We also do not see an absorption feature with the SiO $J$=2-1 line profile in Arp 220W, however, for Arp 220E there is a hint of an absorption feature (Figure \ref{fig:arp220spec}). Since the detection in the Arp 220E is poor, it is difficult to conclude whether this is a real feature. We note that \cite{Tunnard2015a} do not see an absorption feature in SiO $J$=2-1.

\textbf{HNCO (\textit{Isocyanic acid})} and \textbf{CS (\textit{Carbon monosulfide})}: Only part of the HNCO line is observed due to bandwidth constraints and may contain contamination from \ceo\ due to the broad linewidths of Arp 220. \citet{Martin2009} observed \hncofive\ in Arp 220 using the IRAM 30m and measured a flux of $\sim$5.5 Jy \kms, more than twice our measured flux. The emission of HNCO has a horizontal elongation in the vicinity of the two nuclei with very little emission in the surrounding disk. The peak of the emission of HNCO is near the position of the western nucleus (Figure \ref{fig:arp220maps}); however, this may be influenced by the lack of the eastern portion of the line profile. 

\cite{Greve2009} observed \cstwo\ and $J$=5-4 in Arp 220. Our high-resolution observations of \cstwo\ agree with the total flux within uncertainties while the \csfive\ is missing 37$\pm$12$\%$ of the total flux in the single dish map. \cite{Galametz2016} observed the \csfour\ emission from Arp 220 using APEX. The ALMA SV \csfour\ observations total flux agrees with the single dish observations. A comparison of the three CS lines (Figure \ref{fig:arp220spec}), shows small differences in the line profiles. The \csfour\ and $J$=5-4 have a double peak profile similar to what \cite{Greve2009} noted, while the \cstwo\ line profile lacks the peak at higher velocities. The \csfive\ line also has a third peak around an observed frequency of $\sim$ 240.8GHz, but this may be due to line contamination from perhaps CH$_{3}$OH and/or CH$_{2}$NH lines. \cite{Martin2009} observed a similar third peak in the \csfive\ line.

CS is a known tracer of photo-dominated regions (PDRs), where the CS is enhanced through S$^{+}$ chemistry \citep{Sternberg1995}. HNCO, however, is photodissociated easily by UV photons, but is enhanced in regions with shocks \citep{Zinchenko2000}. \citet{Martin2008} proposed that the relative abundance of HNCO to CS can be used as a good diagnostic tool to distinguish between the influence of shocks and molecular clouds illuminated by UV radiation. They determined that an offset region in IC 342 (denoted by IC 342* in their analysis) is dominated by shocks with a CS(5-4)/HNCO(5-4) line ratio of $\sim$0.3. M82 was determined to be dominated by PDRs with a CS(5-4)/HNCO(5-4) line ratio $>$64. In the analysis of Martin et al., Arp 220 is a composite system with contributions from PDRs and shocks. Since we only observe part of the HNCO, our line ratios of CS(5-4)/HNCO(5-4) (Figure \ref{fig:arp220lineratios}) is strictly an upper limit to the true line ratios; however, despite this handicap, we see significant variations of the line ratio between the region of the two nuclei ($\sim$7) and the kiloparsec disk ($\sim$ 12-30). Assuming both lines are optically thin and in LTE, the line ratios trace the relative abundance and this strongly suggests that the ISM in the kiloparsec disk is not strongly influenced by shocks while there is very likely a more significant contribution of shocks, likely from stellar winds, possible outflows as suggested by \citet{Sakamoto2009} and/or supernova explosions, near the two nuclei. Higher spatial resolution imaging is required to properly separate these two regions. 

\textbf{HN$^{13}$C (\textit{Rare isotopologue of hydrogen isocyanide})} - The HN$^{13}$C emission is peaked oddly below the western nucleus by $\sim$0.5 $\pm$ 0.1\arcsec. Nearly the entire emission is originating from the western side of Arp~220 which may be due to sensitivity since the line is weak and the eastern nucleus is the weaker emitter of the two. \citet{Wang2016} observed HN$^{13}$C $J$=1-0 and 3-2 using the IRAM~30m and APEX single dish telescopes. They find that the emission is only detected in the blue component of Arp~220 (i.e. the western nucleus region) similar to what we see and our flux measurement (Table \ref{tab:observations}) agrees with theirs. 
\citet{Tunnard2015a} observed HN$^{13}$C $J$=3-2 in Arp~220 with the PdBI. The peaks of HN$^{13}$C $J$=3-2 are within the positions of the nuclei (both east and west; see Figure 7 of Tunnard et al. 2015a). The flux measured by \citet{Tunnard2015a} is a factor of $\sim$2.5 times higher than that of \citet{Wang2016}; however, the line in the observations of \citet{Tunnard2015a} is near the bandpass edge and as noted by \citet{Wang2016}, the observations may have been difficult to continuum subtract properly. If the case is that the line was poorly continuum subtracted, the peak positions of HN$^{13}$C $J$=3-2 in the PdBI map may not be due to the molecular emission but of the 1.2mm continuum contamination. 
Using the total flux of HNC~$J$~=~1-0 published in \cite{Greve2009} and our flux measurement for HN$^{13}$C (Table \ref{tab:observations}), the line ratio of HNC/HN$^{13}$C is $\sim$ 52. Since HNC is likely optically thick, this measured line ratio is a lower limit to the [$^{12}$C]/[$^{13}$C] abundance. 

\textbf{CH$_{3}$CN (\textit{Methyl cyanide}}) -  Due to bandwidth limitations, only part of the CH$_{3}$CN~(6-5) line is observed. The CH$_{3}$CN is partially blended with the \tcoone\ line and is nearly as strong in emission as \tcoone. The CH$_{3}$CN~(12-11) line, which would be near the \tcotwo, is not included in our bandpass; however a very small part of the CH$_{3}$CN~(12-11) line may be blended in the blue side of the \tcotwo\ line profile. The ALMA CH$_{3}$CN (10-9) observations observed the entire line profile (Figure \ref{fig:arp220spec}). 

Since only the emission on the blue side of the CH$_{3}$CN~(6-5) line is observed, it is not surprising that the emission is peaked near the western nucleus; however, when compared to the CH$_{3}$CN (10-9), the peak positions agree. CH$_{3}$CN is believed to be a tracer of hot cores with the relative ease of detecting the line near hot cores \citep[e.g.][]{Remijan2004,Purcell2006} and is a good ISM thermometer \citep{Guesten1985} if multiple transitions are observed. \citet{Gratier2013} find that CH$_{3}$CN is more abundant in UV illuminated gas and thus may be a PDR tracer. Since U/LIRGs do experience intense massive star formation, CH$_{3}$CN should be relatively bright (and is in Arp~220). Observations of other transitions of CH$_{3}$CN will be relatively easy in the ALMA/NOEMA era as demonstrated by the ALMA SV and early PdBI observations.


\subsection{Molecular Gas Conditions in Arp 220}
The radiative transfer modelling of the molecular gas in Arp 220 is consistent with a warm ($\sim$~40~K), moderately dense (10$^{3.4}$~cm$^{-3}$ molecular gas component. The best fit molecular gas density is a factor of $\sim$4 higher than what was found by \citet{Rangwala2011} using $Herschel$ FTS observations; however, the best fit \nhtwo\ and \tkin\ of the two sets of models are consistent within the 1$\sigma$ uncertainties. We note that this result should not be perceived as a lack of very dense molecular gas as been presented in \citet{Scoville2015}. Our results are averaged over a $\sim$700~pc scale which may be dominated by a less dense medium, most likely from the diffuse kiloparsec disk. 

Within the 1$\sigma$ range, a warm, moderately dense gas component is favoured (Table \ref{tab:results} and Figure \ref{fig:arp220prob}). This result is very similar to other advanced mergers such as VV~114 \citep{Sliwa2013} and NGC~2623 \citep{Sliwa2017a} where, on average over $\sim$1~kpc scales, a warm, moderately dense molecular gas is the best fit solution. This can be explained by feedback from massive star formation, which these mergers will experience intensely over the course of the merger process and possibly outflows. The feedback in the form of stellar winds and supernova from massive stars and the outflows that have been observed \citep[e.g.][]{Cicone2014} are able to push back on the molecular gas, relieving the pressure while decreasing the molecular gas volume density. \citet{Sakamoto2009} have observed a molecular P-Cygni profile suggesting $\sim$100~\kms\ outflows from the nuclei, which can explain our picture of a less dense medium traced by CO. Within early/intermediate stage mergers we should see a denser molecular medium over roughly kiloparsec scales traced by CO and its isotopologues since the molecular gas is likely still inflowing to the central regions of the merging systems thus increasing the pressure and volume density of the molecular gas \citep{Sliwa2017a}. In Arp~220, the volume density and thus thermal pressure on $\sim$kiloparsec scales will likely be on the rise as the two nuclei are on the verge of the final coalescence.

\subsection{\xco\ and \xceo\ Abundance Ratios in Arp 220}
The R$_{10}$ and R$_{21}$ line ratios of Arp 220 (see Table \ref{tab:lineratios}) on a position by position basis are typical for ULIRGs but abnormally high when compared with normal disk galaxies \citep[e.g.][]{Davis2014}  The R$_{10}$ line ratio is higher (at some positions) than other LIRGs such as Arp~299 and Arp~55 \citep{Casoli1999,Sliwa2017a}. However, these high R$_{10}$ line ratios are more common for advanced major mergers such as  VV~114, IRAS~13120-5453 and NGC~2623 \citep{Sliwa2013,Sliwa2017a,Sliwa2017c,Saito2015}. The possible explanations for the higher line ratios include optical depth effects, increased \xco\ abundance ratio via some mechanism or excitation effects \citep[e.g][]{Casoli1992,Henkel1993,Taniguchi1999}.

Our radiative transfer analysis allows the \xco\ and \xceo\ abundances to vary as a free parameter. We find that for Arp 220, the most probable abundance ratio is 90-125, roughly 3 times higher than the ISM value around the Galactic center \citep[e.g.][]{Milam2005}. In our Galaxy, the abundance ratio increases with increasing radius. If Arp 220 had no enhancements in \xco, and had a
similar \xco\ abundance gradient as observed in our Galaxy,
then we would expect values of \xco~$\leq$50 since the molecular gas in Arp~220 is concentrated within $\sim$2-3~kpc. Even for NGC~6240, the most probable abundance ratio is 98 \citep{Tunnard2015b}. 

Optical depth effects have been ruled out by our analysis as the best fit solutions have optical depths $>$ 1 for \co\ and $<<$1 for \tco.

Selective photo-dissociation was thought to be a possible mechanism to drive the unusually high \xco\ abundance ratio we observe; however, this mechanism is ruled out because \ceo, a molecule that should be destroyed via photo-dissociation as fast or faster if less abundant than \tco,  is relatively strong in emission. The likely dominant source of the increased \xco\ and \xceo\ abundance ratios is stellar nucleosynthesis. Short-lived ($\leq$~10~Myr) massive stars produce \car\ and \eo, but \tc\ is produced in the envelopes of long-lived low/intermediate mass during the red giant phase. If extreme starbursts have a top-heavy initial mass function, then
many massive OB stars will be formed in the Arp 220 starbursts,
whose supernovae will enrich the ISM, and we would expect to
see increased abundances of \car\ and \eo. \citet{Matsushita2009} and \citet{Greve2009} show that the \tco/\ceo\ line ratio is $\sim$~1 for Arp 220. The \tco/\ceo\ line ratio in spirals has been shown to be on average $\sim$6 \citep{JD2017}, 6 times higher than for Arp 220. This strongly suggests that stellar nucleosynthesis has enriched the ISM of Arp 220. This is a similar result in other advanced mergers such as NCG 2623 and IRAS 13120-5453 \citep{Sliwa2017c}. 

Fractionation of the \car\ and \tc\ has also been considered as playing a role in the relative carbon isotope abundances that we find in U/LIRGs. The most important reaction to exchange carbon isotopes is
\begin{equation}\label{eqn:frac}
^{13}\rm{C^{+}} + \rm{^{12}CO} \rightleftharpoons \rm{^{12}C^{+}} + \rm{^{13}CO} + 35\rm{K}
\end{equation}
predicted by \citet{Watson1976}. The forward reaction is exothermic and in cold molecular gas can dominate isotope fractionation favouring the formation of more \tco. In hotter temperatures, both the forward and reverse reactions are about equally probable and would not affect the overall abundance ratio greatly. Since the molecular gas in Arp 220 is warm ($\sim$ 40K), we can rule out fractionation as a possible contaminate in our \xco\ abundance. \citet{Langer1984} showed that oxygen fractionation is very small and the oxygen isotope ratios reflect stellar nucleosynthesis processes further supporting the stellar nucleosynthesis enrichment scenario for the observed line ratios. 

\subsection{HCN and \hco\ Line Ratios: AGN or Starburst?}
As stated in the introduction, a linear relation between \hcnone\ and infrared luminosity was found and interpreted to be a direct correlation of dense gas and star formation \citep{Gao2004a}. Within the last decade or so, the nature of \hcnone\ as a true dense gas tracer has been questioned. Studies of samples of AGN and starburst systems have found enhancements of \hcnone\ emission among the AGN systems. Recently, \citet{Privon2015} have shown that composite (i.e. AGN and star forming) and star formation dominated systems can also show a similar enhancement in \hcnone\ emission. \cite{Privon2015} found that for AGN, starbursts and composite systems the H$_{10}$ ratios are 1.84$\pm$0.43, 0.88$\pm$0.28 and 1.14$\pm$0.49, respectively. This overlap of values for the different types of systems has complicated the picture of \hcn\ and \hco\ emission suggesting multiple processes contributing to the line ratio differences. 

Nevertheless, we compare the line ratios of Arp~220 and NGC~6240. We find that the line ratios are significantly different between the two sources. Arp~220 is consistent with AGN-like line ratios while NGC~6240 is consistent with a starburst-like ratio, similar to NGC~4039 and the Antennae Overlap region \citep{Schirm2016}. This result is quite surprising considering the fact that NGC~6240 has two well known AGNs \citep[e.g.][]{Komossa2003} while the presence of an AGN in Arp 220 is still debatable. The low line ratio found for NGC~6240 may reflect the different area filling factors of the two species as our observations are slightly unresolved with our beam; however, this is not the case for Arp~220 as our observations do resolve the emission and therefore, line ratios within one beam element would be filled with emission from both species. We note that our line ratio does not suggest the presence of an AGN in Arp 220 as our angular resolution does not distinguish the individual nuclei and the other possible mechanism affecting the H$_{10}$ line ratios \citep[e.g][]{Privon2015}. Radiative transfer analyses including just these dense gas tracers are required to understand what is driving the line ratios. We also note that other effects such as infrared pumping of HCN and recombination of \hco\ may be factors that need to be taken into consideration.

\section{Conclusions}
We present new PdBI observations of several molecular gas tracers for the two nearby major mergers Arp~220 and NGC~6240. The observations show a wealth of chemistry in Arp~220 and different conditions between the two sources. The main results are summarised as follows:
   \begin{enumerate}
      \item \tcotwo\ / $J$=1-0  line ratio reveal that molecular gas conditions vary across the disk of Arp~220 where the line ratio increases from east to west.
      \item The best fit molecular gas conditions at the peak position of \tcoone\ imply a warm (40~K), moderately dense (10$^{3.4}$~cm$^{-3}$) molecular gas component. This solutions differs in volume density by a factor of $\sim$4 from the analysis of \cite{Rangwala2011} most likely because of the spatial resolution difference.
	\item The \hcn/\hcoone\ line ratio for Arp~220 is greater than 2, while for NGC~6240 the line ratios are below 1.  
	\item The \xco\ abundance ratio for Arp~220 is a factor of $\sim$3 or more higher than the central value in the Milky Way, while the \xceo\ abundance ratio is quite low indicating the enhancement of \ceo\ in the ISM. The likely explanation that can explain both \xco\ and \xceo\ is an enrichment of $^{12}$C and $^{18}$O in the ISM via stellar nucleosynthesis. Higher-resolution and sensitive observations of \tco\ and \ceo\ are needed to look for specific spatial variations as seen in IRAS~13120-5453 \citep{Sliwa2017c}.
	
	These data, along with other sources (see Appendix \ref{sec:release}) will be made available to the public for download with the intention to use for future analyses and comparisons with new observations. 
   \end{enumerate}

\begin{acknowledgements}
We thank the referee for their comments and suggestions that improved this manuscript. We thank K. Sakamoto for passing along the SMA \cothree\ observations of Arp 220. KS thanks Julia Kamenetzky for her MCMC code made available via github. KS also thanks L. Barcos-M\~unoz, A.K. Leroy, E. Schinnerer, F. Walter, and L.K. Zschaechner for fruitful discussions.  Based on observations carried out with the IRAM Interferometer PdBI. IRAM is supported by INSU/CNRS (France), MPG (Germany) and IGN (Spain). The Submillimeter Array is a joint project between the Smithsonian Astrophysical Observatory and the Academia Sinica Institute of Astronomy and Astrophysics and is funded by the Smithsonian Institution and the Academia Sinica.

This paper makes use of the following ALMA data: ADS/JAO.ALMA$\#$2011.0.00018.SV. 
ALMA is a partnership of ESO (representing its member states), NSF (USA) and 
NINS (Japan), together with NRC (Canada) and NSC and ASIAA (Taiwan), and KASI (Republic of Korea), 
in cooperation with the Republic of Chile. The Joint ALMA Observatory is 
operated by ESO, AUI/NRAO and NAOJ.
\end{acknowledgements}


\begin{thebibliography}{101}
\expandafter\ifx\csname natexlab\endcsname\relax\def\natexlab#1{#1}\fi

\bibitem[{{Aalto} {et~al.}(1995){Aalto}, {Booth}, {Black}, \&
  {Johansson}}]{Aalto1995}
{Aalto}, S., {Booth}, R.~S., {Black}, J.~H., \& {Johansson}, L.~E.~B. 1995,
  \aap, 300, 369

\bibitem[{{Aalto} {et~al.}(1991){Aalto}, {Johansson}, {Booth}, \&
  {Black}}]{Aalto1991}
{Aalto}, S., {Johansson}, L.~E.~B., {Booth}, R.~S., \& {Black}, J.~H. 1991,
  \aap, 249, 323

\bibitem[{{Aalto} {et~al.}(2015){Aalto}, {Mart{\'{\i}}n}, {Costagliola},
  {Gonz{\'a}lez-Alfonso}, {Muller}, {Sakamoto}, {Fuller},
  {Garc{\'{\i}}a-Burillo}, {van der Werf}, {Neri}, {Spaans}, {Combes}, {Viti},
  {M{\"u}hle}, {Armus}, {Evans}, {Sturm}, {Cernicharo}, {Henkel}, \&
  {Greve}}]{Aalto2015b}
{Aalto}, S., {Mart{\'{\i}}n}, S., {Costagliola}, F., {et~al.} 2015, \aap, 584,
  A42

\bibitem[{{Aalto} {et~al.}(1997){Aalto}, {Radford}, {Scoville}, \&
  {Sargent}}]{Aalto1997}
{Aalto}, S., {Radford}, S.~J.~E., {Scoville}, N.~Z., \& {Sargent}, A.~I. 1997,
  \apjl, 475, L107

\bibitem[{{Aalto} {et~al.}(2009){Aalto}, {Wilner}, {Spaans}, {Wiedner},
  {Sakamoto}, {Black}, \& {Caldas}}]{Aalto2009}
{Aalto}, S., {Wilner}, D., {Spaans}, M., {et~al.} 2009, \aap, 493, 481

\bibitem[{{Barcos-Mu{\~n}oz} {et~al.}(2015){Barcos-Mu{\~n}oz}, {Leroy},
  {Evans}, {Privon}, {Armus}, {Condon}, {Mazzarella}, {Meier}, {Momjian},
  {Murphy}, {Ott}, {Reichardt}, {Sakamoto}, {Sanders}, {Schinnerer},
  {Stierwalt}, {Surace}, {Thompson}, \& {Walter}}]{Barcos-Munoz2015}
{Barcos-Mu{\~n}oz}, L., {Leroy}, A.~K., {Evans}, A.~S., {et~al.} 2015, \apj,
  799, 10

\bibitem[{{Bryant} \& {Scoville}(1999)}]{Bryant1999}
{Bryant}, P.~M. \& {Scoville}, N.~Z. 1999, \aj, 117, 2632

\bibitem[{{Buchner} {et~al.}(2014){Buchner}, {Georgakakis}, {Nandra}, {Hsu},
  {Rangel}, {Brightman}, {Merloni}, {Salvato}, {Donley}, \&
  {Kocevski}}]{Buchner2014}
{Buchner}, J., {Georgakakis}, A., {Nandra}, K., {et~al.} 2014, \aap, 564, A125

\bibitem[{{Casoli} {et~al.}(1992){Casoli}, {Dupraz}, \& {Combes}}]{Casoli1992}
{Casoli}, F., {Dupraz}, C., \& {Combes}, F. 1992, \aap, 264, 55

\bibitem[{{Casoli} {et~al.}(1999){Casoli}, {Willaime}, {Viallefond}, \&
  {Gerin}}]{Casoli1999}
{Casoli}, F., {Willaime}, M.-C., {Viallefond}, F., \& {Gerin}, M. 1999, \aap,
  346, 663

\bibitem[{{Cicone} {et~al.}(2014){Cicone}, {Maiolino}, {Sturm},
  {Graci{\'a}-Carpio}, {Feruglio}, {Neri}, {Aalto}, {Davies}, {Fiore},
  {Fischer}, {Garc{\'{\i}}a-Burillo}, {Gonz{\'a}lez-Alfonso},
  {Hailey-Dunsheath}, {Piconcelli}, \& {Veilleux}}]{Cicone2014}
{Cicone}, C., {Maiolino}, R., {Sturm}, E., {et~al.} 2014, \aap, 562, A21

\bibitem[{{Davis}(2014)}]{Davis2014}
{Davis}, T.~A. 2014, \mnras, 445, 2378

\bibitem[{{Depoy} {et~al.}(1986){Depoy}, {Becklin}, \&
  {Wynn-Williams}}]{Depoy1986}
{Depoy}, D.~L., {Becklin}, E.~E., \& {Wynn-Williams}, C.~G. 1986, \apj, 307,
  116

\bibitem[{{Downes} \& {Eckart}(2007)}]{Downes2007}
{Downes}, D. \& {Eckart}, A. 2007, \aap, 468, L57

\bibitem[{{Downes} \& {Solomon}(1998)}]{Downes1998}
{Downes}, D. \& {Solomon}, P.~M. 1998, \apj, 507, 615

\bibitem[{{Elston} \& {Maloney}(1990)}]{Elston1990}
{Elston}, R. \& {Maloney}, P. 1990, \apj, 357, 91

\bibitem[{{Feroz} {et~al.}(2009){Feroz}, {Hobson}, \& {Bridges}}]{Feroz2009}
{Feroz}, F., {Hobson}, M.~P., \& {Bridges}, M. 2009, \mnras, 398, 1601

\bibitem[{{Feruglio} {et~al.}(2013{\natexlab{a}}){Feruglio}, {Fiore},
  {Maiolino}, {Piconcelli}, {Aussel}, {Elbaz}, {Le Floc'h}, {Sturm}, {Davies},
  \& {Cicone}}]{Feruglio2013a}
{Feruglio}, C., {Fiore}, F., {Maiolino}, R., {et~al.} 2013{\natexlab{a}}, \aap,
  549, A51

\bibitem[{{Feruglio} {et~al.}(2013{\natexlab{b}}){Feruglio}, {Fiore},
  {Piconcelli}, {Cicone}, {Maiolino}, {Davies}, \& {Sturm}}]{Feruglio2013b}
{Feruglio}, C., {Fiore}, F., {Piconcelli}, E., {et~al.} 2013{\natexlab{b}},
  \aap, 558, A87

\bibitem[{{Galametz} {et~al.}(2016){Galametz}, {Zhang}, {Immer}, {Humphreys},
  {Aladro}, {De Breuck}, {Ginsburg}, {Madden}, {M{\o}ller}, \&
  {Arumugam}}]{Galametz2016}
{Galametz}, M., {Zhang}, Z.-Y., {Immer}, K., {et~al.} 2016, \mnras, 462, L36

\bibitem[{{Gao} \& {Solomon}(2004{\natexlab{a}})}]{Gao2004b}
{Gao}, Y. \& {Solomon}, P.~M. 2004{\natexlab{a}}, \apjs, 152, 63

\bibitem[{{Gao} \& {Solomon}(2004{\natexlab{b}})}]{Gao2004a}
{Gao}, Y. \& {Solomon}, P.~M. 2004{\natexlab{b}}, \apj, 606, 271

\bibitem[{{Garay} {et~al.}(1993){Garay}, {Mardones}, \& {Mirabel}}]{Garay1993}
{Garay}, G., {Mardones}, D., \& {Mirabel}, I.~F. 1993, \aap, 277, 405

\bibitem[{{Genzel} {et~al.}(1998){Genzel}, {Lutz}, {Sturm}, {Egami}, {Kunze},
  {Moorwood}, {Rigopoulou}, {Spoon}, {Sternberg}, {Tacconi-Garman}, {Tacconi},
  \& {Thatte}}]{Genzel1998}
{Genzel}, R., {Lutz}, D., {Sturm}, E., {et~al.} 1998, \apj, 498, 579

\bibitem[{{Gratier} {et~al.}(2013){Gratier}, {Pety}, {Guzm{\'a}n}, {Gerin},
  {Goicoechea}, {Roueff}, \& {Faure}}]{Gratier2013}
{Gratier}, P., {Pety}, J., {Guzm{\'a}n}, V., {et~al.} 2013, \aap, 557, A101

\bibitem[{{Greve} {et~al.}(2009){Greve}, {Papadopoulos}, {Gao}, \&
  {Radford}}]{Greve2009}
{Greve}, T.~R., {Papadopoulos}, P.~P., {Gao}, Y., \& {Radford}, S.~J.~E. 2009,
  \apj, 692, 1432

\bibitem[{{Guesten} {et~al.}(1985){Guesten}, {Walmsley}, {Ungerechts}, \&
  {Churchwell}}]{Guesten1985}
{Guesten}, R., {Walmsley}, C.~M., {Ungerechts}, H., \& {Churchwell}, E. 1985,
  \aap, 142, 381

\bibitem[{{Henkel} {et~al.}(2014){Henkel}, {Asiri}, {Ao}, {Aalto}, {Danielson},
  {Papadopoulos}, {Garc{\'{\i}}a-Burillo}, {Aladro}, {Impellizzeri},
  {Mauersberger}, {Mart{\'{\i}}n}, \& {Harada}}]{Henkel2014}
{Henkel}, C., {Asiri}, H., {Ao}, Y., {et~al.} 2014, \aap, 565, A3

\bibitem[{{Henkel} \& {Mauersberger}(1993)}]{Henkel1993}
{Henkel}, C. \& {Mauersberger}, R. 1993, \aap, 274, 730

\bibitem[{{Herbst} {et~al.}(1990){Herbst}, {Graham}, {Tsutsui}, {Beckwith},
  {Matthews}, \& {Soifer}}]{Herbst1990}
{Herbst}, T.~M., {Graham}, J.~R., {Tsutsui}, K., {et~al.} 1990, \aj, 99, 1773

\bibitem[{{Hinshaw} {et~al.}(2013){Hinshaw}, {Larson}, {Komatsu}, {Spergel},
  {Bennett}, {Dunkley}, {Nolta}, {Halpern}, {Hill}, {Odegard}, {Page}, {Smith},
  {Weiland}, {Gold}, {Jarosik}, {Kogut}, {Limon}, {Meyer}, {Tucker}, {Wollack},
  \& {Wright}}]{Hinshaw2012}
{Hinshaw}, G., {Larson}, D., {Komatsu}, E., {et~al.} 2013, \apjs, 208, 19

\bibitem[{{Houck} {et~al.}(1985){Houck}, {Schneider}, {Danielson},
  {Neugebauer}, {Soifer}, {Beichman}, \& {Lonsdale}}]{Houck1985}
{Houck}, J.~R., {Schneider}, D.~P., {Danielson}, G.~E., {et~al.} 1985, \apjl,
  290, L5

\bibitem[{{Houck} {et~al.}(1984){Houck}, {Soifer}, {Neugebauer}, {Beichman},
  {Aumann}, {Clegg}, {Gillett}, {Habing}, {Hauser}, {Low}, {Miley},
  {Rowan-Robinson}, \& {Walker}}]{Houck1984}
{Houck}, J.~R., {Soifer}, B.~T., {Neugebauer}, G., {et~al.} 1984, \apjl, 278,
  L63

\bibitem[{{Imanishi} {et~al.}(2010){Imanishi}, {Nakanishi}, {Yamada}, {Tamura},
  \& {Kohno}}]{Imanishi2010}
{Imanishi}, M., {Nakanishi}, K., {Yamada}, M., {Tamura}, Y., \& {Kohno}, K.
  2010, \pasj, 62, 201

\bibitem[{{Iono} {et~al.}(2007){Iono}, {Wilson}, {Takakuwa}, {Yun}, {Petitpas},
  {Peck}, {Ho}, {Matsushita}, {Pihlstrom}, \& {Wang}}]{Iono2007}
{Iono}, D., {Wilson}, C.~D., {Takakuwa}, S., {et~al.} 2007, \apj, 659, 283

\bibitem[{{Jim{\'e}nez-Donaire} {et~al.}(2017){Jim{\'e}nez-Donaire}, {Cormier},
  {Bigiel}, {Leroy}, {Gallagher}, {Krumholz}, {Usero}, {Hughes}, {Kramer},
  {Meier}, {Murphy}, {Pety}, {Schinnerer}, {Schruba}, {Schuster}, {Sliwa}, \&
  {Tomicic}}]{JD2017}
{Jim{\'e}nez-Donaire}, M.~J., {Cormier}, D., {Bigiel}, F., {et~al.} 2017,
  \apjl, 836, L29

\bibitem[{{Kamenetzky} {et~al.}(2014){Kamenetzky}, {Rangwala}, {Glenn},
  {Maloney}, \& {Conley}}]{Kamenetzky2014}
{Kamenetzky}, J., {Rangwala}, N., {Glenn}, J., {Maloney}, P.~R., \& {Conley},
  A. 2014, \apj, 795, 174

\bibitem[{{Komossa} {et~al.}(2003){Komossa}, {Burwitz}, {Hasinger}, {Predehl},
  {Kaastra}, \& {Ikebe}}]{Komossa2003}
{Komossa}, S., {Burwitz}, V., {Hasinger}, G., {et~al.} 2003, \apjl, 582, L15

\bibitem[{{K{\"o}nig} {et~al.}(2012){K{\"o}nig}, {Garc{\'{\i}}a-Mar{\'{\i}}n},
  {Eckart}, {Downes}, \& {Scharw{\"a}chter}}]{Konig2012}
{K{\"o}nig}, S., {Garc{\'{\i}}a-Mar{\'{\i}}n}, M., {Eckart}, A., {Downes}, D.,
  \& {Scharw{\"a}chter}, J. 2012, \apj, 754, 58

\bibitem[{{K{\"o}nig} {et~al.}(2016){K{\"o}nig}, {Mart{\'{\i}}n}, {Muller},
  {Cernicharo}, {Sakamoto}, {Zschaechner}, {Humphreys}, {Mroczkowski}, {Krips},
  {Galametz}, {Aalto}, {Vlemmings}, {Ott}, {Meier}, {Fuente},
  {Garc{\'{\i}}a-Burillo}, \& {Neri}}]{Koenig2016b}
{K{\"o}nig}, S., {Mart{\'{\i}}n}, S., {Muller}, S., {et~al.} 2016, ArXiv
  e-prints

\bibitem[{{Langer} {et~al.}(1984){Langer}, {Graedel}, {Frerking}, \&
  {Armentrout}}]{Langer1984}
{Langer}, W.~D., {Graedel}, T.~E., {Frerking}, M.~A., \& {Armentrout}, P.~B.
  1984, \apj, 277, 581

\bibitem[{{Larson} {et~al.}(2016){Larson}, {Sanders}, {Barnes}, {Ishida},
  {Evans}, {U}, {Mazzarella}, {Kim}, {Privon}, {Mirabel}, \&
  {Flewelling}}]{Larson2016}
{Larson}, K.~L., {Sanders}, D.~B., {Barnes}, J.~E., {et~al.} 2016, \apj, 825,
  128

\bibitem[{{Lester} {et~al.}(1988){Lester}, {Harvey}, \& {Carr}}]{Lester1988}
{Lester}, D.~F., {Harvey}, P.~M., \& {Carr}, J. 1988, \apj, 329, 641

\bibitem[{{Lockhart} {et~al.}(2015){Lockhart}, {Kewley}, {Lu}, {Allen},
  {Rupke}, {Calzetti}, {Davies}, {Dopita}, {Engel}, {Heckman}, {Leitherer}, \&
  {Sanders}}]{Lockhart2016}
{Lockhart}, K.~E., {Kewley}, L.~J., {Lu}, J.~R., {et~al.} 2015, \apj, 810, 149

\bibitem[{{Mart{\'{\i}}n} {et~al.}(2016){Mart{\'{\i}}n}, {Aalto}, {Sakamoto},
  {Gonz{\'a}lez-Alfonso}, {Muller}, {Henkel}, {Garc{\'{\i}}a-Burillo},
  {Aladro}, {Costagliola}, {Harada}, {Krips}, {Mart{\'{\i}}n-Pintado},
  {M{\"u}hle}, {van der Werf}, \& {Viti}}]{Martin2016}
{Mart{\'{\i}}n}, S., {Aalto}, S., {Sakamoto}, K., {et~al.} 2016, \aap, 590, A25

\bibitem[{{Mart{\'{\i}}n} {et~al.}(2011){Mart{\'{\i}}n}, {Krips},
  {Mart{\'{\i}}n-Pintado}, {Aalto}, {Zhao}, {Peck}, {Petitpas}, {Monje},
  {Greve}, \& {An}}]{Martin2011}
{Mart{\'{\i}}n}, S., {Krips}, M., {Mart{\'{\i}}n-Pintado}, J., {et~al.} 2011,
  \aap, 527, A36

\bibitem[{{Mart{\'{\i}}n} {et~al.}(2009){Mart{\'{\i}}n},
  {Mart{\'{\i}}n-Pintado}, \& {Mauersberger}}]{Martin2009}
{Mart{\'{\i}}n}, S., {Mart{\'{\i}}n-Pintado}, J., \& {Mauersberger}, R. 2009,
  \apj, 694, 610

\bibitem[{{Mart{\'{\i}}n} {et~al.}(2008){Mart{\'{\i}}n}, {Requena-Torres},
  {Mart{\'{\i}}n-Pintado}, \& {Mauersberger}}]{Martin2008}
{Mart{\'{\i}}n}, S., {Requena-Torres}, M.~A., {Mart{\'{\i}}n-Pintado}, J., \&
  {Mauersberger}, R. 2008, \apj, 678, 245

\bibitem[{{Matsushita} {et~al.}(2009){Matsushita}, {Iono}, {Petitpas}, {Chou},
  {Gurwell}, {Hunter}, {Muller}, {Peck}, {Sakamoto}, {Sawada Satoh}, {Wiedner},
  {Wilner}, \& {Wilson}}]{Matsushita2009}
{Matsushita}, S., {Iono}, D., {Petitpas}, G.~R., {et~al.} 2009, \apj, 693, 56

\bibitem[{{McMullin} {et~al.}(2007){McMullin}, {Waters}, {Schiebel}, {Young},
  \& {Golap}}]{McMullin2007}
{McMullin}, J.~P., {Waters}, B., {Schiebel}, D., {Young}, W., \& {Golap}, K.
  2007, in Astronomical Society of the Pacific Conference Series, Vol. 376,
  Astronomical Data Analysis Software and Systems XVI, ed. R.~A. {Shaw},
  F.~{Hill}, \& D.~J. {Bell}, 127

\bibitem[{{Meijerink} {et~al.}(2013){Meijerink}, {Kristensen}, {Wei{\ss}}, {van
  der Werf}, {Walter}, {Spaans}, {Loenen}, {Fischer}, {Israel}, {Isaak},
  {Papadopoulos}, {Aalto}, {Armus}, {Charmandaris}, {Dasyra}, {Diaz-Santos},
  {Evans}, {Gao}, {Gonz{\'a}lez-Alfonso}, {G{\"u}sten}, {Henkel}, {Kramer},
  {Lord}, {Mart{\'{\i}}n-Pintado}, {Naylor}, {Sanders}, {Smith}, {Spinoglio},
  {Stacey}, {Veilleux}, \& {Wiedner}}]{Meijerink2013}
{Meijerink}, R., {Kristensen}, L.~E., {Wei{\ss}}, A., {et~al.} 2013, \apjl,
  762, L16

\bibitem[{{Milam} {et~al.}(2005){Milam}, {Savage}, {Brewster}, {Ziurys}, \&
  {Wyckoff}}]{Milam2005}
{Milam}, S.~N., {Savage}, C., {Brewster}, M.~A., {Ziurys}, L.~M., \& {Wyckoff},
  S. 2005, \apj, 634, 1126

\bibitem[{{Mori} {et~al.}(2014){Mori}, {Imanishi}, {Alonso-Herrero}, {Packham},
  {Ramos Almeida}, {Nikutta}, {Gonz{\'a}lez-Mart{\'{\i}}n}, {Perlman}, {Saito},
  \& {Levenson}}]{Mori2014}
{Mori}, T.~I., {Imanishi}, M., {Alonso-Herrero}, A., {et~al.} 2014, \pasj, 66,
  93

\bibitem[{{Murphy} {et~al.}(1996){Murphy}, {Armus}, {Matthews}, {Soifer},
  {Mazzarella}, {Shupe}, {Strauss}, \& {Neugebauer}}]{Murphy1996}
{Murphy}, Jr., T.~W., {Armus}, L., {Matthews}, K., {et~al.} 1996, \aj, 111,
  1025

\bibitem[{{Nakanishi} {et~al.}(2005){Nakanishi}, {Okumura}, {Kohno}, {Kawabe},
  \& {Nakagawa}}]{Nakanishi2005}
{Nakanishi}, K., {Okumura}, S.~K., {Kohno}, K., {Kawabe}, R., \& {Nakagawa}, T.
  2005, \pasj, 57, 575

\bibitem[{{Norris}(1988)}]{Norris1988}
{Norris}, R.~P. 1988, \mnras, 230, 345

\bibitem[{{Ohyama} {et~al.}(2003){Ohyama}, {Yoshida}, \& {Takata}}]{Ohyama2003}
{Ohyama}, Y., {Yoshida}, M., \& {Takata}, T. 2003, \aj, 126, 2291

\bibitem[{{Ohyama} {et~al.}(2000){Ohyama}, {Yoshida}, {Takata}, {Imanishi},
  {Usuda}, {Saito}, {Taguchi}, {Ebizuka}, {Iwamuro}, {Motohara}, {Taguchi},
  {Hata}, {Maihara}, {Iye}, {Sasaki}, {Kosugi}, {Ogasawara}, {Noumaru},
  {Mizumoto}, {Yagi}, \& {Chikada}}]{Ohyama2000}
{Ohyama}, Y., {Yoshida}, M., {Takata}, T., {et~al.} 2000, \pasj, 52, 563

\bibitem[{{Papadopoulos} {et~al.}(2012){Papadopoulos}, {van der Werf},
  {Xilouris}, {Isaak}, {Gao}, \& {M{\"u}hle}}]{Papadopoulos2012a}
{Papadopoulos}, P.~P., {van der Werf}, P.~P., {Xilouris}, E.~M., {et~al.} 2012,
  \mnras, 426, 2601

\bibitem[{{Papadopoulos} {et~al.}(2014){Papadopoulos}, {Zhang}, {Xilouris},
  {Weiss}, {van der Werf}, {Israel}, {Greve}, {Isaak}, \&
  {Gao}}]{Papadopoulos2014}
{Papadopoulos}, P.~P., {Zhang}, Z.-Y., {Xilouris}, E.~M., {et~al.} 2014, \apj,
  788, 153

\bibitem[{{Privon} {et~al.}(2015){Privon}, {Herrero-Illana}, {Evans},
  {Iwasawa}, {Perez-Torres}, {Armus}, {D{\'{\i}}az-Santos}, {Murphy},
  {Stierwalt}, {Aalto}, {Mazzarella}, {Barcos-Mu{\~n}oz}, {Borish}, {Inami},
  {Kim}, {Treister}, {Surace}, {Lord}, {Conway}, {Frayer}, \&
  {Alberdi}}]{Privon2015}
{Privon}, G.~C., {Herrero-Illana}, R., {Evans}, A.~S., {et~al.} 2015, \apj,
  814, 39

\bibitem[{{Purcell} {et~al.}(2006){Purcell}, {Balasubramanyam}, {Burton},
  {Walsh}, {Minier}, {Hunt-Cunningham}, {Kedziora-Chudczer}, {Longmore},
  {Hill}, {Bains}, {Barnes}, {Busfield}, {Calisse}, {Crighton}, {Curran},
  {Davis}, {Dempsey}, {Derragopian}, {Fulton}, {Hidas}, {Hoare}, {Lee}, {Ladd},
  {Lumsden}, {Moore}, {Murphy}, {Oudmaijer}, {Pracy}, {Rathborne}, {Robertson},
  {Schultz}, {Shobbrook}, {Sparks}, {Storey}, \& {Travouillion}}]{Purcell2006}
{Purcell}, C.~R., {Balasubramanyam}, R., {Burton}, M.~G., {et~al.} 2006,
  \mnras, 367, 553

\bibitem[{{Rangwala} {et~al.}(2011){Rangwala}, {Maloney}, {Glenn}, {Wilson},
  {Rykala}, {Isaak}, {Baes}, {Bendo}, {Boselli}, {Bradford}, {Clements},
  {Cooray}, {Fulton}, {Imhof}, {Kamenetzky}, {Madden}, {Mentuch}, {Sacchi},
  {Sauvage}, {Schirm}, {Smith}, {Spinoglio}, \& {Wolfire}}]{Rangwala2011}
{Rangwala}, N., {Maloney}, P.~R., {Glenn}, J., {et~al.} 2011, \apj, 743, 94

\bibitem[{{Rangwala} {et~al.}(2015){Rangwala}, {Maloney}, {Wilson}, {Glenn},
  {Kamenetzky}, \& {Spinoglio}}]{Rangwala2015}
{Rangwala}, N., {Maloney}, P.~R., {Wilson}, C.~D., {et~al.} 2015, \apj, 806, 17

\bibitem[{{Remijan} {et~al.}(2004){Remijan}, {Sutton}, {Snyder}, {Friedel},
  {Liu}, \& {Pei}}]{Remijan2004}
{Remijan}, A., {Sutton}, E.~C., {Snyder}, L.~E., {et~al.} 2004, \apj, 606, 917

\bibitem[{{Rieke} {et~al.}(1985){Rieke}, {Cutri}, {Black}, {Kailey}, {McAlary},
  {Lebofsky}, \& {Elston}}]{Rieke1985}
{Rieke}, G.~H., {Cutri}, R.~M., {Black}, J.~H., {et~al.} 1985, \apj, 290, 116

\bibitem[{{Saito} {et~al.}(2015){Saito}, {Iono}, {Yun}, {Ueda}, {Nakanishi},
  {Sugai}, {Espada}, {Imanishi}, {Motohara}, {Hagiwara}, {Tateuchi}, {Lee}, \&
  {Kawabe}}]{Saito2015}
{Saito}, T., {Iono}, D., {Yun}, M.~S., {et~al.} 2015, \apj, 803, 60

\bibitem[{{Sakamoto} {et~al.}(2009){Sakamoto}, {Aalto}, {Wilner}, {Black},
  {Conway}, {Costagliola}, {Peck}, {Spaans}, {Wang}, \&
  {Wiedner}}]{Sakamoto2009}
{Sakamoto}, K., {Aalto}, S., {Wilner}, D.~J., {et~al.} 2009, \apjl, 700, L104

\bibitem[{{Sakamoto} {et~al.}(1999){Sakamoto}, {Scoville}, {Yun}, {Crosas},
  {Genzel}, \& {Tacconi}}]{Sakamoto1999}
{Sakamoto}, K., {Scoville}, N.~Z., {Yun}, M.~S., {et~al.} 1999, \apj, 514, 68

\bibitem[{{Sakamoto} {et~al.}(2008){Sakamoto}, {Wang}, {Wiedner}, {Wang},
  {Peck}, {Zhang}, {Petitpas}, {Ho}, \& {Wilner}}]{Sakamoto2008}
{Sakamoto}, K., {Wang}, J., {Wiedner}, M.~C., {et~al.} 2008, \apj, 684, 957

\bibitem[{{Sanders} {et~al.}(2003){Sanders}, {Mazzarella}, {Kim}, {Surace}, \&
  {Soifer}}]{Sanders2003}
{Sanders}, D.~B., {Mazzarella}, J.~M., {Kim}, D.-C., {Surace}, J.~A., \&
  {Soifer}, B.~T. 2003, \aj, 126, 1607

\bibitem[{{Schirm} {et~al.}(2016){Schirm}, {Wilson}, {Madden}, \&
  {Clements}}]{Schirm2016}
{Schirm}, M.~R.~P., {Wilson}, C.~D., {Madden}, S.~C., \& {Clements}, D.~L.
  2016, \apj, 823, 87

\bibitem[{{Scoville} {et~al.}(2017){Scoville}, {Murchikova}, {Walter},
  {Vlahakis}, {Koda}, {Vanden Bout}, {Barnes}, {Hernquist}, {Sheth}, {Yun},
  {Sanders}, {Armus}, {Cox}, {Thompson}, {Robertson}, {Zschaechner}, {Tacconi},
  {Torrey}, {Hayward}, {Genzel}, {Hopkins}, {van der Werf}, \&
  {Decarli}}]{Scoville2017}
{Scoville}, N., {Murchikova}, L., {Walter}, F., {et~al.} 2017, \apj, 836, 66

\bibitem[{{Scoville} {et~al.}(2015){Scoville}, {Sheth}, {Walter}, {Manohar},
  {Zschaechner}, {Yun}, {Koda}, {Sanders}, {Murchikova}, {Thompson},
  {Robertson}, {Genzel}, {Hernquist}, {Tacconi}, {Brown}, {Narayanan},
  {Hayward}, {Barnes}, {Kartaltepe}, {Davies}, {van der Werf}, \&
  {Fomalont}}]{Scoville2015}
{Scoville}, N., {Sheth}, K., {Walter}, F., {et~al.} 2015, \apj, 800, 70

\bibitem[{{Scoville} {et~al.}(1997){Scoville}, {Yun}, \&
  {Bryant}}]{Scoville1997}
{Scoville}, N.~Z., {Yun}, M.~S., \& {Bryant}, P.~M. 1997, \apj, 484, 702

\bibitem[{{Sliwa} {et~al.}(2017{\natexlab{a}}){Sliwa}, {Wilson}, {Aalto}, \&
  {Privon}}]{Sliwa2017c}
{Sliwa}, K., {Wilson}, C.~D., {Aalto}, S., \& {Privon}, G.~C.
  2017{\natexlab{a}}, ApJL, submitted

\bibitem[{{Sliwa} {et~al.}(2014){Sliwa}, {Wilson}, {Iono}, {Peck}, \&
  {Matsushita}}]{Sliwa2014}
{Sliwa}, K., {Wilson}, C.~D., {Iono}, D., {Peck}, A., \& {Matsushita}, S. 2014,
  \apjl, 796, L15

\bibitem[{{Sliwa} {et~al.}(2013){Sliwa}, {Wilson}, {Krips}, {Petitpas}, {Iono},
  {Juvela}, {Matsushita}, {Peck}, \& {Yun}}]{Sliwa2013}
{Sliwa}, K., {Wilson}, C.~D., {Krips}, M., {et~al.} 2013, \apj, 777, 126

\bibitem[{{Sliwa} {et~al.}(2017{\natexlab{b}}){Sliwa}, {Wilson}, {Matsushita},
  {Peck}, {Petitpas}, {Saito}, \& {Yun}}]{Sliwa2017a}
{Sliwa}, K., {Wilson}, C.~D., {Matsushita}, S., {et~al.} 2017{\natexlab{b}},
  ArXiv e-prints

\bibitem[{{Sliwa} {et~al.}(2012){Sliwa}, {Wilson}, {Petitpas}, {Armus},
  {Juvela}, {Matsushita}, {Peck}, \& {Yun}}]{Sliwa2012}
{Sliwa}, K., {Wilson}, C.~D., {Petitpas}, G.~R., {et~al.} 2012, \apj, 753, 46

\bibitem[{{Soifer} {et~al.}(1984{\natexlab{a}}){Soifer}, {Neugebauer}, {Helou},
  {Lonsdale}, {Hacking}, {Rice}, {Houck}, {Low}, \&
  {Rowan-Robinson}}]{Soifer1984b}
{Soifer}, B.~T., {Neugebauer}, G., {Helou}, G., {et~al.} 1984{\natexlab{a}},
  \apjl, 283, L1

\bibitem[{{Soifer} {et~al.}(1984{\natexlab{b}}){Soifer}, {Rowan-Robinson},
  {Houck}, {de Jong}, {Neugebauer}, {Aumann}, {Beichman}, {Boggess}, {Clegg},
  {Emerson}, {Gillett}, {Habing}, {Hauser}, {Low}, {Miley}, \&
  {Young}}]{Soifer1984}
{Soifer}, B.~T., {Rowan-Robinson}, M., {Houck}, J.~R., {et~al.}
  1984{\natexlab{b}}, \apjl, 278, L71

\bibitem[{{Sternberg} \& {Dalgarno}(1995)}]{Sternberg1995}
{Sternberg}, A. \& {Dalgarno}, A. 1995, \apjs, 99, 565

\bibitem[{{Sugai} {et~al.}(1997){Sugai}, {Malkan}, {Ward}, \&
  {McLean}}]{Sugai1997}
{Sugai}, H., {Malkan}, M.~A., {Ward}, M.~J., \& {McLean}, I.~S. 1997, \apj,
  481, 186

\bibitem[{{Tacconi} {et~al.}(1999){Tacconi}, {Genzel}, {Tecza}, {Gallimore},
  {Downes}, \& {Scoville}}]{Tacconi1999}
{Tacconi}, L.~J., {Genzel}, R., {Tecza}, M., {et~al.} 1999, \apj, 524, 732

\bibitem[{{Taniguchi} {et~al.}(1999){Taniguchi}, {Ohyama}, \&
  {Sanders}}]{Taniguchi1999}
{Taniguchi}, Y., {Ohyama}, Y., \& {Sanders}, D.~B. 1999, \apj, 522, 214

\bibitem[{{Tunnard} {et~al.}(2015{\natexlab{a}}){Tunnard}, {Greve},
  {Garcia-Burillo}, {Graci{\'a} Carpio}, {Fischer}, {Fuente},
  {Gonz{\'a}lez-Alfonso}, {Hailey-Dunsheath}, {Neri}, {Sturm}, {Usero}, \&
  {Planesas}}]{Tunnard2015a}
{Tunnard}, R., {Greve}, T.~R., {Garcia-Burillo}, S., {et~al.}
  2015{\natexlab{a}}, \apj, 800, 25

\bibitem[{{Tunnard} {et~al.}(2015{\natexlab{b}}){Tunnard}, {Greve},
  {Garcia-Burillo}, {Graci{\'a} Carpio}, {Fuente}, {Tacconi}, {Neri}, \&
  {Usero}}]{Tunnard2015b}
{Tunnard}, R., {Greve}, T.~R., {Garcia-Burillo}, S., {et~al.}
  2015{\natexlab{b}}, \apj, 815, 114

\bibitem[{{U} {et~al.}(2012){U}, {Sanders}, {Mazzarella}, {Evans}, {Howell},
  {Surace}, {Armus}, {Iwasawa}, {Kim}, {Casey}, {Vavilkin}, {Dufault},
  {Larson}, {Barnes}, {Chan}, {Frayer}, {Haan}, {Inami}, {Ishida},
  {Kartaltepe}, {Melbourne}, \& {Petric}}]{U2012}
{U}, V., {Sanders}, D.~B., {Mazzarella}, J.~M., {et~al.} 2012, \apjs, 203, 9

\bibitem[{{van der Tak} {et~al.}(2007){van der Tak}, {Black}, {Sch{\"o}ier},
  {Jansen}, \& {van Dishoeck}}]{vanderTak2007}
{van der Tak}, F.~F.~S., {Black}, J.~H., {Sch{\"o}ier}, F.~L., {Jansen}, D.~J.,
  \& {van Dishoeck}, E.~F. 2007, \aap, 468, 627

\bibitem[{{van der Werf} {et~al.}(1993){van der Werf}, {Genzel}, {Krabbe},
  {Blietz}, {Lutz}, {Drapatz}, {Ward}, \& {Forbes}}]{vanderWerf1993}
{van der Werf}, P.~P., {Genzel}, R., {Krabbe}, A., {et~al.} 1993, \apj, 405,
  522

\bibitem[{{Veilleux} {et~al.}(2009){Veilleux}, {Rupke}, {Kim}, {Genzel},
  {Sturm}, {Lutz}, {Contursi}, {Schweitzer}, {Tacconi}, {Netzer}, {Sternberg},
  {Mihos}, {Baker}, {Mazzarella}, {Lord}, {Sanders}, {Stockton}, {Joseph}, \&
  {Barnes}}]{Veilleux2009}
{Veilleux}, S., {Rupke}, D.~S.~N., {Kim}, D.-C., {et~al.} 2009, \apjs, 182, 628

\bibitem[{{Wang} {et~al.}(2016){Wang}, {Zhang}, {Zhang}, {Shi}, \&
  {Fang}}]{Wang2016}
{Wang}, J., {Zhang}, Z.-Y., {Zhang}, J., {Shi}, Y., \& {Fang}, M. 2016, \mnras,
  455, 3986

\bibitem[{{Wang} {et~al.}(1991){Wang}, {Scoville}, \& {Sanders}}]{Wang1991}
{Wang}, Z., {Scoville}, N.~Z., \& {Sanders}, D.~B. 1991, \apj, 368, 112

\bibitem[{{Watson} {et~al.}(1976){Watson}, {Anicich}, \&
  {Huntress}}]{Watson1976}
{Watson}, W.~D., {Anicich}, V.~G., \& {Huntress}, Jr., W.~T. 1976, \apjl, 205,
  L165

\bibitem[{{Wilson} {et~al.}(2008){Wilson}, {Petitpas}, {Iono}, {Baker}, {Peck},
  {Krips}, {Warren}, {Golding}, {Atkinson}, {Armus}, {Cox}, {Ho}, {Juvela},
  {Matsushita}, {Mihos}, {Pihlstrom}, \& {Yun}}]{Wilson2008}
{Wilson}, C.~D., {Petitpas}, G.~R., {Iono}, D., {et~al.} 2008, \apjs, 178, 189

\bibitem[{{Wilson} {et~al.}(2014){Wilson}, {Rangwala}, {Glenn}, {Maloney},
  {Spinoglio}, \& {Pereira-Santaella}}]{Wilson2014}
{Wilson}, C.~D., {Rangwala}, N., {Glenn}, J., {et~al.} 2014, \apjl, 789, L36

\bibitem[{{Wilson} \& {Scoville}(1990)}]{Wilson1990}
{Wilson}, C.~D. \& {Scoville}, N. 1990, \apj, 363, 435

\bibitem[{{Wu} {et~al.}(2005){Wu}, {Evans}, {Gao}, {Solomon}, {Shirley}, \&
  {Vanden Bout}}]{Wu2005}
{Wu}, J., {Evans}, II, N.~J., {Gao}, Y., {et~al.} 2005, \apjl, 635, L173

\bibitem[{{Zinchenko} {et~al.}(2000){Zinchenko}, {Henkel}, \&
  {Mao}}]{Zinchenko2000}
{Zinchenko}, I., {Henkel}, C., \& {Mao}, R.~Q. 2000, \aap, 361, 1079

\bibitem[{{Zschaechner} {et~al.}(2016){Zschaechner}, {Ott}, {Walter}, {Meier},
  {Momjian}, \& {Scoville}}]{Zsch2016}
{Zschaechner}, L.~K., {Ott}, J., {Walter}, F., {et~al.} 2016, \apj, 833, 41

\end{thebibliography}
\bibliographystyle{aa.bst}

\begin{appendix}
\section{Continuum}
\begin{figure*}[!htbp] 
\centering
$\begin{array}{c@{\hspace{0.5in}}c}
\includegraphics[scale=0.3]{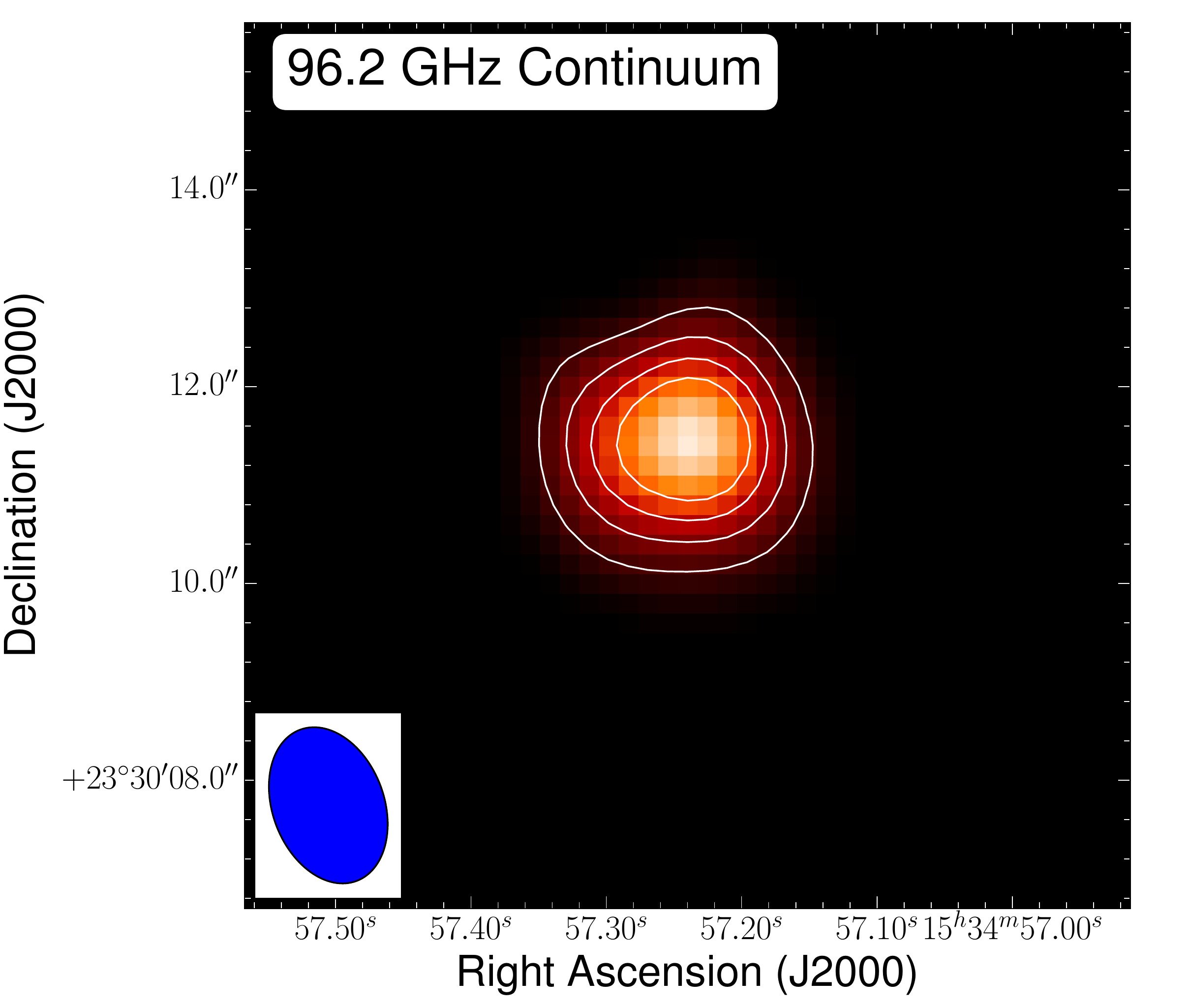}  & \includegraphics[scale=0.3]{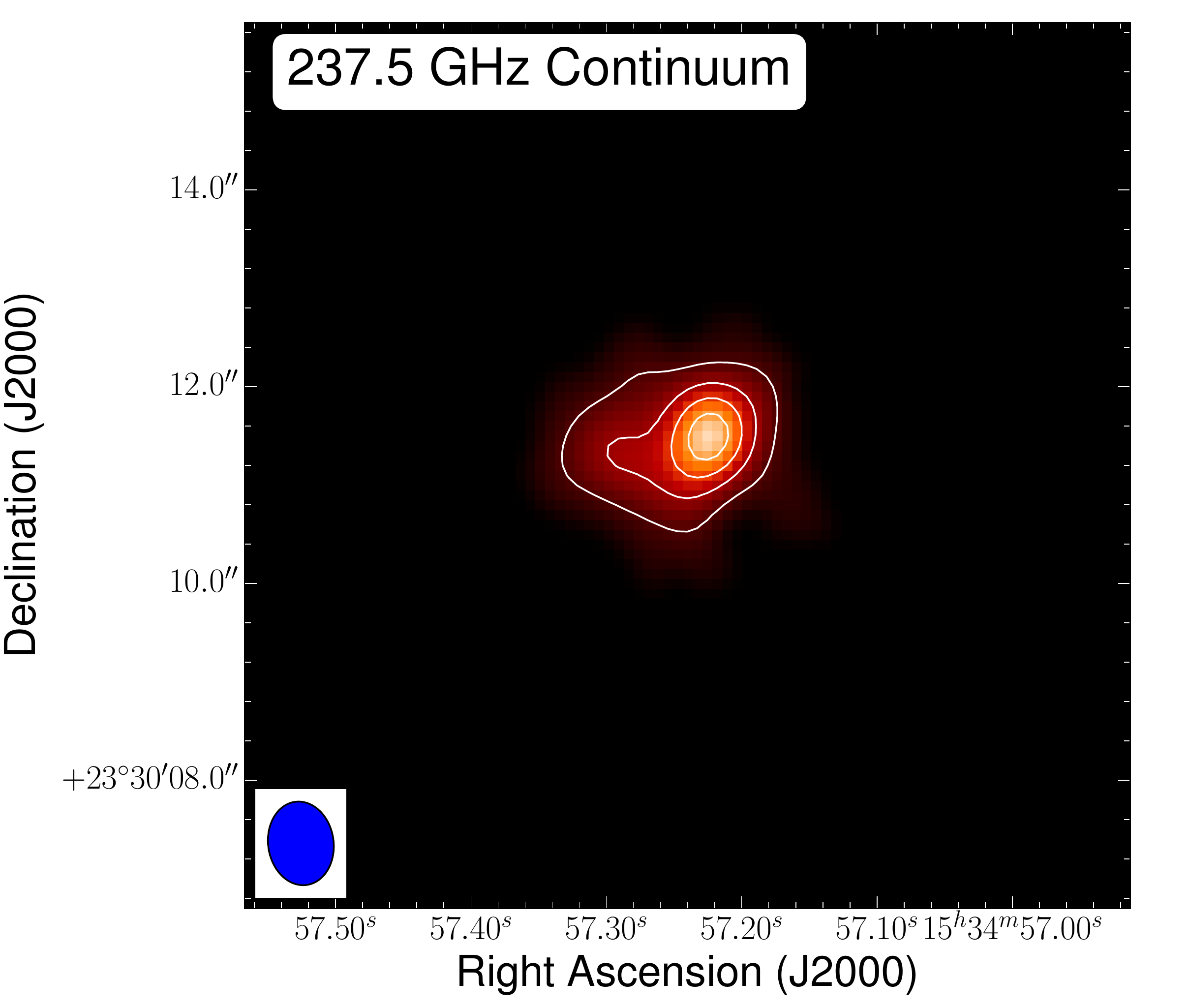} \\
\end{array}$
\includegraphics[scale=0.3]{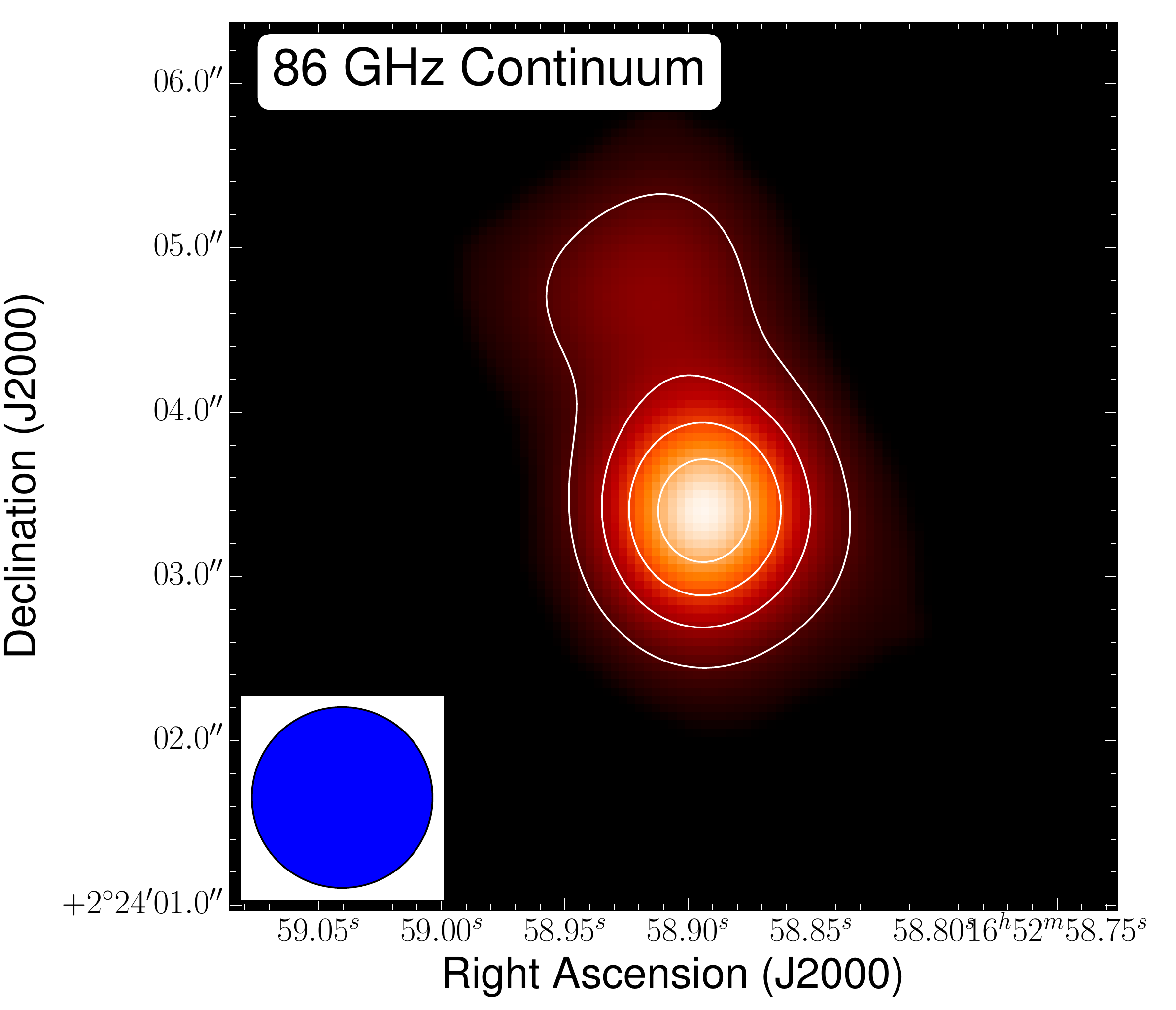}
\caption[]{ (TOP) Continuum at 96.2~GHz and 237.5~GHz for Arp~220 with contours corresponding to [3, 6, 9, 12] $\times$  1.0 and 4.4 mJy beam$^{-1}$, respectively. . (BOTTOM) Continuum at 86~GHz for NGC~6240 with contours corresponding to [3,6,9,12] $\times$ 0.32 mJy beam$^{-1}$}
\label{fig:continuum}
\end{figure*}

In this Section, we present the continuum data. Table \ref{tab:contemiss} presents the continuum measurements for Arp~220 and NGC~6240. Figure \ref{fig:continuum} presents the 3~mm and 1~mm continuum for Arp~220 and 3~mm for NGC~6240. 

\begin{table*}
\caption{Continuum Emission}\label{tab:contemiss}
\begin{center}
\begin{tabular}{lcccccc}
\hline \hline
		& \multicolumn{2}{c}{\underline{Continuum Position}}& &\multicolumn{2}{c}{\underline{Flux Density}} & Resolution\\ 
Source 	& RA 	& Dec 	& Frequency 			&Peak	& Total & \\
		& (J2000)		& (J2000)		&(GHz)	&(mJy beam$^{-1}$) 	& (mJy)  & (arcsec)\\
\hline 
Arp 220		&15 34 57.24 	& +23 30 11.2	&87.0	&23.0	&31.0		&1.3 $\times$ 0.8\\
			&			&			&96.2	&17.8	&26.0 		&1.6 $\times$  1.1	\\
			&			&			&105.1	&22.0	&25.0 		&2.0 $\times$  1.3\\
			&			&			&113.2	&30.0	&41.0 		&2.0 $\times$  1.3\\
			&			&			&219.5	&42.0	&99.0 		&0.9 $\times$  0.7\\
			&			&			&237.5	&62.0	&150.0 		&0.9 $\times$  0.7\\
			&			&			&		&		& \\

NGC 6240S	&16 52 58.893 	&+02 24 03.4 	&86.5	&4.6		&5.1 & 1.1 $\times$  1.1\\
NGC 6240N	&16 52 58.918	&+02 24 04.8	&86.5	&1.7		&1.8 &1.1 $\times$  1.1\\
NGC 6240	&	&	&86.5	&1.7		&6.9 & 1.1 $\times$  1.1\\
\hline
\end{tabular}
\end{center}
\tablefoot{Errors: Position = $\pm$0\arc.1, 3mm fluxes = $\pm$5$\%$, 1mm fluxes = $\pm$10$\%$}
\end{table*}

\section{Release of Data} \label{sec:release}
The PdBI data presented in this paper will be released via the \textbf{IRAM Large Program Archive} and NASA's Extragalactic Database apart of the PdBI U/LIRG Survey (PULS). Several other U/LIRGs were also observed (see Table \ref{tab:sources}) and will be released as well.  The data taken in between 2003 - 2008 and for most sources, the PdBI was setup to observe \hcnone\ at 3~mm and \tcotwo\ at 1.4~mm, simultaneously. In the weaker sources, \cotwo\ was observed instead of \tcotwo. The six 15~m antennas were arranged with spacings from 24~m to 400~m. The long baselines, observed in winter, had rms phase errors $\leq$ 40$^{\circ}$ at 1.4~mm, and $\leq$ 15$^{\circ}$ at 3~mm. Short spacings ($\leq$ 80~m), observed in summer at 3~mm, had rms phase errors $\leq$20$^{\circ}$.

A paper on the analysis of Arp~193 and VII~Zw~31 is forth coming (Sliwa \& Downes in preparation).
\begin{table*}
\caption{Sources Observed\label{tab:sources}}
\begin{center}
\begin{tabular}{cl}
\hline \hline
Source	& Lines \\
\hline 
IRAS 00057+4021 & HCN (1-0), \co\ (1-0), \co\ (2-1) \\
VII Zw 31 & HCN (1-0), \co\ (1-0), \co\ (2-1), \tco\ (1-0), \tco\ (2-1) \\
UGC 5101 & HCN (1-0), \co\ (1-0), \co\ (2-1) \\
IRAS 10565+2448 & HCN (1-0), \co\ (2-1) \\
Mrk 231 & HCN (1-0), HCO$^{+}$ (1-0), \co\ (2-1) \\
Arp 193 & HCN (1-0), \co\ (1-0), \co\ (2-1), \tco\ (2-1) \\
Mrk 273 & HCN (1-0), \co\ (1-0), \co\ (2-1), \tco\ (2-1) \\
IRAS 17208-0014 & HCN (1-0), \co\ (1-0), \co\ (2-1) \\
IRAS 23365+3604 & HCN (1-0), \co\ (1-0), \co\ (2-1) \\
\hline
\end{tabular}
\end{center}
\end{table*}

\end{appendix}

\end{document}